\documentclass[12pt]{article}
\usepackage{graphicx,amsthm,amsmath,amssymb,mathrsfs,url,hyperref,tikz}
\usepackage{float} 
\usepackage{placeins}   
\usepackage{mathtools}
\usepackage{caption}
\usepackage{booktabs}
\usepackage{siunitx}
\usepackage{graphicx}
\usepackage{caption}
\usepackage{subcaption}
\usepackage{comment}
\usepackage[T1]{fontenc}
\usepackage{xcolor}
\definecolor{StrongBlue}{RGB}{0,130,250}        
\definecolor{navylink}{RGB}{20,60,140}
\definecolor{RefBlue1}{RGB}{0,90,160}
\definecolor{RefTeal}{RGB}{0,110,130}
\hypersetup{
    colorlinks=true,
    linkcolor=RefBlue1,
    citecolor=RefBlue1,
    urlcolor=RefBlue1
}
\usepackage{makecell}
\usepackage[utf8]{inputenc} 

\title{Detection Time Distribution Predicted Using Absorbing Boundary Conditions and Imaginary Potentials}
\author{
Alireza Jozani\footnote{Fachbereich Physik,
	Eberhard-Karls-Universit\"at T\"ubingen, Auf der Morgenstelle 10,
	72076 T\"ubingen, Germany.
	E-mail: alireza.jozani@uni-tuebingen.de and jozani.alireza@gmail.com}~~~and~~Roderich Tumulka\footnote{Fachbereich Mathematik,
	Eberhard-Karls-Universit\"at T\"ubingen, Auf der Morgenstelle 10,
	72076 T\"ubingen, Germany.
	E-mail: roderich.tumulka@uni-tuebingen.de}
}
\date{June 25, 2026}

\usepackage{color}     
\definecolor{BKorange}{HTML}{D55E00}

\renewcommand{\Re}{\operatorname{Re}}
\renewcommand{\Im}{\operatorname{Im}}
\newcommand{\vsigma}{\boldsymbol{\sigma}}
\newcommand{\vj}{\boldsymbol{j}}
\newcommand{\vk}{\boldsymbol{k}}
\newcommand{\vn}{\boldsymbol{n}}
\newcommand{\vr}{\boldsymbol{r}}

\newcommand{\vv}{\boldsymbol{v}}
\newcommand{\vQ}{\boldsymbol{Q}}
\newcommand{\vR}{\boldsymbol{R}}
\newcommand{\vX}{\boldsymbol{X}}

\newcommand{\fr}{\mathrm{free}}
\newcommand{\be}{\begin{equation}}
\newcommand{\ee}{\end{equation}}

\newcommand{\EEE}{\mathbb{E}}
\newcommand{\PPP}{\mathbb{P}}
\newcommand{\RRR}{\mathbb{R}}
\theoremstyle{definition}
\newtheorem{rem}{Remark}

\begin{document}
\maketitle
\begin{abstract}
There are several inequivalent proposals in the literature for how to compute the probability distribution of the time that a detector registers for the arrival of a quantum particle. For three of these proposals, based on two kinds of absorbing boundary conditions and imaginary potentials, we compute the predicted distribution for an experimental setup 
involving a single non-relativistic quantum particle with spin 0 or 1/2 in a wave guide along the $z$ axis with the detector waiting downstream. We find that the distribution shows signs of partial reflection of the wave function off of the detector; for a spin-1/2 wave function, it is independent of the initial spin orientation for the parameters tested but does depend, for boundary conditions coupling to the spin, on the width of the wave guide. We also compare our predictions with the competing ones of Das and D\"urr [Sci.\ Rep.\ 9: 2242 (2019)].

\medskip

\end{abstract}

\section{Introduction}

Consider a detector waiting for a quantum particle to arrive. There has been a long debate over how to compute the probability distribution of the time of detection (e.g., \cite{All69b,ML00,VHD13}), as the standard quantum formalism fails to provide a rule for it (see Remark~\ref{rem:OQM} at the end of this section). Several mutually inequivalent proposals have been made, e.g., \cite{ML00,AB61,Kij74,detect-rule,DD19}; they tend to agree in the scattering (far-field) regime \cite{Daumer,KT26} but not in general. Experimental tests of some proposals appear to be within reach in the near future. Here, we compute numerically the predictions of the two most promising proposed rules, based on absorbing boundary conditions (ABCs) \cite{detect-rule}, \cite[Sec.~5.2]{Tum22} and complex absorbing potentials (CAPs) \cite{All69b}, for an experimental setup suggested by Das and D\"urr \cite{DD19}. We also compare these predictions to those that Das and D\"urr \cite{DD19} proposed starting from a naive application of Bohmian mechanics. Our predictions are particularly different in the case of a spin-1/2 particle.

The ABC and CAP proposals are distinguished from most others (such as \cite{DD19}) by their property that the distribution of detection time $T$ and place $\vR$ that they predict is given by a positive-operator-valued measure (POVM) $E$, i.e., that
\be\label{TXPOVM}
\mathbb{P}\Bigl[ (T,\vR)\in B \Bigr] = \langle \psi|E(B)|\psi\rangle
\ee
for any set $B$ and $\psi$ the initial wave function of the quantum particle \cite{GTZ23,GTZ24}. This is significant in view of the general theorem \cite{DGZ04}, \cite[Sec.~5.1]{Tum22}, \cite{BL25} that for any quantum experiment, there is a POVM $E$ such that, when the experiment is carried out on a system with wave function $\psi$, the probability distribution of the outcome is $\langle\psi|E(B)|\psi\rangle$. That is, most other proposals are excluded by theoretical reasons, while ABCs and CAPs are not. A further relevant theoretical consideration is that proposals that are not of the POVM form imply the possibility of superluminal signaling \cite{Gi89,GTZ24}, while superluminal signaling is excluded for quantum mechanics \cite{Ebe78}, as well as for quantum mechanics with ABC and/or CAP \cite{TZZ24,PTZT26}. (Likewise for Bohmian mechanics, as its empirical predictions agree with those of standard quantum mechanics.)

The ABC rule was suggested in \cite{detect-rule} and uses a boundary condition considered before in \cite{Wer87}; further studies of it can be found in \cite{detect-dirac,Tum19,DBD20,FTT25,Fro25}. Imaginary potentials have been used early on \cite{Bet40} for modeling absorption and at least since 1969 \cite{All69b} for modeling detectors. Imaginary potentials behave like ``soft'' detectors that take a while to notice a particle traveling in the detector volume, while an absorbing boundary is intended as a ``hard'' detector along a surface $\Sigma$ that gets triggered as soon as the particle reaches $\Sigma$. ABCs can be obtained as a suitable limit of CAPs \cite{Tum19}. 

Our aim here is to obtain concrete predictions from ABC and CAP that can be compared to future experiments. We follow a proposal of Das and D\"urr \cite{DD19} for how such an experiment could be set up: a particle can move in a 3d waveguide whose axis will be called the $z$ axis, bounded by a Dirichlet wall (infinite potential step) at $z=0$ and a detector at $z=L>0$. In between, the Hamiltonian is given by
\be\label{Hdef}
H=-\tfrac{\hbar^2}{2m}\nabla^2 + \tfrac{m}{2}\omega^2 (x^2+y^2)\,,
\ee
featuring a harmonic wave guide potential. The particle gets released near $z=0$ from a trap between $z=0$ and $z=d$, where $d$ is much smaller than $L$. Specifically, the initial wave function $\Psi_0$ is prepared as the ground state $\sin(\pi z/d)$ of an infinite potential well in the interval $[0,d]$ (i.e., of the free Schr\"odinger operator with Dirichlet boundary conditions at $z=0$ and $z=d$), times the ground state $g(x,y)$ of the transverse wave guide potential. We study here the cases of spin-0 and spin-1/2, and in the latter case the strength of the dependence of the detection time distribution \eqref{TXPOVM} on the initial spin state. We study three detector models, one CAP and two specific ABCs: 
\begin{itemize}
\item[(i)] an imaginary potential in the region $z_0<z<L$ (where $z_0$ is much larger than $d$); 
\item[(ii)] the \emph{spin-decoupled ABC} (or briefly ``spinless ABC'') \cite{detect-rule}
\be\label{nospinABC}
\vn \cdot \nabla \Psi = i\kappa \Psi
\ee
with $\vn$ the outer unit normal vector to the boundary and $\kappa>0$ a constant characterizing the type of detector (wave number of maximal sensitivity); and \item[(iii)] the \emph{spin-coupled ABC} (or briefly ``spinor ABC'') \cite{detect-dirac}
\be\label{spinABC}
\vsigma\cdot\nabla\Psi = i\kappa \vn\cdot\vsigma \Psi
\ee
with $\vsigma=(\sigma_x,\sigma_y,\sigma_z)$ the triple of Pauli matrices and $\kappa>0$ and $\vn$ as before. 
\end{itemize}
This finite absorbing termination of the $z$ axis leads to a non-self-adjoint initial--boundary value problem, which is studied via numerical solutions of the time-dependent Schr\"odinger equation (TDSE). For elucidation of why CAPs and ABCs are reasonable as models of detectors, see, e.g., \cite{detect-rule,detect-several,Fro25}, \cite[Sec.~2.3]{KT26}. It would be of interest to determine of common types of real-world detectors whether they are better regarded as hard or soft, and with which parameters.

The motivation for \eqref{nospinABC} is that it is the simplest ABC, and that detectors are not designed to couple to spin, the motivation for \eqref{spinABC} that it is the non-relativistic limit of the simplest ABC for the Dirac equation that possesses a non-relativistic limit \cite{detect-dirac}; its spin-dependence arises from the fact that in the Dirac equation, already the free time evolution couples to spin. In addition, for a spin-1/2 particle, \eqref{spinABC} is more natural from the point of view of Bohmian mechanics, as it represents a detecting surface that is ideal in the sense that it registers the particle precisely when and where it first arrives (as explained in Remark~\ref{rem:TWID} in Section~\ref{sec:Bohmian}). 

It can be derived that for an imaginary potential and \eqref{nospinABC}, the detection time distribution does not depend on the initial spinor. For \eqref{spinABC}, we find that the dependence is weak and hardly noticeable, 
in contrast to a pronounced spin dependence in the prediction of Das and D\"urr \cite{DD19}.

The numerical solution of the TDSE for complex-valued or spinor-valued (two-component) wave functions is obtained by means of a Crank--Nicolson method \cite{CN47} (an implicit Runge--Kutta method) in a 3d box in Cartesian coordinates. For the comparison with the predictions of Das and D\"urr, which are based on the unwarranted hypothesis that the detection time agrees with the time at which the Bohmian trajectory would arrive at $z=L$ \emph{in the absence of detectors}, we also computed the Bohmian trajectories in the absence of detectors (with results in agreement with those reported by Das and D\"urr).

Apart from the dependence on the initial spinor, we also pay attention to the dependence of the detection time distribution on the width of the wave guide, as controlled by the parameter $\omega$ of the harmonic confining potential in \eqref{Hdef}. There is no dependence for CAP and \eqref{nospinABC}, but a relevant dependence for \eqref{spinABC} that we describe in Section~\ref{subsec:spinorABC} and  study further in a separate work \cite{JT26b}.

\begin{rem}\label{rem:OQM}
    Concerning theoretical approaches to the detection time distribution: The question of determining this distribution can be studied in both Bohmian and orthodox quantum mechanics in two ways: either by trying to guess a formula/operator/POVM for it, or by trying to analyze a microscopic quantum-mechanical model of the detector particles and their interaction with the observed particle. The former approach has led to numerous proposals, e.g.,  \cite{All69b,ML00,AB61,Kij74,detect-rule,DD19}, and different interpretations of quantum mechanics have inspired different guesses. The latter approach should in principle decide the question on theoretical grounds, but it is difficult to carry out (for example, because decoherence plays a role but is not sufficiently accounted for in highly simplified models using a single particle as a detector \cite{AOPRU98}), so while some progress has been made \cite{All69b,AOPRU98,Hal99,detect-rule}, a conclusive analysis is still missing. The collapse of the wave function is (intended to be) effectively included in the CAP and ABC proposals, while it is neglected in \cite{DD19}.\hfill$\diamond$
\end{rem}

\bigskip

The remainder of this paper is organized as follows. In Section~\ref{sec:setup}, we describe in more detail the equations that we have simulated. In Section~\ref{sec:results}, we show figures of our results. In Section~\ref{sec:experimental}, we comment on possible experimental realizations, and in Section~\ref{sec:outlook} we conclude. In Appendices~\ref{app:TDSE}--\ref{app:arrival}, we describe the numerical methods used, and in Appendices~\ref{app:transform} and \ref{app:rate} we collect further calculations. In Appendix~\ref{app:Bohmianresults}, we show simulation results of what the Bohmian trajectories look like, and in Appendix~\ref{app:variation} we include some further observations about how the probability  distribution of the detection time changes when the parameters are varied.

\section{Setup: Geometry, Parameters, and Initial State}
\label{sec:setup}

We first describe the mathematical model in Section~\ref{sec:model}, then explain how the relevant distribution is related to the probability current (Section~\ref{sec:current}) and the Bohmian trajectories (Section~\ref{sec:Bohmian}),  add a few comments about the numerical implementation (Section~\ref{sec:setupnum}), and finally in Section~\ref{sec:comparisoncurve} define the curve $\rho^\fr(t)$ that we use for comparison.

\subsection{Mathematical Model}
\label{sec:model}

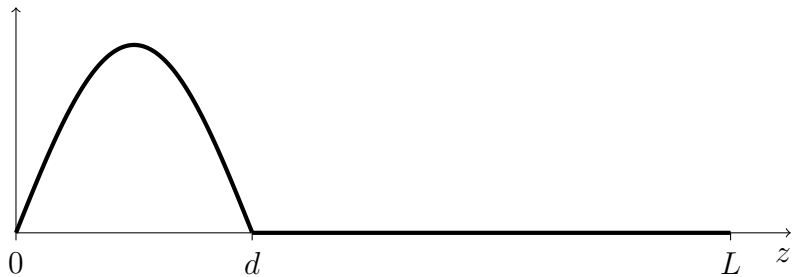
\begin{figure}[ht]
\begin{center}
\begin{tikzpicture}
\draw[->] (0,-0.1) -- (0,3);
\draw[->] (0,0) -- (10.3,0);
\node at (0,-0.4) {0};
\draw (3.14,-0.1) -- (3.14,0);
\node at (3.14,-0.4) {$d$};
\draw (9.5,-0.1) -- (9.5,0);
\node at (9.5,-0.4) {$L$};
\node at (10.2,-0.3) {$z$};
\draw[ultra thick,domain=0:3.14, samples=101] plot(\x,{2.5*sin(\x r))});
\draw[ultra thick] (3.14,0) -- (9.5,0);
\end{tikzpicture}
\end{center}
\caption{Graph of the function $1_{0<z<d} \, \sin(\pi z/d)$, which is the $z$ dependence of the initial wave function $\Psi_0$ as in \eqref{initialpsi}}
\label{fig:initialpsi}
\end{figure}

The region available to the particle's wave function is
\be
\Omega= \RRR^2 \times [0,L]\,,
\ee
and the wave function $\Psi(\vr)$ with $\vr=(x,y,z)$ is taken to be either, in the spin-0 case, complex-valued or, in the spin-1/2 case, spinor-valued (having two components) as in 
\be
\Psi(\vr)=
\begin{pmatrix}
\psi_\uparrow(\vr)\\[3pt]
\psi_\downarrow(\vr)
\end{pmatrix} \,.
\ee
The initial wave function at $t=0$ is assumed to be, in the spin-1/2 case, 
\be\label{initialpsi}
\Psi_0(\vr) = |\chi \rangle \sqrt{\frac{2m\omega}{\pi\hbar d}} \, \exp\Bigl(-\frac{m\omega}{2\hbar}(x^2+y^2)\Bigr) \: 1_{0<z<d} \: \sin\Bigl(\frac{\pi z}{d}\Bigr)
\ee
with
\be\label{chidef}
|\chi\rangle = \begin{pmatrix}
\chi_\uparrow\\[3pt]
\chi_\downarrow
\end{pmatrix} 
= \begin{pmatrix}
\cos(\theta/2)\\[3pt]
\sin(\theta/2)e^{i\phi}
\end{pmatrix} \in \mathbb{C}^2
\ee
a constant unit spinor with Bloch angles $\theta\in[0,\pi]$ and $\phi\in[0,2\pi)$; in other words, $\theta$ and $\phi$ are the spherical coordinates of the direction vector $\langle\chi|\vsigma|\chi\rangle \in \RRR^3$. In the spin-0 case, \eqref{initialpsi} applies without the spinor factor $|\chi\rangle$. Eq.~\eqref{initialpsi} means that the initial state is prepared analogously to the laboratory baseline used in the literature: the transverse harmonic ground state multiplied by the longitudinal ground state in the slab $0<z<d$, depicted in Figure~\ref{fig:initialpsi}.

The time evolution will be different for the three detector models we consider, as follows.

\subsubsection{Imaginary Potential}
\label{sec:CAP}

\begin{figure}[t]
  \centering
  \includegraphics[width=0.65\linewidth]{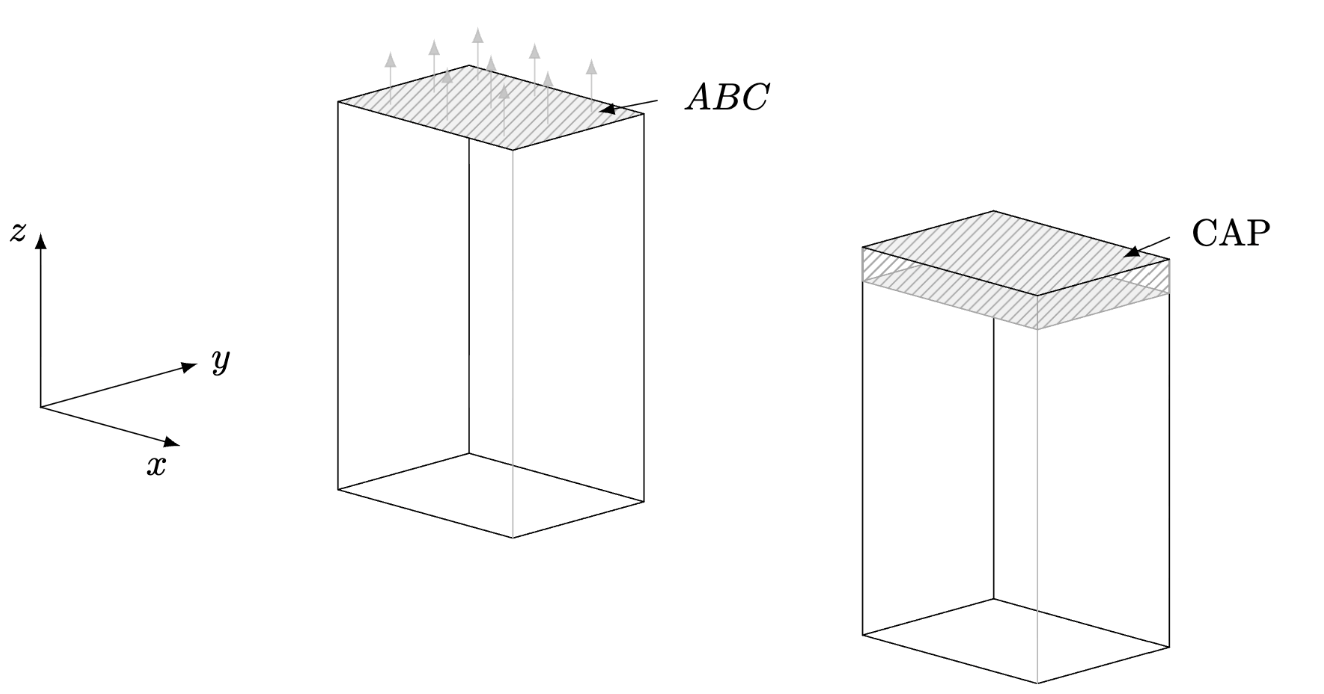}
  \caption{Schematic depiction of detector geometry: in models (ii) and (iii), an absorbing boundary condition (ABC) is imposed at the roof $z=L$, in model (i) a complex absorbing potential (CAP) in a layer near the roof.}
  \label{fig:abc_schematic_cap}
\end{figure}

The potential $V$ in the Schr\"odinger equation
\be\label{Schr}
    i\hbar\,\partial_t \Psi(\vr,t)
    =\left(-\dfrac{\hbar^2}{2m}\nabla^2+V(\vr)\right)\Psi(\vr,t)
\ee
for $\vr\in\Omega$ and $t\geq 0$ contains two contributions, a confining wave guide potential (harmonic in $x$ and $y$) and a negative-imaginary absorbing potential,
\be\label{Vdef}
    V(x,y,z) = \frac{m}{2} \omega^2 (x^2+y^2) -i W(z)
\ee
with $W(z)\geq 0$ the detector strength; it vanishes (approximately) outside the detector region, which we take to be the layer $z_0\leq z \leq L$ with $z_0$ near $L$ (see Figure~\ref{fig:abc_schematic_cap}). We might take $W(z)$ to be constant within this layer, but a smooth step (a ramp over a finite width such as tanh in \eqref{Wtanh} or a polynomial piece in \eqref{Wcubic}) is perhaps more realistic and advantageous as it suppresses broadband reflections better than a discontinuous step.

The Schr\"odinger equation \eqref{Schr} is complemented by (reflecting) Dirichlet boundary conditions at $z=0$ and $z=L$,
\be\label{Dir0L}
    \Psi_t(x,y,0)=0\,, ~~\Psi_t(x,y,L)=0
\ee
for all $x,y$, and $t$.
This time evolution can be applied to either spin-0 or spin-1/2 wave functions; in the latter case, the Hamiltonian does not couple to the spin, so $\Psi_t(\vr)$ is always of the form $|\chi\rangle \, \psi_t(\vr)$ with scalar-valued $\psi_t(\vr)$.

As appropriate for CAPs, we take the joint probability distribution of the detection time $T$ and place $\vR$ to be given by \cite{All69b,detect-several}
\be\label{PTR}
    \PPP(T\in dt, \vR \in d^3\vr) = \frac{2}{\hbar} \, W(z) \, |\Psi_t(\vr)|^2 \, dt \, d^3\vr\,.
\ee
As a consequence, the distribution of $T$ alone is given by
\be\label{PT}
    \PPP(T\in dt) = \frac{2}{\hbar} dt \int_\Omega d^3\vr \, W(z) \, |\Psi_t(\vr)|^2  =: \rho_T(t) \, dt\,.
\ee

Since an imaginary potential $-iW(\vr,t)$ leads to an additional term in the continuity equation,
\be\label{continuityW}
\partial_t \rho(\vr,t) +\nabla \cdot \vj(\vr,t) =- \frac{2}{\hbar} W(\vr,t)  \, \rho(\vr,t)
\ee
with $\rho=|\Psi|^2$ and probability current
\be\label{jdefnospin}
\vj = \frac{\hbar}{m} \Im(\Psi^* \nabla \Psi)\,,
\ee
the time evolution in the presence of a negative-imaginary potential is not unitary but satisfies instead
\be\label{normderivative}
    \frac{d}{dt}\|\Psi_t\|^2 = - \frac{2}{\hbar} \int_\Omega d^3\vr \, W(\vr,t) |\Psi_t(\vr)|^2 \leq 0\,,
\ee
where
\be\label{normdef}
    \|\Psi\|^2 := \int_\Omega d^3\vr \, |\Psi(\vr)|^2\,.
\ee
Thus, the probability density of $T$ can also be expressed as
\be\label{rhoTt}
    \rho_T(t) = -\frac{d}{dt}\|\Psi_t\|^2 \,,
\ee
while $\|\Psi_t\|^2$ equals the probability that the particle has not been detected yet,
\be\label{PTt}
    \PPP(T>t) = \|\Psi_t\|^2\,.
\ee

\subsubsection{Spinless ABC}

In our second model, the Schr\"odinger equation \eqref{Schr} is used without the imaginary potential, retaining only the confining potential
\be\label{harmonicV}
V(x,y,z) =\frac{m}{2} \omega^2 (x^2+y^2)\,.
\ee
It is complemented by a Dirichlet boundary condition at $z=0$,
\be\label{Dir0}
\Psi_t(x,y,0) = 0
\ee
and the spinless ABC \eqref{nospinABC} on the plane $z=L$ (so that $\vn=\hat{\boldsymbol{z}}$), which means explicitly
\be\label{nospinABCzL}
\partial_z \Psi_t(x,y,L) = i\kappa \Psi_t(x,y,L)
\ee
for all $x,y$, and $t$. Again, this time evolution can be applied to either spin-0 or spin-1/2 wave functions; in the latter case, the Hamiltonian does not couple to the spin, so $\Psi_t(\vr)$ is always of the form $|\chi\rangle \, \psi_t(\vr)$ with scalar-valued $\psi_t(\vr)$ (and $|\chi\rangle$ independent of $\vr$ and $t$).

In this setup, the joint probability distribution of the detection time and place is given by \cite{detect-rule}
\be\label{PTRnospin}
\PPP(T\in dt, \vR\in d^3\vr) = \frac{\hbar\kappa}{m} \, |\Psi_t(\vr)|^2 \, \delta(z-L) \, dt \, d^3\vr \,,
\ee
so that the distribution of $T$ alone is given by
\be\label{PTnospin}
\PPP(T \in dt) = \frac{\hbar\kappa}{m}dt \int dx \int dy\, |\Psi_t(x,y,L)|^2 =: \rho_T(t) \, dt \,.
\ee

Again, the time evolution is not unitary; rather, it satisfies
\begin{align}\label{normderivativenospin}
\frac{d}{dt}\|\Psi_t\|^2 
&= -\int dx \int dy \, \frac{\hbar}{m} \, \Im(\Psi^*_t \partial_z \Psi_t)(x,y,L)\\
&\!\! \stackrel{\eqref{nospinABCzL}}{=} - \frac{\hbar\kappa}{m}\int dx \int dy \, |\Psi_t(x,y,L)|^2 \leq 0 \,.
\end{align}
As a consequence, \eqref{rhoTt} and \eqref{PTt} are valid also in this case.

\subsubsection{Spinor ABC}

This version works only with spinor-valued wave functions and replaces the ABC \eqref{nospinABCzL} with the spinor ABC \eqref{spinABC} on the plane $z=L$ (so that $\vn=\hat{\boldsymbol{z}}$); it can be rewritten more explicitly as
\be\label{spinABCzL}
\begin{aligned}
\partial_z \psi_\uparrow &= -\big(\partial_x - i\,\partial_y\big)\,\psi_\downarrow + i\,\kappa \,\psi_\uparrow,\\
\partial_z \psi_\downarrow &= \big(\partial_x + i\,\partial_y\big)\,\psi_\uparrow + i\,\kappa \,\psi_\downarrow.
\end{aligned}
\ee
Here, the boundary acts as a \emph{matrix} impedance.
Eq.s \eqref{PTRnospin}, \eqref{PTnospin}, \eqref{rhoTt}, and \eqref{PTt} are valid also in this case.

\subsection{Probability Current}
\label{sec:current}

The distribution of $T$ and $\vR$ can also be expressed in terms of the \emph{probability current} $\vj$. In the spin-0 case, $\vj$ is given by \eqref{jdefnospin}, in the spin-1/2 case by the \emph{Pauli current}
\be\label{Paulij}
\vj^\mathrm{P}=\frac{\hbar}{m} \Im(\Psi^\dagger \nabla \Psi) + \frac{\hbar}{2m} \nabla \times (\Psi^\dagger \vsigma \Psi)
\ee
(obtained as the non-relativistic limit of the Dirac current $\overline{\psi}\gamma^\mu\psi$). The first term in \eqref{Paulij} is sometimes called the convective current
\be\label{jCdef}
\vj^\mathrm{C}= \frac{\hbar}{m}\Im(\Psi^\dagger \nabla \Psi)\,,
\ee
the second the curl term or Gordon term. 

For the spinless ABC \eqref{nospinABC} or \eqref{nospinABCzL}, the distribution \eqref{PTRnospin} of $T$ and $\vR$ can equivalently be written as (for either spin-0 or spin-1/2 wave functions)
\be\label{PTRj}
\PPP(T\in dt,\vR \in d^3\vr) = \vn \cdot \vj^\mathrm{C} \, \delta(z-L) \, dt \, d^3\vr\,,
\ee
which is shown in Appendix~\ref{app:transform} to agree with \eqref{PTRnospin}. Likewise, for the spinor ABC \eqref{spinABC} or \eqref{spinABCzL}, the distribution of $T$ and $\vR$ can equivalently be written as
\be\label{PTRjP}
\PPP(T\in dt,\vR \in d^3\vr) = \vn \cdot \vj^\mathrm{P} \, \delta(z-L) \, dt \, d^3\vr\,,
\ee
as shown also in Appendix~\ref{app:transform}. This is connected to the mathematical fact that the spinless ABC \eqref{nospinABC} forces $\vj^\mathrm{C}$ to be outward-pointing (but not necessarily, for a spin-1/2 wave function, $\vj^\mathrm{P}$), whereas the spinor ABC \eqref{spinABC} forces $\vj^\mathrm{P}$ to be outward-pointing (but not necessarily $\vj^\mathrm{C}$).

\begin{rem}\label{rem:jCjPequalrhoT}

It is worth noting that when the $z$ component of \eqref{Paulij} is integrated over a plane parallel to the $xy$ plane, the curl term vanishes:
\be\label{vanishcurl}
\int dx \int dy \, j_z^\mathrm{P}(x,y,c) = \int dx \int dy \, j^\mathrm{C}_z(x,y,c)
\ee
for any constant $c\in\RRR$. (That is because the $z$ component of the curl consists of $\partial_x(\Psi^\dagger \sigma_y\Psi)-\partial_y(\Psi^\dagger \sigma_x\Psi)$; the first term, when integrated over $x$, just yields the values of $\Psi^\dagger \sigma_y\Psi$ at $x=\pm\infty$ (which are 0 because $\Psi$ vanishes at infinity); the second term, when integrated over $y$, vanishes for the same reason.) 

As a consequence of \eqref{vanishcurl}, for both the spinless and the spinor ABC, the distribution of $T$ can be expressed using either $\vj^\mathrm{C}$ or $\vj^\mathrm{P}$:
\be\label{rhoTjCjP}
\rho_T(t) = \int dx \int dy \, j^\mathrm{C}_z(x,y,L,t) = \int dx \int dy \, j^\mathrm{P}_z(x,y,L,t)\,.
\ee
Note, however, that this formula yields different functions $\rho_T(t)$ for the spinless and the spinor ABC because the two ABCs lead to different wave functions $\Psi$ (to which then the formula \eqref{Paulij} or \eqref{jCdef} can be applied).
\hfill$\diamond$
\end{rem}

\subsection{Bohmian Trajectories}
\label{sec:Bohmian}

In the case of a detecting surface $\Sigma$ (such as the plane $z=L$), the distribution of detection time $T$ and place $\vR$ can also be expressed as the distribution of the arrival time and place of the Bohmian trajectories at $\Sigma$ (provided the wave function $\Psi$ used evolves with ABC, unlike the freely evolving wave function used by Das and D\"urr \cite{DD19}).

Bohmian trajectories $\vQ(t)= (Q_x(t),Q_y(t),Q_z(t))$ are the solutions of Bohm's equation of motion \cite{Bohm52,DGZ13}
\be\label{Bohm}
\frac{d\vQ}{dt}(t)=\vv(\vQ(t),t)
\ee
using the velocity vector field
\be\label{vdef}
\vv(\vr,t) = \frac{\vj(\vr,t)}{|\Psi(\vr,t)|^2}
\ee
and initial condition $\vQ(0)$ sampled from the Born distribution $|\Psi(\vr,0)|^2$. It follows that at every time $t$, $\vQ(t)$ has Born distribution $|\Psi(\vr,t)|^2$ \cite{Bohm52,DGZ13}. For spin-1/2 particles, the Pauli current $\vj^\mathrm{P}$ should be used for $\vj$ because it is the non-relativistic limit of the Dirac current. The arrival time of the Bohmian trajectory at a surface $\Sigma$ is given by
\be\label{eqn:tau}
\tau= \inf \{t>0:\vQ(t)\in\Sigma\}\,,
\ee
which of course is a function $\tau(\vQ_0)$ of the initial position $\vQ_0=\vQ(0)$; the arrival place is $\vQ(\tau)\in\Sigma$. The probability distribution of the arrival time and place at $\Sigma$ is
\begin{align}
&\PPP(\tau \in dt, \vQ(\tau) \in d^3\vr)= \nonumber\\
&\qquad = dt\, d^3\vr \int_\Omega d^3\vQ_0 \, |\Psi(\vQ_0,0)|^2 \, \delta(t-\tau(\vQ_0)) \, \delta^3(\vr-\vQ(\tau(\vQ_0)))\\
&\qquad = dt \, d^3\vr \, \delta_\Sigma(\vr) \, \vn \cdot \vj(\vr,t) \, f(\vr,t)\,,\label{arrivalf}
\end{align}
where $f(\vr,t)=1$ if the trajectory passing through $\vr$ at $t$ is crossing $\Sigma$ for the first time and $f(\vr,t)=0$ if it is crossing $\Sigma$ for the second or later time.

As is well known \cite{detect-rule} and was mentioned after \eqref{PTRjP}, ABCs force suitable currents $\vj$ to point outward on $\Sigma$, i.e., $\vn\cdot \vj \geq 0$. As a consequence, no trajectory can intersect $\Sigma$ multiple times, and $f=1$, so the $f$ factor can be dropped in \eqref{arrivalf}. For $\Sigma=\{z=L\}$, we thus obtain that
\be\label{TWID}
\PPP(\tau \in dt, \vQ(\tau) \in d^3\vr) = \vn \cdot \vj \, \delta(z-L) \, dt\, d^3\vr
\ee
with
\be\label{jCorP}
\vj=\begin{cases}\vj^\mathrm{C}&\text{for \eqref{nospinABC}}\\ \vj^\mathrm{P}&\text{for \eqref{spinABC},}
\end{cases}
\ee
which agrees in both cases with the distribution of the detection time and place, according to \eqref{PTRj} and \eqref{PTRjP}.

\begin{rem}\label{rem:TWID}
This fact has consequences for the question \cite{GTZ24} whether the detection time agrees with the arrival time; or more generally, whether the time and place of detection (denoted in \cite{GTZ24} by $T_D$ and $\vX_D$, here $T$ and $\vR$) agree with the arrival time and place of the Bohmian trajectory (denoted in \cite{GTZ24} by $T_{WID}$ and $\vX_{WID}$ for ``with detector,'' here $\tau$ and $\vQ(\tau)$): While there are several reasons why, even in a Bohmian universe, $T=T_D$ may differ from $\tau=T_{WID}$ (detectors might not immediately register the particle, have limited resolution or other inaccuracies, or probability might flow back from post-detection to pre-detection configurations), it would be desirable for good detectors that they click when the particle arrives, i.e., that $T=\tau$. However, according to \eqref{TWID} and \eqref{jCorP}, whether the distribution of $(T,\vR)$ agrees with that of $(\tau,\vQ(\tau))$ depends on which ABC, \eqref{nospinABC} or \eqref{spinABC}, gets combined with which current in the equation of motion, \eqref{Paulij} or \eqref{jCdef}. In this sense, \eqref{nospinABC} is appropriate for spin-0 particles, but \eqref{spinABC} appears more appropriate than \eqref{nospinABC} for spin-1/2 particles.\footnote{\label{fn:vanishcurl}It might seem that \eqref{vanishcurl} implies that when $\vR=\vX_D$ and $\vQ(\tau)=\vX_{WID}$ are integrated out and only $T=T_D$ and $\tau=T_{WID}$ are considered, their distributions coincide; that is not the case because, in the presence of the spinless ABC \eqref{nospinABC}, $j_z^\mathrm{P}$ at $z=L$ can be negative, so that the $f$ factor in \eqref{arrivalf} is not always 1, so \eqref{arrivalf} integrated over $\vr$ is not the same expression as \eqref{vanishcurl}, and \eqref{rhoTjCjP} not the same as the density of $\tau=T_{WID}$.}
\hfill$\diamond$
\end{rem}

Also for CAPs, Bohmian trajectories can be used for sampling the distribution of $T$ and $\vR$ by following a trajectory and using that the detection rate is $\tfrac{2}{\hbar}W(Q_z(t))$. This means that the trajectory that we consider begins at $t=0$ and ends at a random time $T$ (at which the particle gets detected and will no longer be considered in the ensemble of pre-detection particles). In particular, the detection occurs not on the surface $z=L$ but at a random place in the region $z_0\leq z \leq L$.
We verify in Appendix~\ref{app:rate} that for any $t$ the probability that the trajectory has not ended yet and is located in $d^3\vr$ is given by $|\Psi_t(\vr)|^2$, and the joint distribution of the time $T$ and place $\vR$ where it ends agrees with the formula given earlier in \eqref{PTR} for the detection time and place.

\subsection{Numerical Implementation}
\label{sec:setupnum}

For the numerical implementation, we used Cartesian rather than cylindrical coordinates. This proved easier and more accurate for the finite-difference method we employed. That is, we used an evenly spaced 3D lattice $(\varepsilon_x \mathbb{Z})\times (\varepsilon_y \mathbb{Z})\times(\varepsilon_z \mathbb{Z})$ of grid points, restricted to the rectangular domain
\be
\Omega=[-L_x/2,L_x/2]\times[-L_y/2,L_y/2]\times[0,L_z],
\ee
with $L_z=L$.
We also impose Dirichlet conditions at the lateral walls $x=\pm L_x/2$ and $y=\pm L_y/2$; however, due to the confining potential, the wave function should not reach these walls, which was confirmed a posteriori in the simulations. Since the width of the wave guide is of order $\sqrt{\hbar/m\omega}$, we choose $L_x$ and $L_y$ proportional to (but much larger than) $\sqrt{\hbar/m\omega}$, and adjust $\varepsilon_x,\varepsilon_y$ proportionally (and much smaller than $\sqrt{\hbar/m\omega}$) so as to keep the number of grid points in the $x$ and $y$ direction constant.

Imposing the boundary conditions described above, we solve the TDSE using a second-order Crank--Nicolson (CN) finite-difference scheme. For self-adjoint discretizations, CN is unconditionally stable, in the sense that it has no CFL-type time-step stability restriction, and it is norm preserving. In the presence of absorptive, non-self-adjoint elements---such as complex absorbing potentials or absorbing boundary conditions---CN is no longer unitary but remains contractive, yielding a non-increasing discrete norm consistent with probability removal. This robustness makes CN well suited to mixed boundary conditions on different faces of the computational box and to the boundary-induced spin coupling arising in the spinor ABC model \cite{CN47,Str04,HW96,vDT07}.

It is useful to contrast this with the standard split-operator Fourier approach, which is often highly efficient for Schr\"odinger evolution in simpler geometries.
Split-operator Fourier (pseudo-spectral/FFT) schemes achieve spectral accuracy in space and \(O(N\log N)\) scaling per step for constant-coefficient kinetic terms on periodic (or effectively periodic) uniform grids. However, non-periodic boundary conditions such as Dirichlet or Robin conditions, as well as localized absorbers, typically require additional machinery (for example sine/cosine transforms, masks or filters, or specialized transparent boundaries), and may suffer from wrap-around artifacts or residual reflections if not carefully tuned \cite{FFS82,KK83}. In our setting, moreover, the spin dependence of the spinor ABC model enters through the absorbing boundary condition at the detector surface rather than through a bulk potential term. As a result, FFT-based propagation would still require additional treatment of the boundary-induced coupling, increasing the implementation complexity. For these reasons, despite its higher per-step computational cost, CN is the most suitable and robust choice for our setting. Implementation details are given in Appendix~\ref{app:TDSE}.

To obtain detection-time statistics we use two independent procedures: (i) evaluating the surface-flux formula from the quantum probability current, and (ii) sampling Bohmian trajectories, with the Pauli current used in the spinor case for both the ABC and CAP models, see Appendices~\ref{app:Bohm}, \ref{app:arrival}, and \ref{app:rate}. 
We consistently find very good agreement between the two procedures.
Representative numerical convergence tests, together with a discussion of the stability of the chosen discretization parameters, are reported at the end of Appendix~\ref{app:arrival}.

Additional animations of the time evolution of the full 3-dimensional probability density $|\Psi_t(\vr)|^2$ for representative runs are available in the accompanying repository at 
\cite{JT26_code}, illustrating the propagation, absorption, and reflection dynamics.

\begin{rem}\label{rem:mean}
Whenever we report an \emph{average} of $T$ (or $\tau$), the estimator for it was obtained as follows. Since our numerical simulations end at a cut-off time $t_\mathrm{cutoff}$ (such as $t_\mathrm{cutoff}=20$ in Figure~\ref{fig:Arrival_CAP}), the parts of the distribution density $\rho_T(t)$ to the right of $t_\mathrm{cutoff}$ are not known. We use the \emph{restricted mean value} 
\be\label{mudef}
\mu^*(T) := \EEE \min(T,t_\mathrm{cutoff}) = \int_0^{t_\mathrm{cutoff}} dt \, t \, \rho_T(t) + t_\mathrm{cutoff} \, \Biggl( 1- \int_0^{t_\mathrm{cutoff}} dt  \, \rho_T(t) \Biggr)
\ee
in the place of $\EEE T$, which satisfies $\mu^*(T) \leq \EEE T$.\hfill$\diamond$
\end{rem}

\subsection{Comparison Curve}
\label{sec:comparisoncurve}

In our diagrams of the probability density $\rho_T$ such as Figure~\ref{fig:Arrival_CAP}, we include for comparison (in orange) the curve $\rho^\fr$ describing the probability loss of a spin-0 wave function \emph{in the absence of detectors}, which is defined as follows. Let $\Psi^\fr$ be the (spinless, scalar-valued) solution of the Schr\"odinger equation \eqref{Schr} in the half space $\Omega^\fr=\RRR^2\times [0,\infty) = \{\vr=(x,y,z)\in \RRR^3: z\geq 0\}$ with $V$ only the usual wave guide potential \eqref{harmonicV} (i.e., $W=0$), Dirichlet boundary condition \eqref{Dir0} at $z=0$, and the usual initial wave function \eqref{initialpsi}. This time evolution is unitary. Define
\be\label{rhofrdef1}
\rho^\fr(t) = -\frac{d}{dt} \Bigl\|1_{z<L}\Psi_t^\fr \Bigr\|^2=-\frac{d}{dt}\int_0^L dz \int dx \int dy \, |\Psi^\fr_t(x,y,z)|^2
\ee
or, equivalently,
\be\label{rhofrdef2}
\rho^\fr(t) = \int dx \int dy \, j_z^\fr(x,y,L,t) \,.
\ee
It turns out that $\rho^\fr(t)\geq 0$ for all $t$, in fact 
\be\label{jfrpos}
j^\fr_z(x,y,L,t)\geq 0
\ee
for all $x,y,t$. (This can be understood as follows. Since the $z$ direction decouples from $x$ and $y$, $\Psi_t(x,y,z) = g(x,y) \, \psi_t(z)$ factorizes; in the 1d problem for the $z$ coordinate, since $L\gg d$, we are in the scattering (far field) regime, in which, as is well known \cite{BH59,New82,Daumer,KT26}, the current is everywhere outward-pointing, $j_z \geq 0$.)  

As a consequence of \eqref{jfrpos}, the $f$ factor in \eqref{arrivalf} is always 1, so it can be dropped, and $\rho^\fr(t)$ coincides also with the probability density of the arrival time of the Bohmian trajectory in the absence of detectors. (This is the distribution proposed by Das and D\"urr in the spin-0 case.)

\section{Results}
\label{sec:results}

\subsection{CAP}

\begin{figure}[h]
  \centering
    \includegraphics[width=0.8\linewidth]{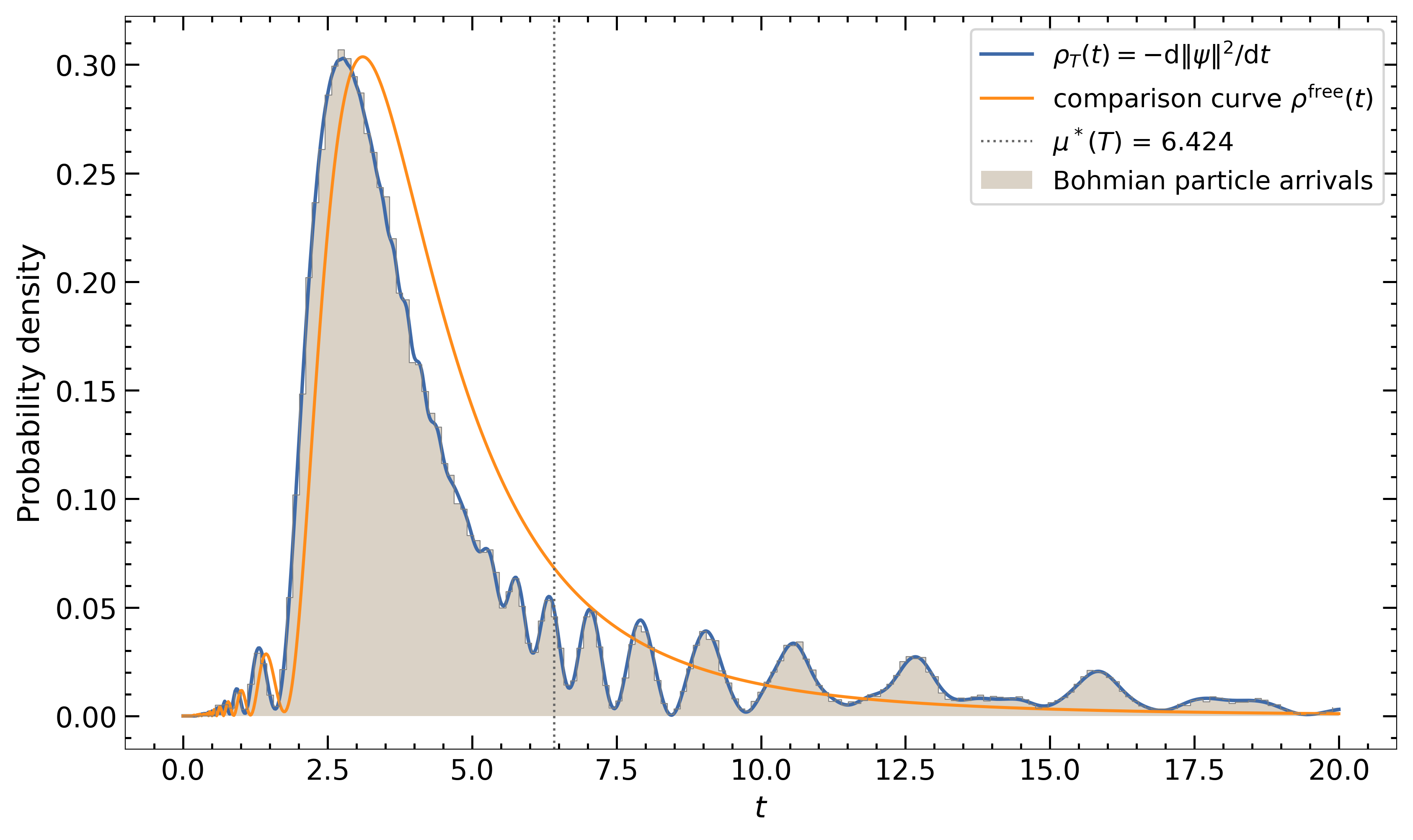}

  \caption{Detection time distribution $\rho_T(t)$ for complex absorbing potential (CAP), computed once using \eqref{rhoTt} (dark blue curve) and once using Bohmian trajectories (light brown histogram); and comparison curve $\rho^\fr(t)$ (orange) as in \eqref{rhofrdef1}. The vertical dotted line indicates the (restricted) mean of $T$ as in \eqref{mudef}. Parameters are $z_0=10$, $L=11$, $a= 0.165$, $W_{\max}= 40$, $\omega=100$  in units with $d=1$, $\hbar=1$, $m=1$. (Realistic values in SI units are given in Section~\ref{sec:experimental}.) The detection fraction (= area under $\rho_T(t)$ = fraction of particles detected by $t=20$) is 91\%.}
  \label{fig:Arrival_CAP}
\end{figure}

For the soft detector modeled by an imaginary potential as described in Section~\ref{sec:CAP}, the distribution density $\rho_T(t)$ of the detection time $T$ is shown in Figure~\ref{fig:Arrival_CAP}.

\begin{figure}[h]
\begin{center}
\begin{tikzpicture}
\draw[->] (0,-0.1) -- (0,3);
\draw[->] (0,0) -- (10.3,0);
\node at (0,-0.4) {0};
\draw (6.5,-0.1) -- (6.5,0);
\node at (6.5,-0.4) {$z_0$};
\draw (9.5,-0.1) -- (9.5,0);
\node at (9.5,-0.4) {$L$};
\node at (10.2,-0.3) {$z$};
\draw[<->] (6.5,0.5) -- (7.5,0.5);
\node at (7,0.8) {$a$};
\draw (0,2) -- (-0.1,2);
\node at (-0.6,2) {$W_{\max}$};
\draw[ultra thick,domain=-3.5:3, samples=101] plot({\x+6.5},{1+tanh(\x)});
\draw[ultra thick] (0,0) -- (3,0);
\end{tikzpicture}
\end{center}
\caption{Graph of the function \eqref{Wtanh}}
\label{fig:tanh}
\end{figure}
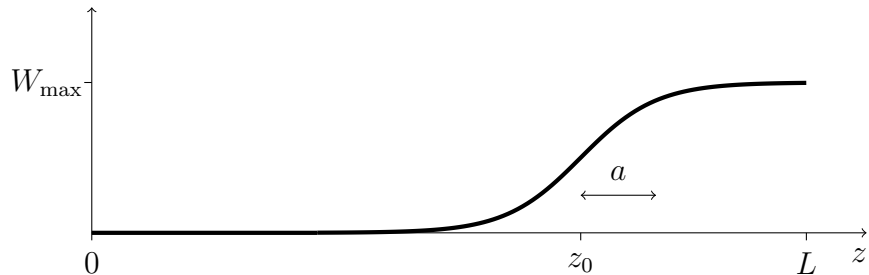

For this figure, we used
\be\label{Wtanh}
W(z) = \Bigl[1+\tanh\Bigl( \frac{z-z_0}{a}\Bigr) \Bigr] \frac{W_{\max}}{2} \,.
\ee
a smooth step centered at $z_0$ and smoothed over the width $a$ as depicted in Figure~\ref{fig:tanh}.

Apart from the consistency check that the distribution computed using Bohmian trajectories (light brown) and the one using \eqref{rhoTt} (dark blue curve) agree very well with each other (they agree exactly in theory, as pointed out in Section~\ref{sec:Bohmian}), we can observe from these figures that $\rho_T(t)$ follows the overall profile of the comparison curve $\rho^\fr(t)$ that represents the transfer of probability across the plane $z=L$ in the absence of detectors at $z=L$, as described in Section~\ref{sec:comparisoncurve}. 

The main difference to $\rho^\fr(t)$ consists in oscillations of $\rho_T$ not present in $\rho^\fr(t)$. They are presumably due to interference between the incident and a reflected wave. The fact that imaginary potentials will partly reflect the incident wave is well known from the fact, noted by Allcock \cite{All69b} and similar to the quantum Zeno effect, that if we increase without bound the strength of the imaginary potential, $W(z)\to \infty$ for $z>z_0$, then \emph{all} of the incident wave will be reflected. It would be of interest to determine whether any reflected waves and corresponding oscillations can be seen in empirical data (see also \cite{CD25}). (On the other hand, there exist shapes of complex potentials, not considered here, with less reflection \cite{PMS98,Man02}. Therefore, if a future experiment showed little or no oscillatory tail, that would not by itself rule out CAP-based detector models; the detector might be describable by a CAP outside the simple family studied here. Part of the reflection in our simulation is also due to the hard wall behind the imaginary potential; e.g., the limit $W(z)\to 0$ would lead to no detection but complete reflection.)

Another observation about Figure~\ref{fig:Arrival_CAP} is that parts (before the main peak) of the dark blue curve $\rho_T$ lie slightly to the left of the orange comparison curve $\rho^\fr$; this presumably represents the fact that with CAP, particles get detected already slightly before reaching $z=L$.

\begin{figure}[ht]
  \centering
  \includegraphics[width=0.8\linewidth]{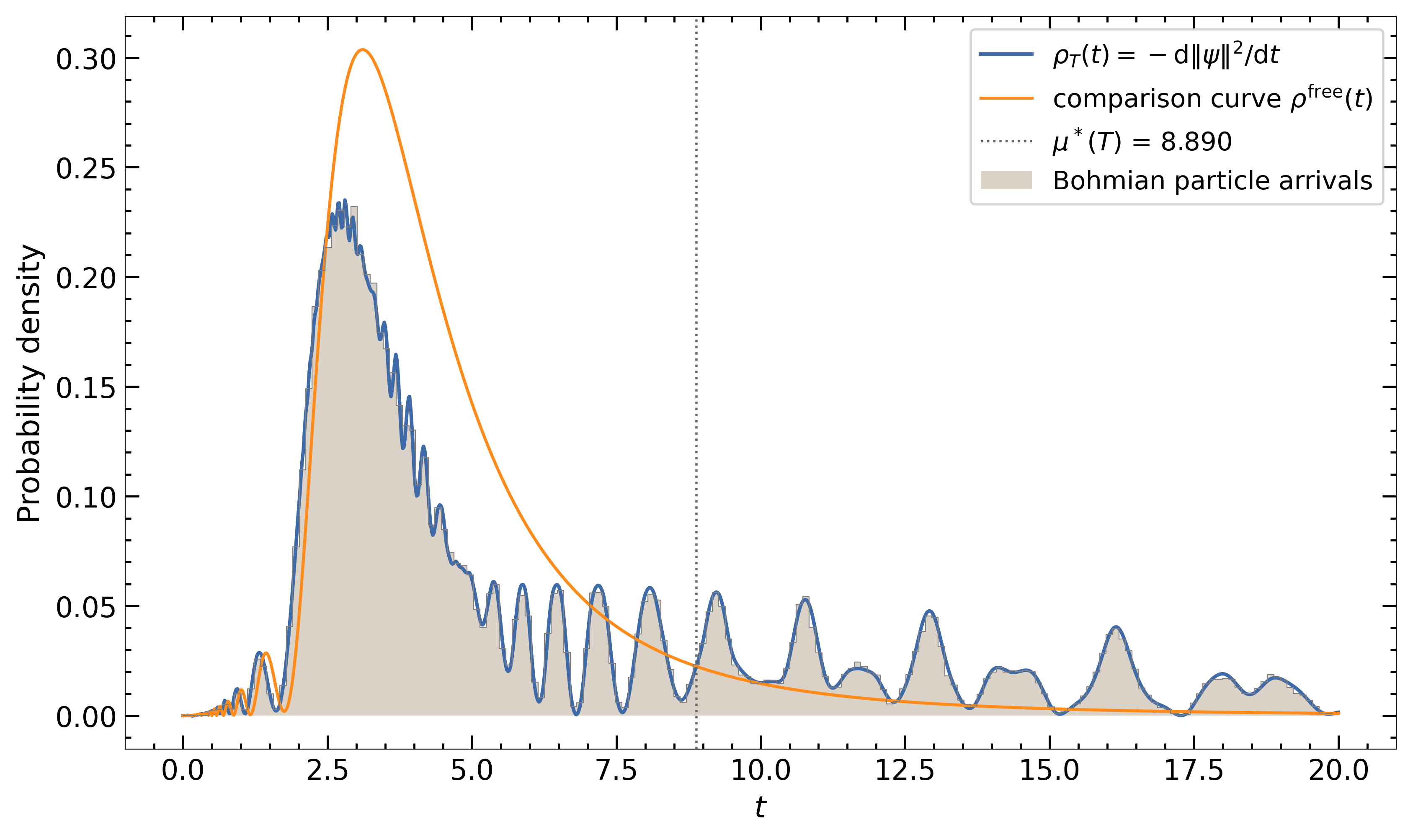}

  \caption{The same as Figure~\ref{fig:Arrival_CAP}, but for a sudden increase (discontinuity) in the imaginary potential $W(z)$ as in \eqref{Wsharp}. Parameters are $z_0=10$, $L=11$, $W_{\max}=40$, $\omega=100$; detection fraction = 81\%.}
  \label{fig:Arrival_CAP_sharp}
\end{figure}

A variation of the setup is shown in Figure~\ref{fig:Arrival_CAP_sharp}, this time with a sharp (discontinuous) step
\be\label{Wsharp}
W(z) = 1_{z_0<z<L} \, W_{\max}
\ee
in the CAP. We observe stronger oscillations, which suggests stronger reflection. Indeed, it is known \cite{Visser}, \cite[Sec.~3]{GGLT11} that for soft \emph{real} potential steps, the reflection increases with the sharpness of the step; this suggests that the same happens for imaginary potentials. Moreover, we observed in further simulations reported in Appendix~\ref{app:variation} that with increasing sharpness of the imaginary step, the distribution develops stronger oscillatory tails, the probability of detection within our time window ($t\leq 20$) decreases, and the mean detection time increases, all consistent with stronger reflection.

\begin{figure}[H]
  \centering
    \includegraphics[width=0.8\linewidth]{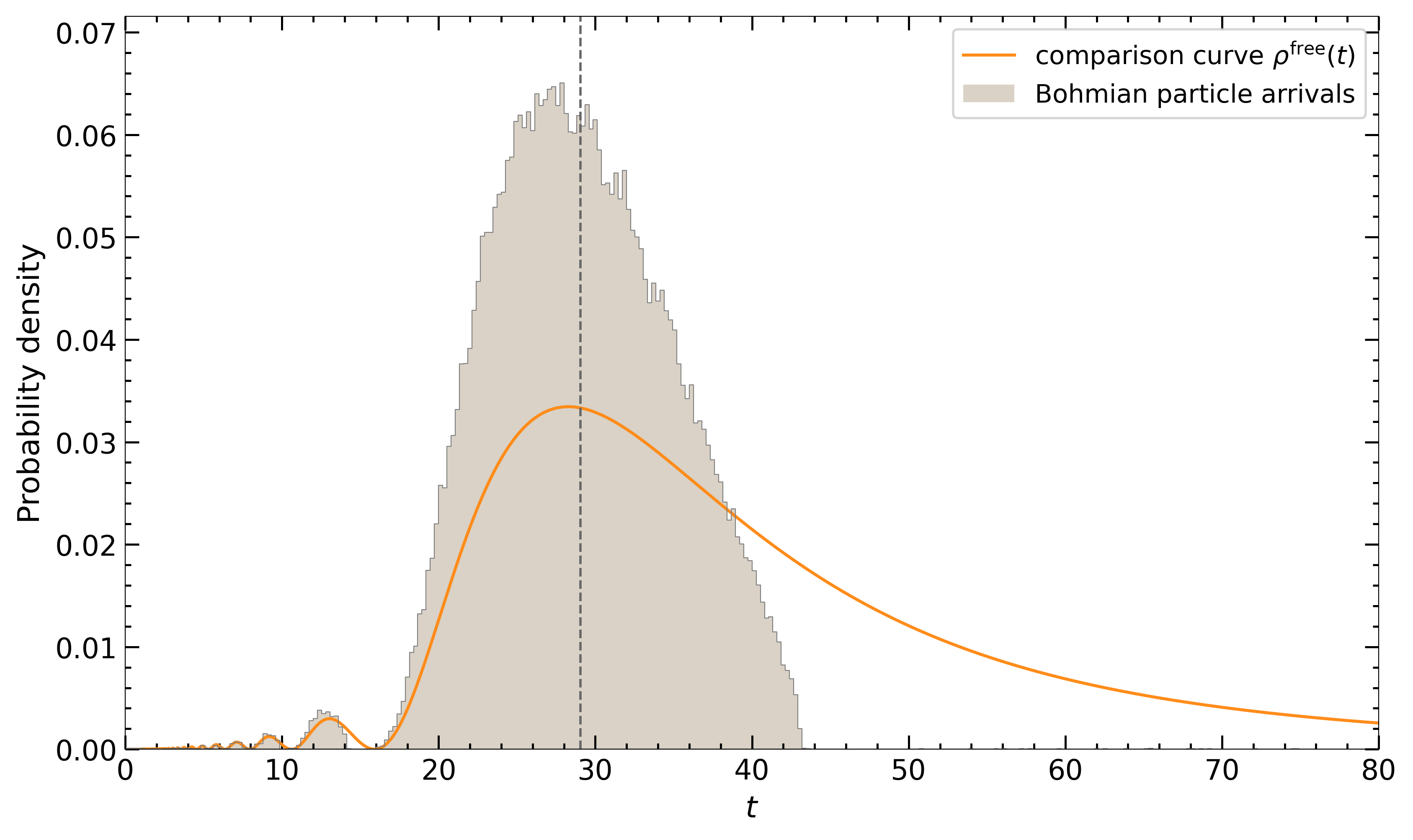}
  \caption{Bohmian arrival times at $z=L$ in the absence of detectors (i.e., neither CAP nor ABC, the domain of $\Psi$ is $\{z\geq 0\}$) for spin-1/2 particles, here for $\theta=\pi/2$, $L=100$, $\omega=500$. Das and D\"urr \cite{DD19} predicted that detectors will find this distribution. The distribution has no weight after $\tau_{\max}\approx 43$; in particular, the detection fraction is 100\%. The orange curve is again $\rho^\fr(t)$.}
  \label{fig:Das}
\end{figure}

\paragraph{Spin 1/2.}
For a spin-1/2 particle, the CAP Hamiltonian is spin-blind, so the detection-time density $\rho_T(t)$ is independent of the initial spinor $|\chi\rangle$. This sharply contrasts with the proposal of Das and D\"urr \cite{DD19}, whose predicted distribution depends strongly on $|\chi\rangle$: for $\theta=0$ in \eqref{chidef} it agrees with the comparison curve $\rho^\fr(t)$, whereas for $\theta=\pi/2$ and large $\omega$ it differs strongly, see Figure~\ref{fig:Das}. (A similar figure was given in \cite[Fig.~2]{DD19}.) Moreover, in that latter case Das and D\"urr predict a distribution supported on a finite interval $[0,\tau_{\max}]$, while both $\rho^\fr(t)$ and the CAP distribution $\rho_T(t)$ have tails to infinity.

\subsection{Spinless ABC}
\label{subsec:scalarABC}

\begin{figure}[H]
  \centering
  \includegraphics[width=0.8\linewidth]{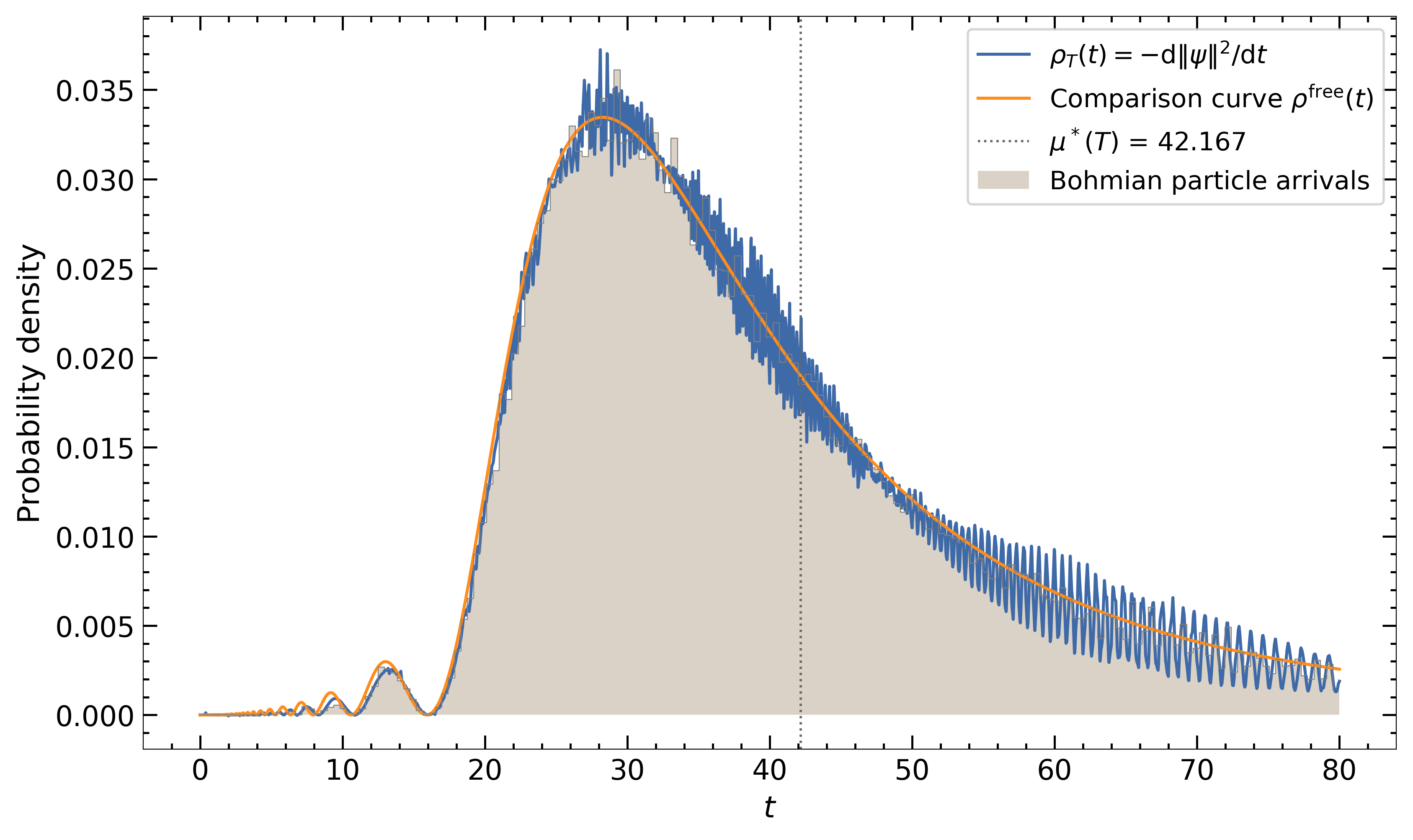}
  \caption{Detection time distribution $\rho_T(t)$ for spinless ABC \eqref{nospinABC} for spin-0 particles, computed once using \eqref{rhoTt} (dark blue curve) and once using arrival times of Bohmian trajectories (light brown histogram); and comparison curve $\rho^\fr(t)$ (orange) as in \eqref{rhofrdef1}. The vertical dotted line marks the (restricted) mean of $T$. Parameters are $L=100$, $\kappa=\pi$, 10,000 trajectories, detection fraction = 90\%.}
  \label{fig:scalarABC1}
\end{figure}

\begin{figure}[H]
  \centering
    \includegraphics[width=0.8\linewidth]{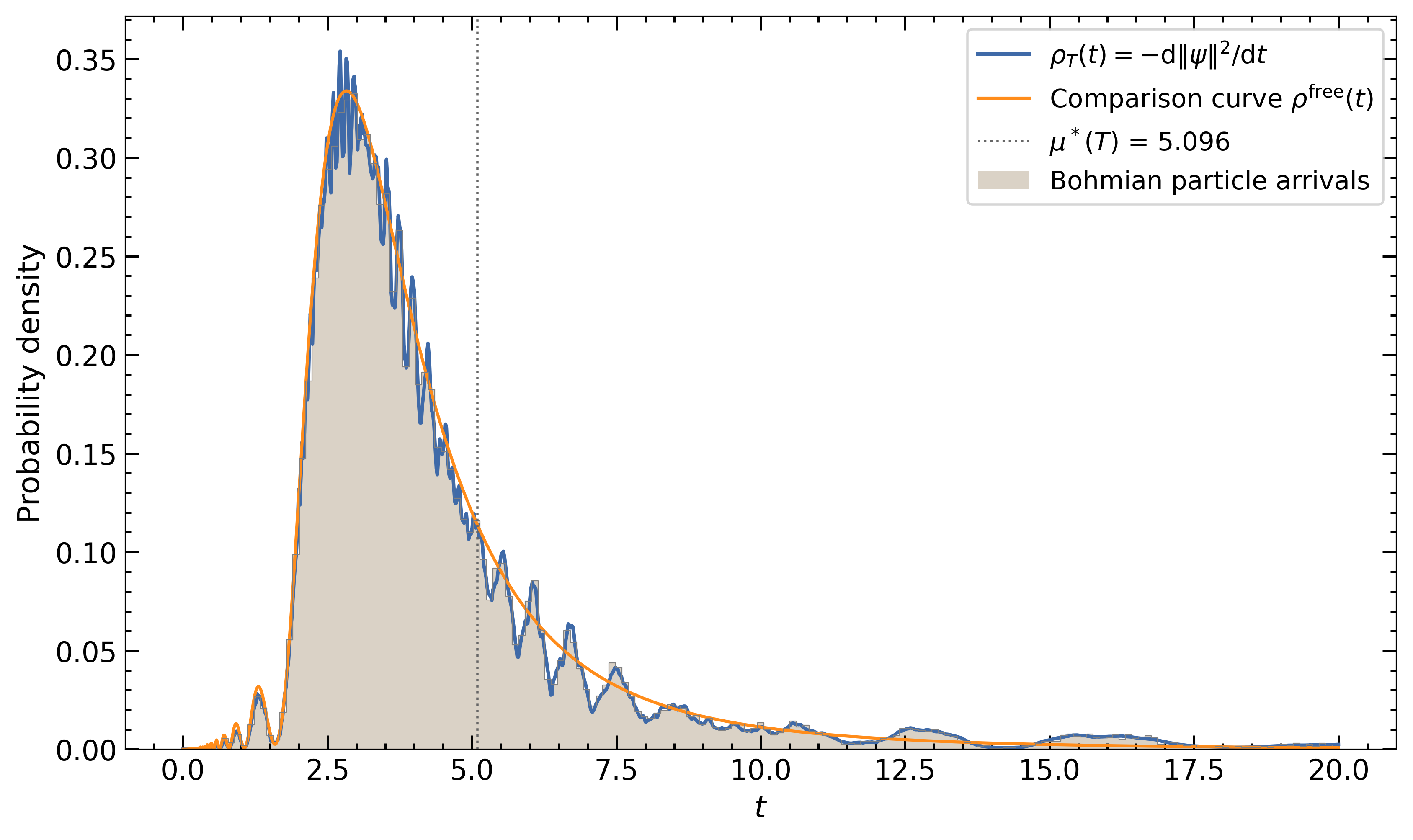}
    
  \caption{The same as Figure~\ref{fig:scalarABC1}, but with $L=10$ instead of $L=100$; detection fraction = 96\%.}
  \label{fig:scalarABC2}
\end{figure}

Now we consider the spinless ABC \eqref{nospinABC} (or \eqref{nospinABCzL}) as a model of a hard detector. The results are shown in Figures~\ref{fig:scalarABC1} and \ref{fig:scalarABC2} for different values of $L/d$.

It is visible in these figures that the distribution density $\rho_T(t)$ is again roughly similar to the comparison curve $\rho^\fr(t)$, and again the main difference consists in oscillations of $\rho_T(t)$ not present in $\rho^\fr(t)$. A comparison of Figures~\ref{fig:scalarABC1} and \ref{fig:scalarABC2} shows that the frequency of these oscillations depends on $L/d$. (The parameter $\kappa=\pi$ was chosen for maximal absorption, which is known \cite{detect-rule} to occur for the spinless ABC at $k=\kappa$, and for $\Psi_0$ as in \eqref{initialpsi} the dominant $k$ value for the $z$ direction is $\pi/d$, with $d=1$ in the figures; for the spinor ABC, we do not know the optimal value of $\kappa$; the choice $W_{\max}=40$ in Figures~\ref{fig:Arrival_CAP} and \ref{fig:Arrival_CAP_sharp} was arbitrary.)

The Hamiltonian with the ABC is again spin-blind, so again the distribution $\rho_T(t)$ does not depend on whether we use spin-0 or spin-1/2 wave functions, and in the latter case does not depend on the spinor $|\chi\rangle$. The distribution does not depend on $\omega$ either because the Hamiltonian for the $x$ and $y$ coordinates decouples from the one for $z$, and only the latter is relevant to $\rho_T(t)$.

\begin{rem}\label{rem:nospinABCjP}
    As pointed out in Remark~\ref{rem:TWID}, when the spinless ABC is used for spin-1/2 wave functions, then the detection time and place agree with the Bohmian arrival time and place for the equation of motion \eqref{Bohm} using $\vj=\vj^\mathrm{C}$ but not using $\vj^\mathrm{P}$. It is visible in the simulations how the distribution of the arrival time $\tau=T_{WID}$ using $\vj^\mathrm{P}$ is different from the distribution of\footnote{For the reasons pointed out in Footnote~\ref{fn:vanishcurl}, \eqref{vanishcurl} does not imply that the two distributions coincide.} $T=T_D$, see Figure~\ref{fig:arrival_hist}. (Note that the former is not the distribution predicted by Das and D\"urr, though, because we are using a $\Psi$ that is subject to an ABC.)

\begin{figure}[H]
  \centering
    \includegraphics[width=0.8\linewidth]{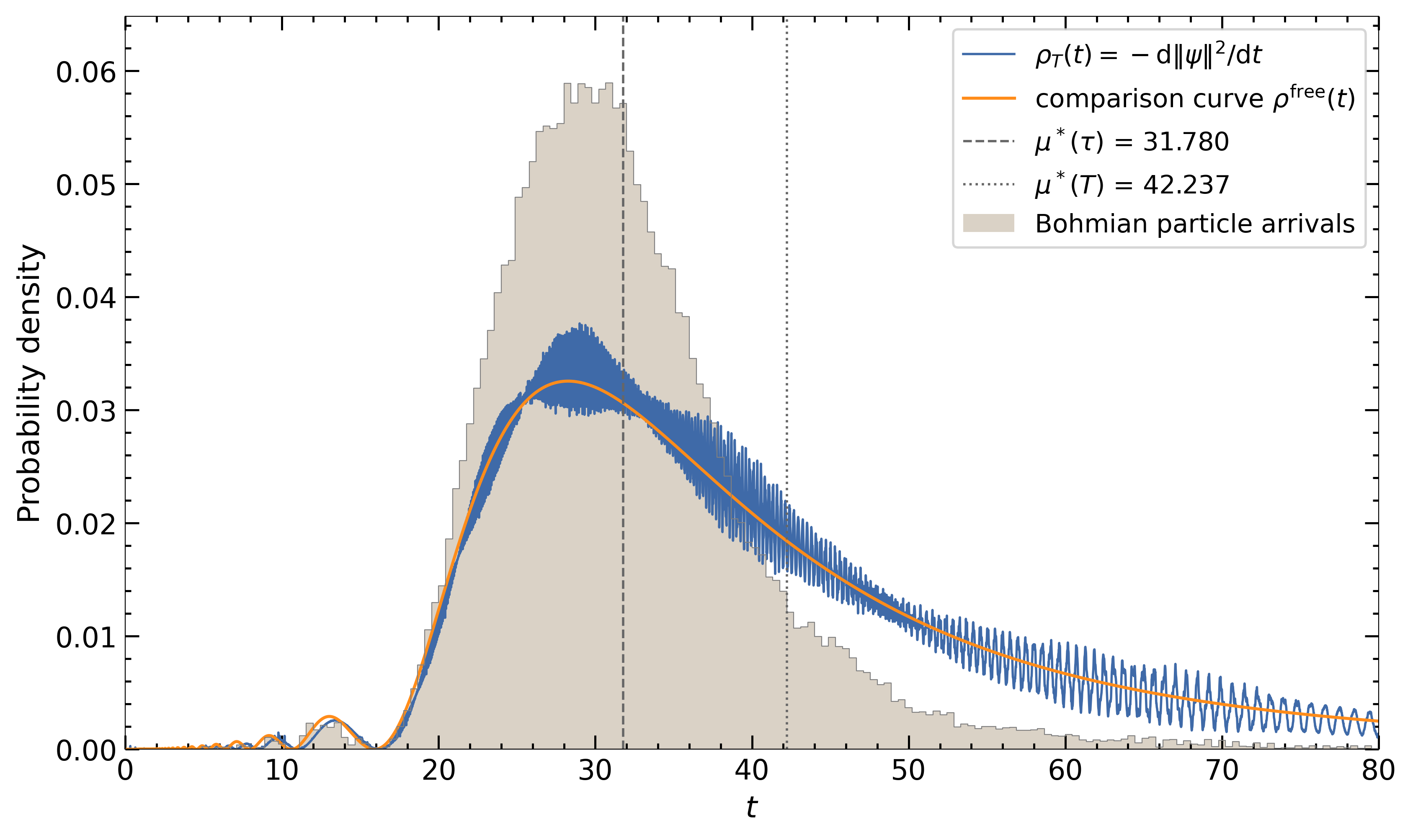}%

  \caption{The distributions discussed in Remark~\ref{rem:nospinABCjP}. Dark blue curve: detection time distribution $\rho_T(t)$ for spinless ABC \eqref{nospinABC} imposed on a spin-1/2 wave function; light brown histogram: arrival times $\tau=T_{WID}$ of Bohmian trajectories defined using $\vj^\mathrm{P}$, whereas $\vj^\mathrm{C}$ would reproduce the light brown histogram of Figure~\ref{fig:scalarABC1}. Parameters are $\theta=\pi/2$, $L=100$, $\omega=500$. The light brown histogram depends on $\theta$, the dark blue curve does not; the dark blue curve is the same as in Figure~\ref{fig:scalarABC1} (which assumed spin 0). Note that the histogram has a tail to infinity not present in Figure~\ref{fig:Das}; detection fraction = 99\%.}
  \label{fig:arrival_hist}
\end{figure}

    The arrival time using $\vj^\mathrm{P}$ depends on the spinor $|\chi\rangle$, even though the Hamiltonian does not couple to $|\chi\rangle$ so $\Psi_t(\vr)= \psi_t(\vr) |\chi\rangle$, because the curl term in \eqref{Paulij} becomes
    \be\label{curlchi}
    \nabla \times (\Psi^\dagger \vsigma \Psi) = -\langle\chi|\vsigma|\chi\rangle \times \nabla (|\psi|^2) \,.
    \ee    
    As a consequence, for $\theta=0$ or $\theta=\pi$ (i.e., $\langle \chi|\vsigma|\chi\rangle = \pm \hat{\boldsymbol{z}}$), the curl term lies in the $xy$ plane and $j_z^\mathrm{P}=j_z^\mathrm{C}$, so the distribution of $(\tau,\vQ(\tau))$ (i.e., in the notation of \cite{GTZ24}, $(T_{WID},\vX_{WID})$) coincides with that of $(T,\vR)=(T_D,\vX_D)$; in particular, for $\theta=0$ and $\theta=\pi$ the light brown histogram (arrival time statistics) agrees exactly with the dark blue curve in Figure~\ref{fig:arrival_hist}; this is not so for other values of $\theta$, neither theoretically  nor in our simulations, see Appendix~\ref{app:variation}.  For $0\neq \theta \neq \pi$, the distribution of $(\tau,\vQ(\tau))$ also depends on $\omega$, as visible from the fact that the $z$ component of \eqref{curlchi} then involves $\partial_x|\psi|^2$ and/or $\partial_y|\psi|^2$. We could also confirm this dependence in the simulations, see Appendix~\ref{app:variation}, particularly Figure~\ref{fig:arrival_hist3} and the plot of the mean of $\tau$  as a function of $\omega$ in Figure~\ref{fig:Spinless_fit}. The behavior of the Bohmian trajectories is investigated further in Appendix~\ref{app:Bohmianresults}. Note also that in contrast to the histogram of Figure~\ref{fig:Das}, which has compact support, that of Figure~\ref{fig:arrival_hist} exhibits a long tail extending to arbitrarily large times, highlighting a qualitative effect of the presence of detectors on the Bohmian arrival times.\hfill$\diamond$
\end{rem}

\subsection{Spinor ABC}
\label{subsec:spinorABC}

\begin{figure}[H]
  \centering
    \includegraphics[width=0.8\linewidth]{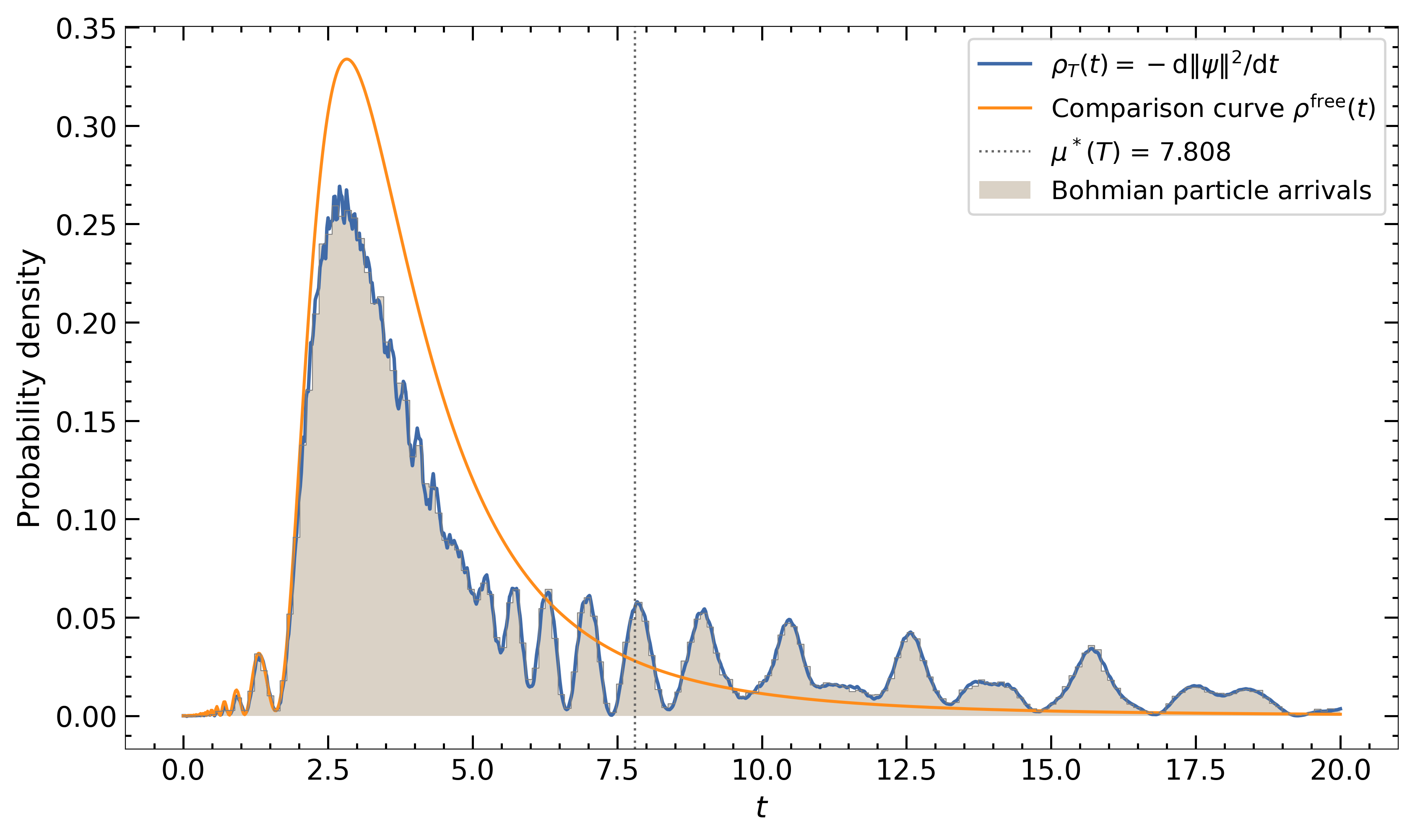}

  \caption{The same as Figures~\ref{fig:Arrival_CAP} and \ref{fig:scalarABC1}, but for the spinor ABC \eqref{spinABC}. Parameters are $L=10$, $\kappa=\pi$, $\omega=50$, $\theta=0$; detection fraction = 85\%.}
  \label{fig:spinor}
\end{figure}

Here we present the results obtained with the spinor absorbing boundary condition, see Figure~\ref{fig:spinor}.

We find again that the distribution is not far from the comparison curve, that there are oscillations (indicative of some amount of reflection), and that the detection time is delayed compared to the comparison curve.

\begin{figure}[ht]
  \centering

  \begin{minipage}[t]{0.48\linewidth}
    \centering
    \includegraphics[width=\linewidth]{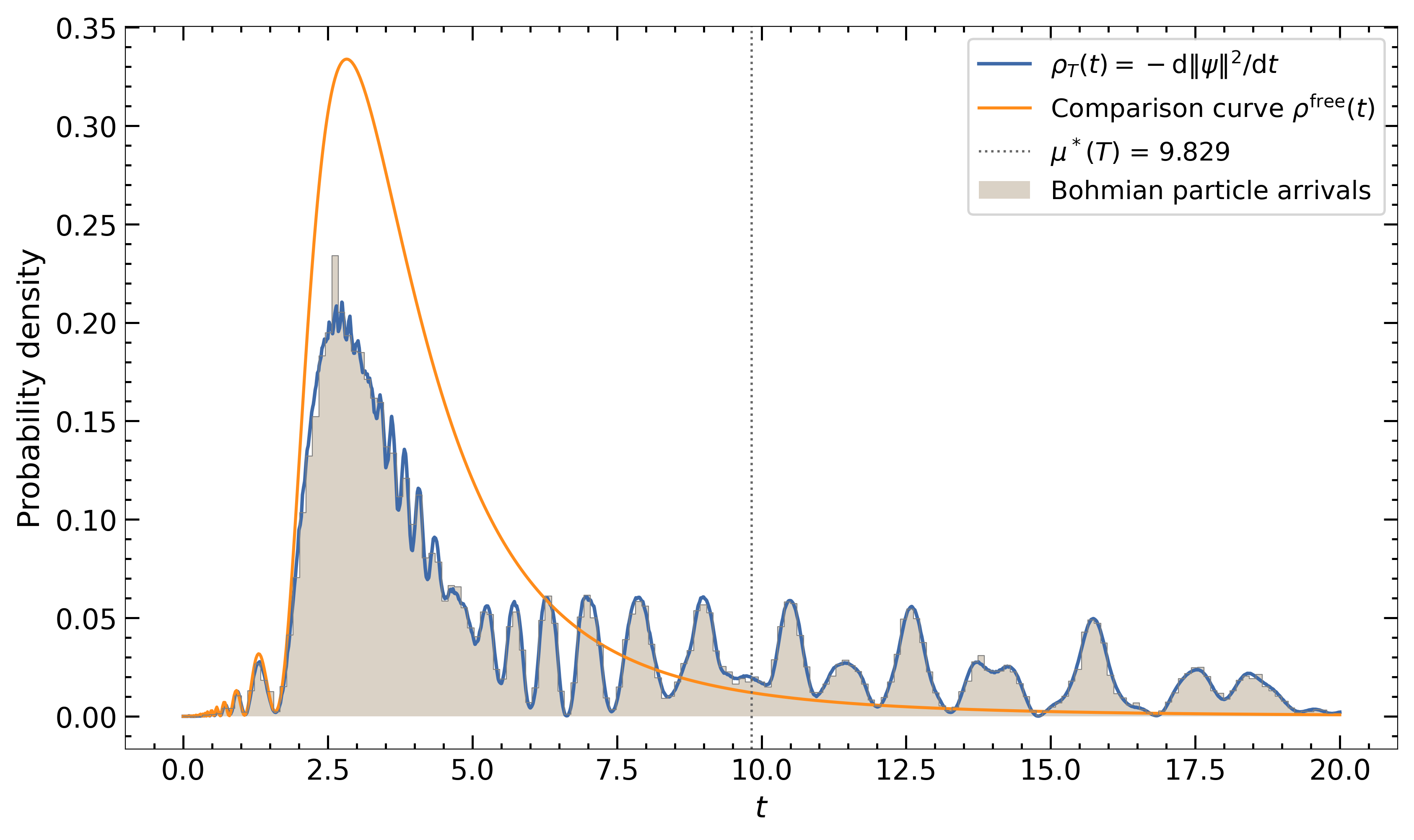}
    \par\vspace{0.3em}
    {\small (a) $\theta=0$}
  \end{minipage}\hfill
  \begin{minipage}[t]{0.48\linewidth}
    \centering
    \includegraphics[width=\linewidth]{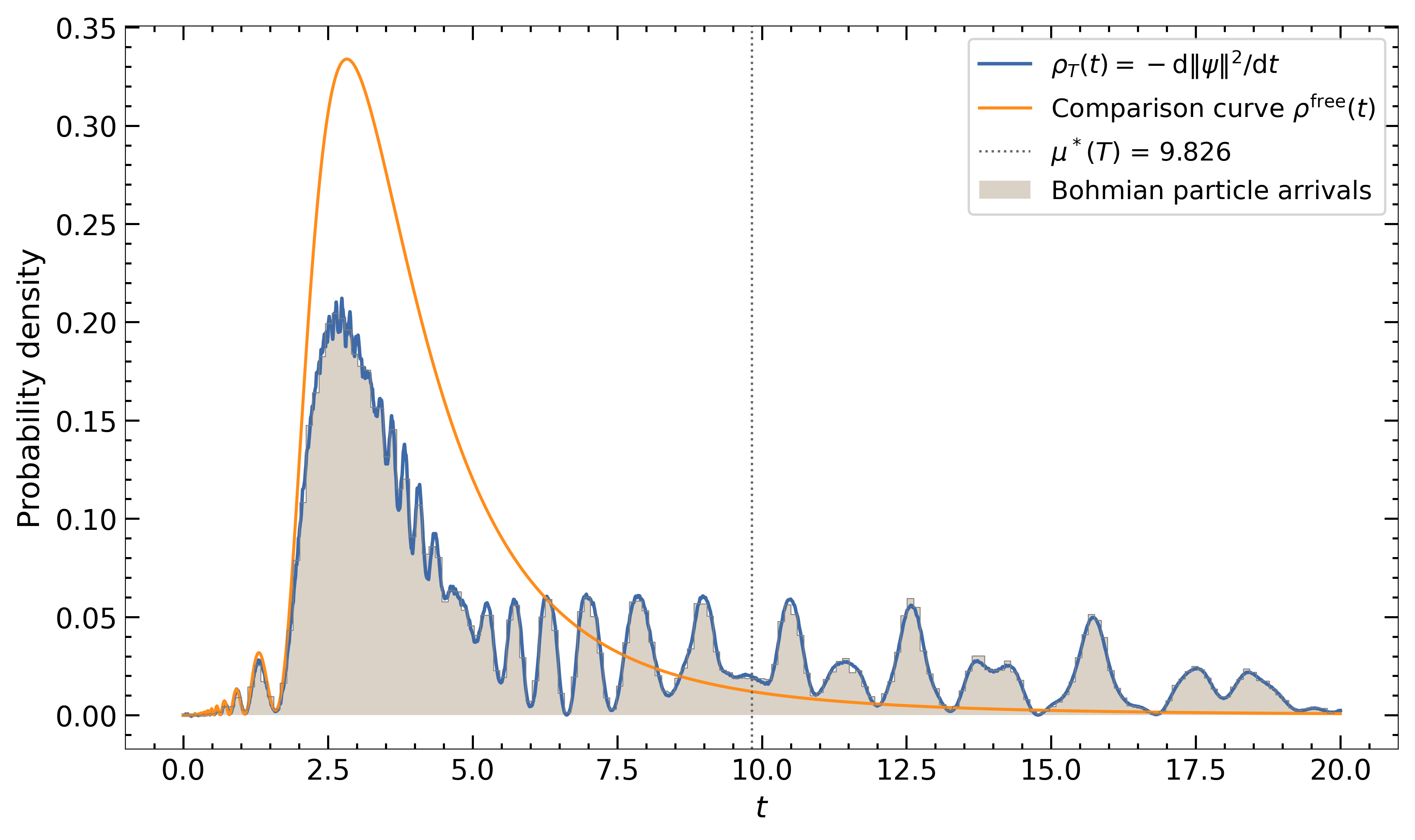}
    \par\vspace{0.3em}
    {\small (b) $\theta=\pi/2$}
  \end{minipage}

    \par\vspace{0.8em}

   \begin{minipage}[t]{0.48\linewidth}
    \centering
    \includegraphics[width=\linewidth]{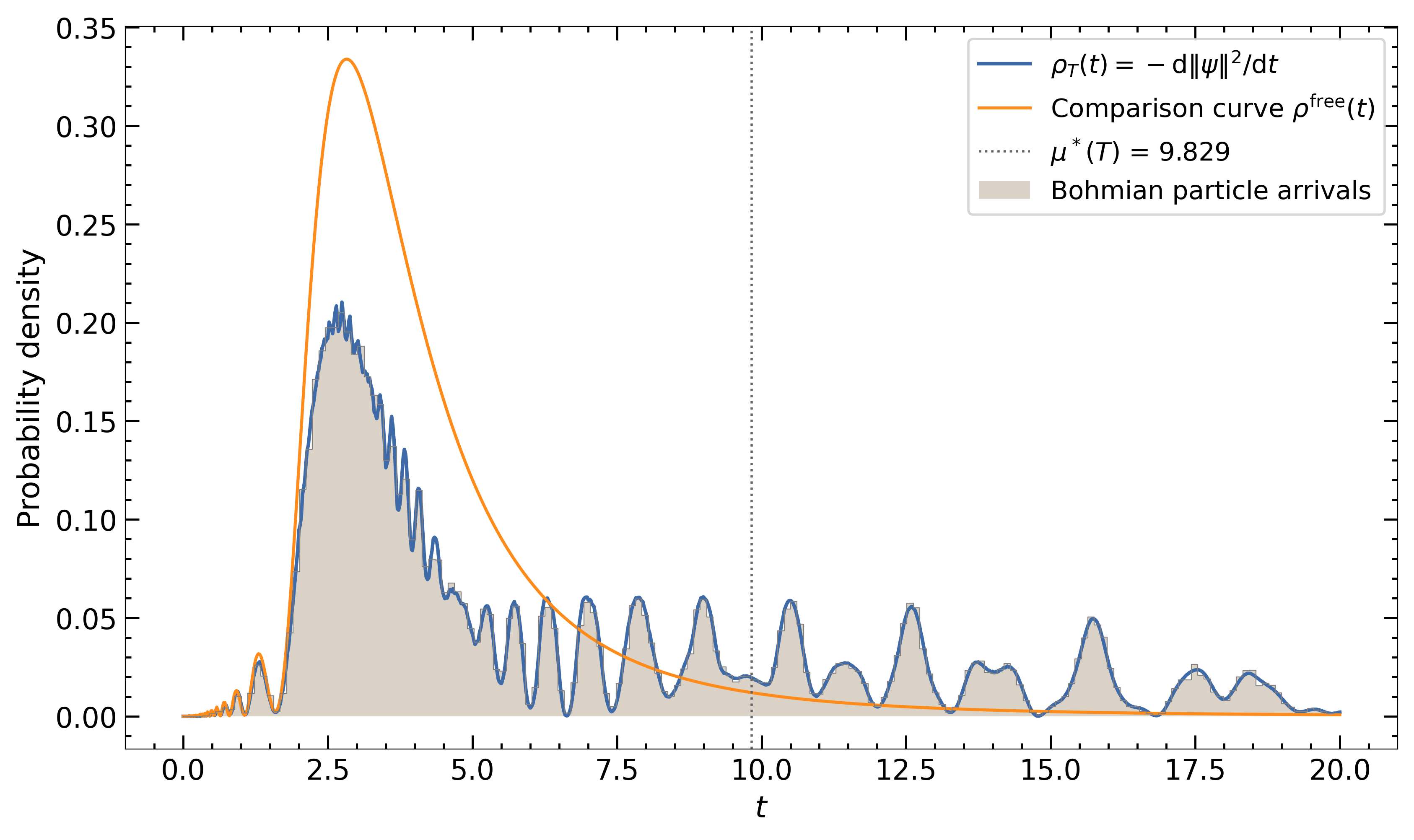}
    \par\vspace{0.3em}
    {\small (c) $\theta=\pi/3$}
  \end{minipage}

  \caption{Plots of $\rho_T(t)$ for the spinor ABC \eqref{spinABC}, which in principle can depend on $\theta$, for 3 different values of $\theta$, showing practically identical curves. Parameters are $L=10$, $\kappa=\pi$, $\omega=100$; detection fraction = 76\%.}
  \label{fig:omega100_pair}
\end{figure}

\paragraph{Spin dependence.} Since the boundary condition couples to the spin, the time evolution couples to the spin, so in theory it is no longer the case that $\Psi_t$ is proportional to $|\chi\rangle$; as a consequence, the curve $\rho_T(t)$ in principle depends on $\theta$. However, in our simulations this dependence was so weak as to be not noticeable, if it exists at all; see Figure~\ref{fig:omega100_pair}.  Again, we note the contrast to the prediction of Das and D\"urr, which features a strong dependence on $\theta$ as shown in Figure~\ref{fig:Das}.

\begin{figure}[ht]
  \centering
  \begin{minipage}[t]{0.48\linewidth}
    \centering
    \includegraphics[width=\linewidth]{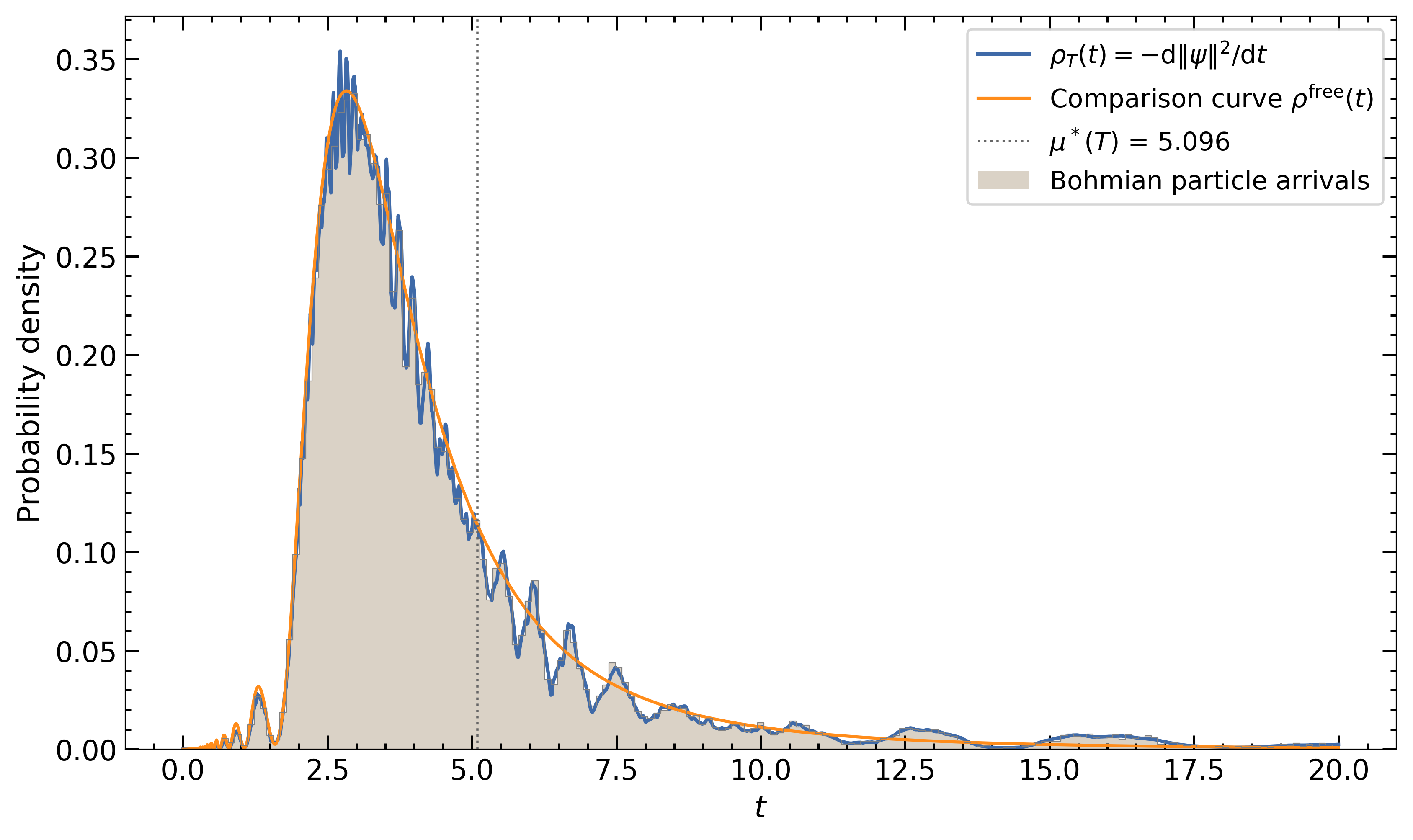}
    \par\vspace{0.3em}
    {\small (a) $\omega=1$, detection fraction = 96\%}
  \end{minipage}\hfill
  \begin{minipage}[t]{0.48\linewidth}
    \centering
    \includegraphics[width=\linewidth]{Fig11b.png}
    \par\vspace{0.3em}
    {\small (b) $\omega=100$, detection fraction = 76\%}
  \end{minipage}\hfill
  \vspace{0.8em}
  \begin{minipage}[t]{0.48\linewidth}
    \centering
    \includegraphics[width=\linewidth]{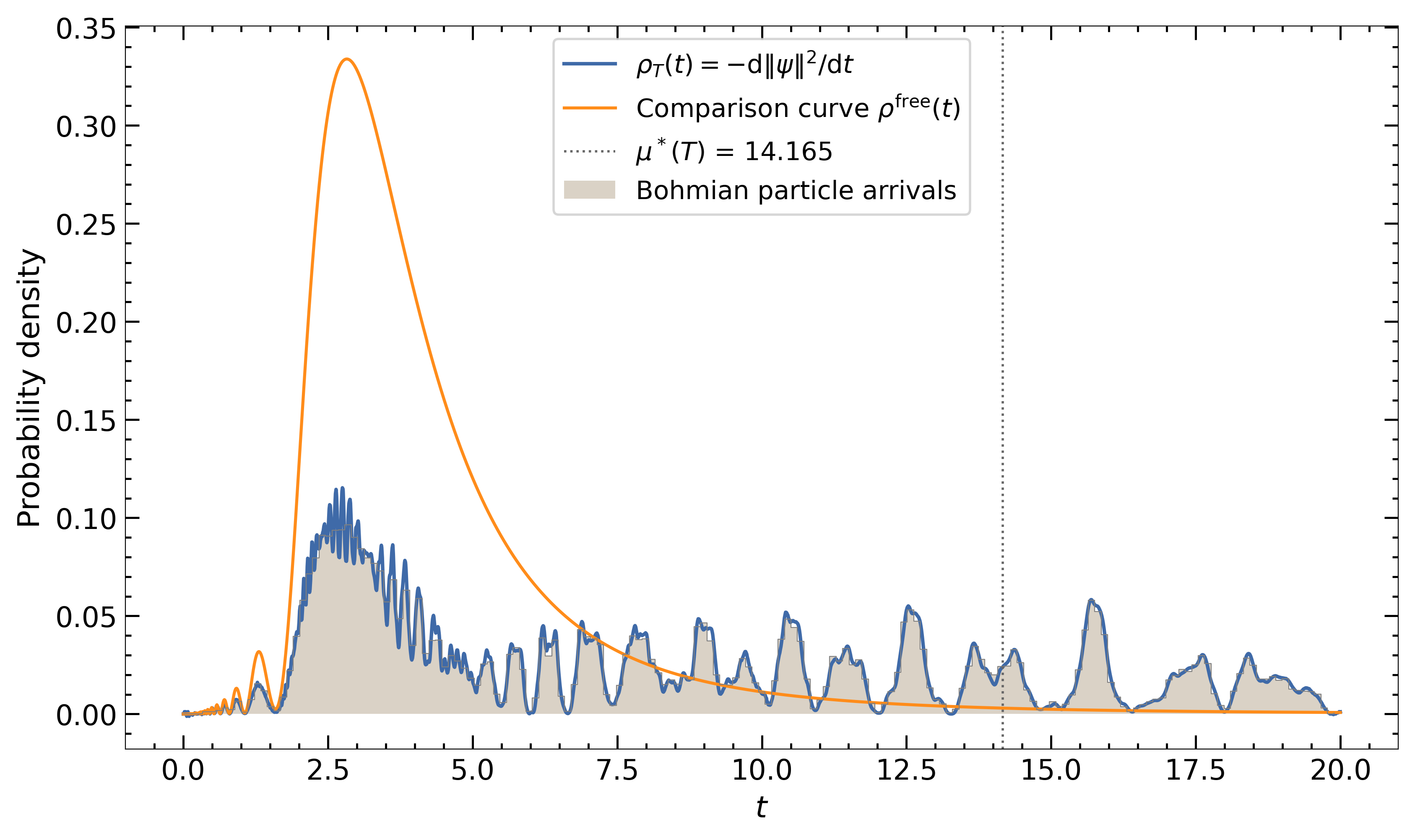}
    \par\vspace{0.3em}
    {\small (c) $\omega=300$, detection fraction = 49\%}
  \end{minipage}

  \caption{Pronounced $\omega$ dependence of the curve $\rho_T(t)$ for the spinor ABC \eqref{spinABC}. Parameters are $L=10$, $\kappa=\pi$; note that Figure~\ref{fig:spinor} also fits into this sequence between (a) and (b) at $\omega=50$.}
  \label{fig:omega_sweep_widC}
\end{figure}

\paragraph{$\omega$ dependence.}
Another way in which the spinor ABC differs from the spinless one is that the $z$ degree of freedom no longer decouples from $x$ and $y$; as a consequence of that, $\rho_T(t)$ also depends on the parameter $\omega$ of the harmonic potential that confines the particle to a neighborhood of the $z$ axis; note that $\sqrt{\hbar/m\omega}$ is the width of the wave guide. Specifically, larger $\omega$ implies a narrower initial wave function in the $x$ and $y$ directions and thus larger $\partial_x \Psi$ and $\partial_y\Psi$ at the boundary $z=L$; it is particularly visible in the form \eqref{spinABCzL} of the boundary condition that this has an effect on the spin and $\partial_z\Psi$ at $z=L$. In our simulations, we find a strong $\omega$ dependence, shown in Figure~\ref{fig:omega_sweep_widC}.

\begin{figure}[h]
  \centering

  \begin{minipage}[t]{0.48\linewidth}
    \centering
    \includegraphics[width=\linewidth]{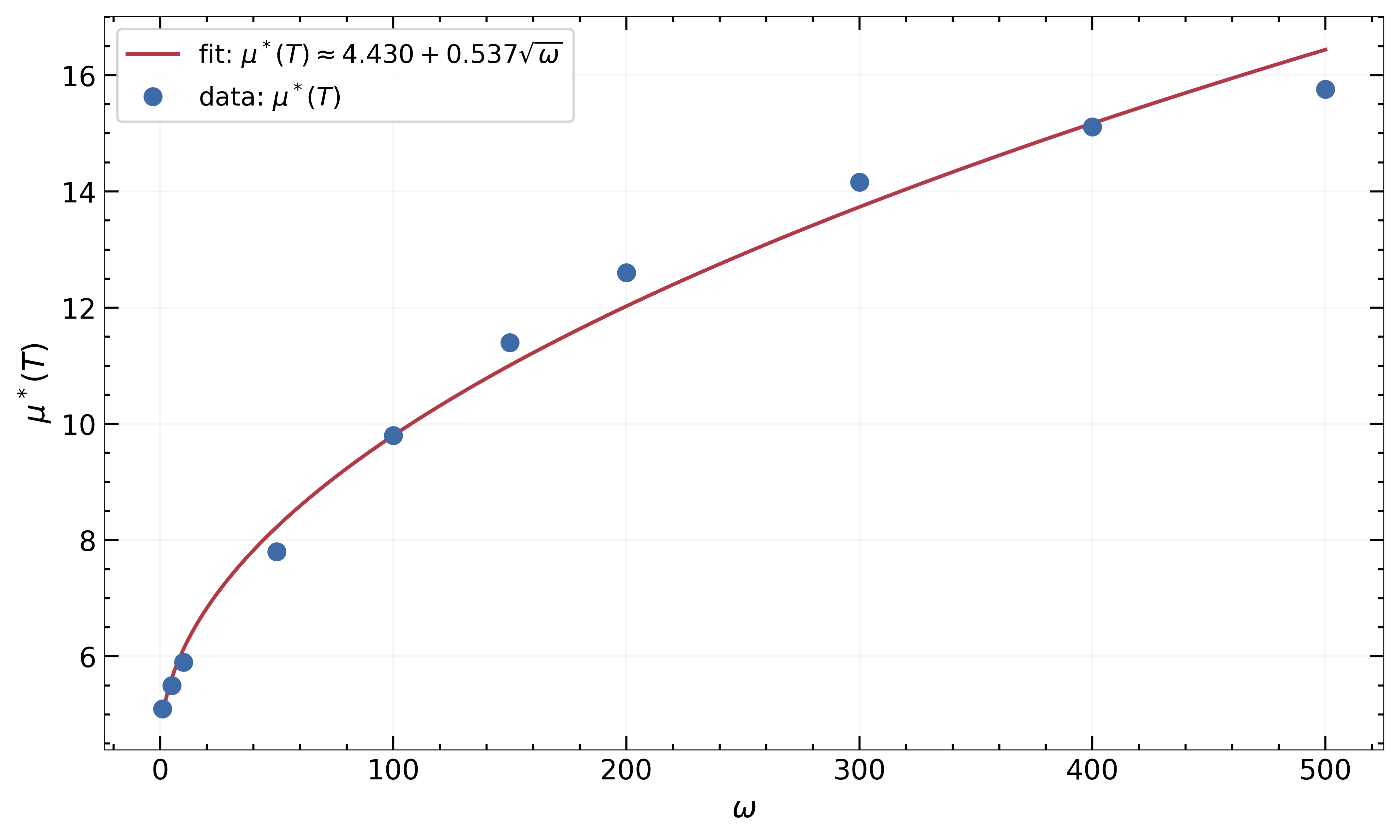}
    \par\vspace{0.3em}
    {\small (a)}
  \end{minipage}\hfill
  \begin{minipage}[t]{0.48\linewidth}
    \centering
    \includegraphics[width=\linewidth]{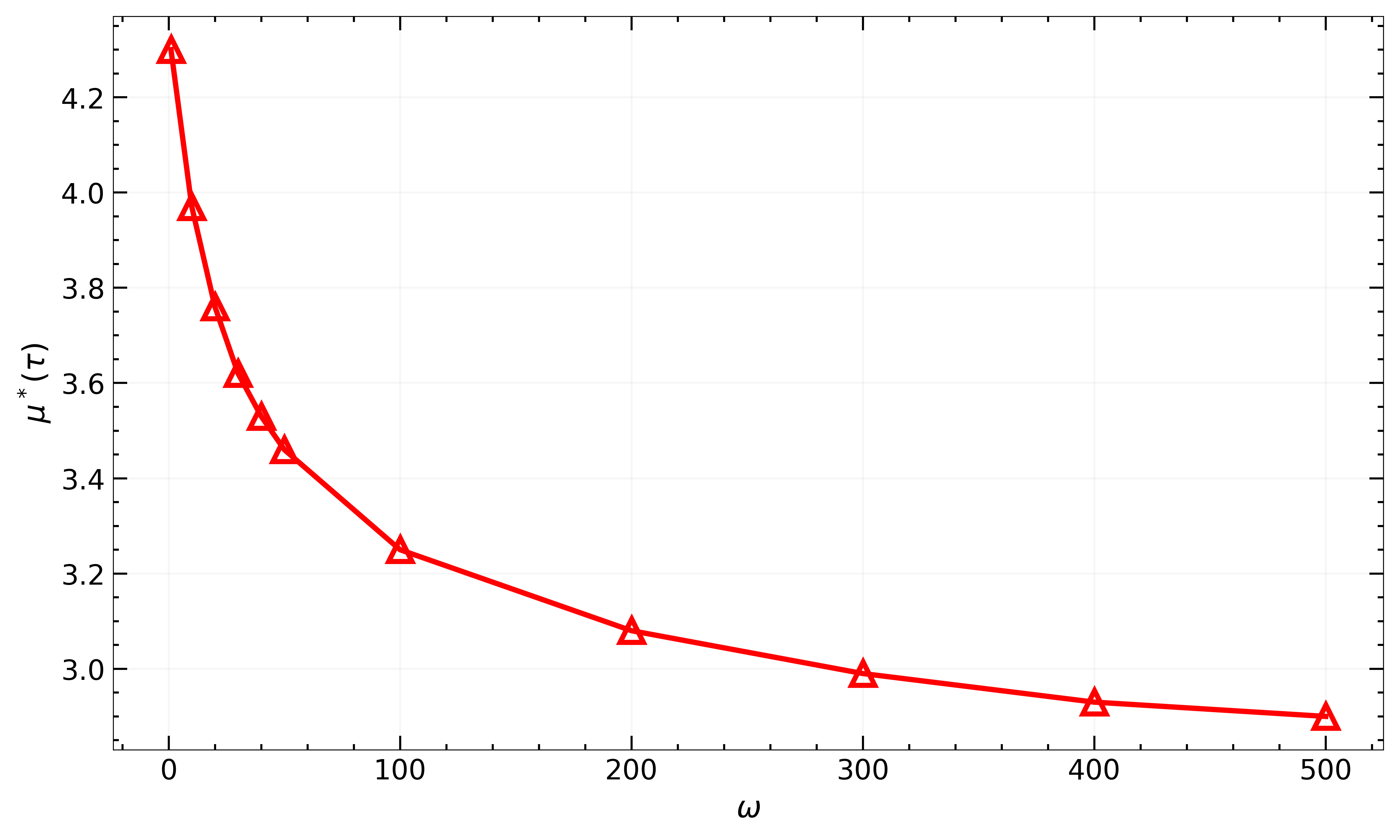}
    \par\vspace{0.3em}
    {\small (b)}
  \end{minipage}
  \caption{Average detection time \eqref{mudef} as a function of $\omega$: (a) from $\rho_T(t)$ using the spinor ABC \eqref{spinABC}; (b) according to the prediction by Das and D\"urr. (A similar figure was given in \cite[Fig.~4b]{DD19}, where the mean of $\tau$ is denoted $\langle \tau \rangle$.)   Parameters are $L=10$ in both cases, $\kappa=\pi$ in (a).}
  \label{fig:omega_vs_tau}
\end{figure}

Specifically, a larger $\omega$ tends to delay the detection time when other parameters are kept fixed; it leads to a distribution $\rho_T(t)$ that has less weight in the region of the main peak of the comparison curve $\rho^\fr(t)$ and more weight to the right of it. As a consequence, the average detection time
\be\label{ETdef}
\EEE T=\int_0^\infty dt \, t \, \rho_T(t)
\ee
increases with $\omega$, see Figure~\ref{fig:omega_vs_tau}. (See Remark~\ref{rem:mean} about the use of $\mu^*(T)$ for estimating $\EEE T$ from the data.)

In our numerical experiments, we observe an $\omega$ dependence of the form
\be\label{ETomega}
\EEE T= A+ B\sqrt{\omega}
\ee
with positive coefficients $A,B$ (see Figure~\ref{fig:omega_vs_tau}a). For comparison, Das and D\"urr \cite{DD19} found that, within their prediction scheme for the detection-time distribution, the mean \emph{decreases} with $\omega$. Recall that for the CAP and the spinless ABC, the distribution of $T$ (and thus also its mean) was independent of $\omega$. An analytic derivation of 
\eqref{ETomega} and a discussion of the reasons behind this interesting $\omega$-dependence will be given in a separate work \cite{JT26b}.

\section{Possible Experimental Implementation and Physical Scales}
\label{sec:experimental}

Following the experimental scenario proposed by Das and D\"urr, we consider a particle initially prepared in the longitudinal slab $0<z<d$ and released into a guided geometry toward a downstream detector. As a concrete realization of this scenario, we focus on single ${}^{87}\mathrm{Rb}$ atoms in atom-chip waveguides \cite{Hei10}, which are a natural candidate for implementing such a setup since they allow one to confine and guide atoms along a prescribed direction while also providing local control and detection near the end of the guide. Alternatively, an optical guide could be used to reduce spin-dependent
trapping effects.  Moreover, ${}^{87}\mathrm{Rb}$ offers a natural effective spin-1/2 degree of freedom, since a suitable pair of internal states can be used to encode a two-level system. (Note however, that then $\vj^\mathrm{C}$ rather than $\vj^\mathrm{P}$ should be the appropriate current.)  The clearest experimentally testable signatures are the presence or absence of pronounced late-time tails, oscillatory structure associated with partial reflection, the dependence on or independence of the initial spin state, and, for the spinor ABC, the pronounced dependence on the transverse confinement parameter $\omega$.

\paragraph{Physical scales.}
To translate the dimensionless numerical results into physical units, we choose a physical value $d_{\mathrm{phys}}$ for the width of the initial longitudinal slab and identify $m_{\mathrm{phys}}$ with the particle mass, here the mass of $^{87}\mathrm{Rb}$. Our simulations are performed in dimensionless units with $d=1$, $\hbar=1$, and $m=1$. Thus, the slab width sets the basic unit of length, so that once $d_{\mathrm{phys}}$ is fixed, every spatial quantity in the numerics acquires a direct physical meaning. More precisely, if $\tilde x,\tilde y,\tilde z,\tilde t$ denote the dimensionless variables used in the simulation, then the corresponding physical variables are defined by
\be
x=d_{\mathrm{phys}}\tilde x,\qquad
y=d_{\mathrm{phys}}\tilde y,\qquad
z=d_{\mathrm{phys}}\tilde z,\qquad
t=T_*\tilde t,
\ee
where the associated unit of time is
\be
T_*=\frac{m_{\mathrm{phys}}d_{\mathrm{phys}}^2}{\hbar}.
\ee
This is the natural Schr\"odinger time scale obtained by restoring the physical constants in the kinetic term. It follows immediately that a dimensionless length $a$ in the numerics corresponds to the physical length $a\,d_{\mathrm{phys}}$, while a dimensionless time $t$ corresponds to the physical time $t\,T_*$. In the same way, all other parameters inherit their physical units from these basic scales. In particular, since an angular frequency has units of inverse time, the dimensionless trap parameter $\omega$ translates into the physical transverse angular frequency according to
\be
\Omega_\perp=\frac{\omega}{T_*},
\ee
Likewise, the absorbing-boundary parameter $\kappa$ has the dimension of inverse length. Indeed, starting from the physical boundary condition $\partial_z\Psi=i\kappa_{\mathrm{phys}}\Psi$
and using $z=d_{\mathrm{phys}}\tilde z$, one has
\be
\partial_z=\frac{1}{d_{\mathrm{phys}}}\partial_{\tilde z},
\ee
so that the dimensionless boundary condition becomes $\partial_{\tilde z}\Psi=i(\kappa_{\mathrm{phys}}d_{\mathrm{phys}})\Psi.$ Hence the numerical parameter is $\kappa=\kappa_{\mathrm{phys}}d_{\mathrm{phys}}$, or equivalently
\be
\kappa_{\mathrm{phys}}=\frac{\kappa}{d_{\mathrm{phys}}}.
\ee
Therefore, once $d_{\mathrm{phys}}$ and $m_{\mathrm{phys}}$ are fixed, all spatial, temporal, trapping, and detector parameters appearing in our simulations are translated uniquely into experimentally meaningful units.

As a representative reference choice, we take single ${}^{87}\mathrm{Rb}$ atoms and choose the
physical width of the initial longitudinal slab as
\be\label{dphys}
d_{\mathrm{phys}}=2~\mu\mathrm{m} \,.
\ee
This choice is not unique, but provides a convenient intermediate scale for us. It gives the corresponding device dimensions in the tens-to-hundreds of micrometers range while placing the characteristic evolution times in the millisecond regime. With the mass of ${}^{87}\mathrm{Rb}$, the natural time unit becomes
\be
T_*=\frac{m_{\mathrm{Rb}}d_{\mathrm{phys}}^2}{\hbar}\approx 5.47~\mathrm{ms}.
\ee


\paragraph{The case of CAP.}
In the CAP model,  probability is removed in the layer \(z_0\le z\le L\)  according to \eqref{PTR}
with the local physical rate
\begin{equation}
   \gamma_{\rm loss}(z)=\frac{2W_{\rm phys}(z)}{\hbar}.
\end{equation}
For cold atoms, such a position-dependent loss rate can be realized by a
localized, shaped optical detection field near the downstream end of the guide.
The field may couple the guided atomic state to a fluorescing, ionizing, or
untrapped channel, and the first time-resolved fluorescence, ionization, or
loss signal is then the detector click.  Conditional on no click having occurred,
eliminating the lossy channel gives an effective non-Hermitian potential
\begin{equation}
   V_{\rm eff}(z)=V_{\rm LS}(z)-iW_{\rm phys}(z),
\end{equation}
where \(V_{\rm LS}\) is the light shift and \(W_{\rm phys}(z)\) is controlled by
the local laser intensity, equivalently by the squared Rabi-frequency profile for
fixed detuning and decay parameters
~\cite{RMH09,MPNE04,NEGH03,JEHM03}.
Here we assume that \(V_{\rm LS}\) is negligible or compensated.  For the
spin-\(\frac12\) comparison, we also assume that the optical loss, and any
uncompensated light shift, act equally on the two internal states; otherwise the
experiment would realize a spin-dependent CAP rather than the scalar,
spin-blind CAP studied here.

In the dimensionless variables used in the simulations,
\begin{equation}
   W_{\rm phys}(z)=\frac{\hbar}{T_*}W(z),\qquad
   \gamma_{\rm loss}(z)=\frac{2W(z)}{T_*}.
\end{equation}

For \(T_*\simeq5.47\,{\rm ms}\) and \(W_{\max}=40\), this gives
\begin{equation}
   \gamma_{\max}\simeq1.46\times10^4\,{\rm s}^{-1}.
\end{equation}
With \(d_{\rm phys}=2\,\mu{\rm m}\), the representative geometry
\(z_0=10\), \(L=11\) corresponds to a \(2\,\mu{\rm m}\) absorbing layer, and
the cubic ramp of Appendix~\ref{app:variation} has physical turn-on length
\(w(L-z_0)d_{\rm phys}\).  Thus \(w=1\) and \(w=0.5\) correspond to optical
turn-on lengths \(2\,\mu{\rm m}\) and \(1\,\mu{\rm m}\), respectively.  The
case \(w=0.01\), corresponding to \(20\,{\rm nm}\), should instead be regarded
as a numerical sharp-step limit rather than as a directly realizable optical
ramp, since an experimental ionization profile cannot be varied on a scale much
shorter than the relevant optical wavelength.

Over the observation window $t\le 20$, i.e., about $109\,\mathrm{ms}$, the corresponding detected fraction is about $93\%$ for $w=1$, $88\%$ for $w=0.5$, and $80\%$ for $w=0.01$.  At the same time, the (restricted) mean detection time 
shifts from
$\mu^\ast(T)\approx 5.921$ to $7.191$ and $8.909$, which in physical units gives
\be
\mu^\ast(T)\approx 32.4\,\mathrm{ms},\qquad 39.3\,\mathrm{ms},\qquad 48.7\,\mathrm{ms}.
\ee

\paragraph{The case of the spinless ABC.}
The spinless absorbing boundary condition can be interpreted as an idealized hard detector placed at the downstream end surface of the guide. Unlike the CAP, which models detection in a finite absorbing region, the spinless ABC represents the limiting situation in which the detector clicks as soon as the atom reaches the terminal surface. Experimentally, this idealized hard-detector behavior could be approximated by a sufficiently narrow, state-insensitive detection zone near the downstream end of the guide, for example using localized fluorescence detection on an atom chip~\cite{Hei10} or, more sharply, a small optical cavity mode intersecting the guided atomic beam~\cite{Haa06}. 

\bigskip

For the spinor ABC, a concrete experimental realization is more subtle; a discussion of possible implementations and of the $\omega$-dependence will be given elsewhere \cite{JT26b}.

\section{Conclusion and Outlook}
\label{sec:outlook}

We have computed predictions for the probability distribution of the detection time on the basis of complex absorbing potentials (CAPs) and absorbing boundary conditions (ABCs) for a setup intended to be experimentally accessible. It would be of interest to carry out such experiments and compare their outcomes to these predictions. Part of the question is which real-world detectors are better described by CAPs and which by ABCs, and with which parameters.

There are several open questions about these predictions that further theoretical studies could explore. First, spinor ABCs and in particular the $\omega$ dependence of their $\rho_T(t)$ are investigated further in a separate work \cite{JT26b}. Second, it would be of interest to also study analytic solutions of the Schr\"odinger equation with ABCs (steps in this direction were taken in \cite{CD25}). Third, there are further ABCs, such as \eqref{nospinABC} with complex $\kappa$ (with $\Re \kappa>0$) \cite{detect-rule}, and the question arises which one a given real-world detector is closest to. Fourth, other initial states $\Psi_0$ may be better suited for concrete experimental setups; they might involve Gaussian shapes in $z$, higher transverse modes, or finite-temperature mixtures. Fifth, the question arises in which ways a direct treatment of the Dirac equation \cite{detect-dirac} would yield different results. Finally, the presence of explicit spin dynamics in the bulk Hamiltonian (such as external magnetic fields or Rashba terms representing spin--orbit coupling) would affect the detection time distribution $\rho_T(t)$.

\appendix

\section{Numerical Solution of the TDSE in a Finite Box}
\label{app:TDSE}

We numerically solve the two-component (spin-$\tfrac12$) time-dependent Schr\"odinger equation (TDSE) in the finite
box $\Omega=[0,L_x]\times[0,L_y]\times[0,L_z]$ with $L_z=L$,
\begin{equation}
    i\,\partial_t \Psi(\vr,t)
=\Bigl(-\tfrac12\nabla^2 + V(\vr)\Bigr)\Psi(\vr,t),\qquad
\Psi=\begin{pmatrix}\psi_\uparrow\\ \psi_\downarrow\end{pmatrix},
\label{eq:Schr\"odinger}
\end{equation}
with Dirichlet conditions at $x=0,L_x$, $y=0,L_y$, and on the bottom face $z=0$, and an absorbing
(spinor Robin) boundary at the roof $z=L$:
\be
(\boldsymbol\sigma\!\cdot\!\nabla)\Psi
= i\,\kappa\,(\vn\!\cdot\!\boldsymbol\sigma)\Psi
\quad\text{on } z=L,\qquad \vn=\hat{\boldsymbol{z}}.
\ee
The potential is $V(\vr)=\tfrac12\omega^2\bigl[(x-\tfrac{L_x}{2})^2+(y-\tfrac{L_y}{2})^2\bigr]$.

\paragraph{Grid and ordering.}
We discretize $\Omega$ on a uniform Cartesian grid with nodes
$x_i = i\,h_x$ for $i=0,\dots,N_x-1$ (with $h_x = L_x/N_x$), and analogously
$y_j = j\,h_y$ and $z_k = k\,h_z$. We use \texttt{endpoint=false} in $z$, so the last
$z$-node lies one step inside the domain: $z_{N_z-1} = L_z - h_z$.
Each spin component $\psi_\uparrow$ and $\psi_\downarrow$ is stored on the same
$(N_x,N_y,N_z)$ grid. We flatten the 3D array to a single 1D vector by lexicographic ordering $(i,j,k)$ with linear index
\be
\ell \;=\; i\,N_yN_z \;+\; j\,N_z \;+\; k,
\ee
so for fixed $(i,j)$ the entries with $k=0,\dots,N_z\!-\!1$ are consecutive in memory\footnote{Here ``lexicographic'' means we sweep $z$ fastest, then $y$, then $x$.} and each component becomes a length-$N$ vector with $N = N_x N_y N_z$. In other words, $\ell$ is the 1-dimensional (linear) index of the grid point $(i,j,k)$ in the flattened state vector used by the Crank--Nicolson step and the GMRES solver. For any fixed time $t$, the grid values of the spinor components are
\be
\psi_{\uparrow,\ell}(t) := \psi_\uparrow(x_i,y_j,z_k,t),\quad
\psi_{\downarrow,\ell}(t) := \psi_\downarrow(x_i,y_j,z_k,t),
\ee
for $0 \le \ell \le N-1$, where $\ell$ is the linear index associated with $(i,j,k)$ via the mapping above.
The corresponding flattened spinor is then assembled by stacking the two components ($\mathbb{C}^N \oplus \mathbb{C}^N \cong \mathbb{C}^{2N}$):
\be\label{psiflatdef}
\psi_{\rm flat}(t)
=
\bigl(
\psi_{\uparrow,0}(t),\dots,\psi_{\uparrow,N-1}(t),
\psi_{\downarrow,0}(t),\dots,\psi_{\downarrow,N-1}(t)
\bigr)^\top \in \mathbb{C}^{2N}.
\ee

\paragraph{Second derivatives and finite-difference stencils.}
All interior second derivatives are discretized by standard \emph{second-order central differences}. Along a given axis with spacing $h$, the centered finite-difference stencil for the 1D second derivative is $\bigl[\tfrac{1}{h^2},\ -\tfrac{2}{h^2},\ \tfrac{1}{h^2}\bigr].$ Accordingly, the stencil for the kinetic operator $-\tfrac12\,\partial^2_{x}$ is $
\bigl[-\tfrac{1}{2h^2},\ \tfrac{1}{h^2},\ -\tfrac{1}{2h^2}\bigr].$ Near Dirichlet faces, the ghost value is eliminated using the boundary value $\psi=0$, which yields a modified one-sided stencil at the first interior node (e.g.\ at $i=1$ along $x$):
\be
-\tfrac12\,\partial^2_{x}\psi(x_1)
\;\approx\;
-\frac{1}{2h_x^2}\Bigl(\psi_{0}-2\psi_{1}+\psi_{2}\Bigr)
=
\frac{1}{h_x^2}\psi_{1}-\frac{1}{2h_x^2}\psi_{2}
\ee
writing $x_i$ for $ih_x$ and $\psi_i$ for $\psi(x=x_i)$,
and analogously at the last interior node $x_{N_x-2}$,
\be
\begin{aligned}
-\tfrac12\,\partial^2_{x}\psi(x_{N_x-2})
&\approx -\frac{1}{2h_x^2}\Bigl(\psi_{N_x-3}-2\psi_{N_x-2}+\psi_{N_x-1}\Bigr)\\
&= -\frac{1}{2h_x^2}\Bigl(\psi_{N_x-3}-2\psi_{N_x-2}+0\Bigr)\\
&= -\frac{1}{2h_x^2}\psi_{N_x-3}+\frac{1}{h_x^2}\psi_{N_x-2}.
\end{aligned}
\ee
The same construction is used along $y$. Along $z$, interior nodes again use second--order central differences. At $z=0$ we impose
Dirichlet exactly as above. At the roof $z=L_z$ we apply a forward one--sided ghost--point
elimination for the scalar Robin-type boundary condition $\partial_z\psi - i\kappa\,\psi=0$.
Approximating $\partial_z\psi|_{z_{N_z-1}}\approx(\psi_N-\psi_{N-1})/h_z$ gives
$\psi_N=(1+i\kappa h_z)\,\psi_{N-1}$. Substituting this into the central stencil at the last
interior node yields
\be
\begin{aligned}
-\tfrac12\,\partial^2_{z}\psi(z_{N_z-1})
&\approx -\frac{1}{2h_z^2}\Bigl(\psi_{N_z-2}-2\psi_{N_z-1}+\psi_{N}\Bigr)\\
&= -\frac{1}{2h_z^2}\psi_{N_z-2}
\;+\;\Bigl(\frac{1}{2h_z^2}-\frac{i\kappa}{2h_z}\Bigr)\psi_{N_z-1}\,,
\end{aligned}
\ee
this modifies only the roof row; the interior remains second--order accurate in space.

\paragraph{Spinor absorbing boundary on the roof.}
At the physical roof \(z=L\) we impose the spinor ABC \eqref{spinABC}, which we repeat here in the form \eqref{spinABCzL}:
\be
\partial_z \psi_\uparrow = -(\partial_x-i\partial_y)\psi_\downarrow + i\kappa\,\psi_\uparrow,
\qquad
\partial_z \psi_\downarrow =  (\partial_x+i\partial_y)\psi_\uparrow + i\kappa\,\psi_\downarrow .
\ee
To \emph{implement} this numerically, we eliminate the ghost value at \(z=L_z\) by a forward,
one--sided (ghost--point) discretization at the last interior plane \(k=N_z-1\). Substituting the
eliminated ghost into the central \(z\)--second--derivative stencil at \(k=N_z-1\) yields, for example,
\be
-\tfrac{1}{2}\,\partial^2_{z}\psi_\uparrow\Big|_{N_z-1}
\;\approx\;
-\frac{1}{2h_z^2}\,\psi_{\uparrow,N_z-2}
\;+\;
\Bigl(\frac{1}{2h_z^2}-\frac{i\kappa}{2h_z}\Bigr)\,\psi_{\uparrow,N_z-1}
\;+\;
\frac{1}{2h_z}\,(\partial_x-i\partial_y)\,\psi_{\downarrow,N_z-1}.
\ee
In words: each spin component acquires the same \emph{diagonal} roof--row factor
\(\bigl(\tfrac{1}{2h_z^2}-\tfrac{i\kappa}{2h_z}\bigr)\) as in the scalar case, plus
\emph{off--diagonal} cross--spin couplings on the roof plane proportional to the transverse
gradients of the opposite component. Discretizing \(\partial_x\) on that plane by \emph{central differences} gives the difference operator $D_x \psi|_{x_i}=\frac{1}{2h_x}(\psi_{i+1}-\psi_{i-1})$ (accurate to second order) with rows touching Dirichlet edges handled in the standard way described above; likewise for $D_y$.  We denote the off diagonal roof plane coupling operators by
\be
C_{\downarrow\uparrow}\coloneqq -\frac{1}{2h_z}\,(D_x+iD_y)\big|_{k=N_z-1},
\qquad
C_{\uparrow\downarrow}\coloneqq +\frac{1}{2h_z}\,(D_x-iD_y)\big|_{k=N_z-1}\,.
\ee

Thus, the \(C\)--blocks implement \(-\tfrac{1}{2h_z}(\partial_x\!\pm\! i\,\partial_y)\)
on the roof plane, and \(C_{\uparrow\downarrow}=C_{\downarrow\uparrow}^\dagger\).
All nonzero entries of these blocks are confined to the \emph{top interior layer} $k = N_z-1$
(the on--grid counting plane $z_{\rm det} = z_{N_z-1}$), while the ABC itself is imposed
at $z = L_z$ via a forward ghost point at index $k = N_z$ which is then eliminated.

\paragraph{Hamiltonian assembly.}
Define the (spin--blind) scalar Laplacian as the Kronecker sum
\be
\mathcal L_{\rm sc}
=
\mathcal{L}_x\!\otimes\!I_y\!\otimes\!I_z
\;+\;
I_x\!\otimes\!\mathcal{L}_y\!\otimes\!I_z
\;+\;
I_x\!\otimes\!I_y\!\otimes\!\mathcal{L}_z,
\ee
where $\mathcal{L}_x$, $\mathcal{L}_y$, and $\mathcal{L}_z$ are the 1D lattice Laplacians with Dirichlet boundaries. Now let $\psi_{\rm flat}(t)\in\mathbb{C}^{2N}$ denote the flattened grid representation of $\Psi(\cdot,t)$ defined in \eqref{psiflatdef}.
With the diagonal matrices given by the real potential $V$ and (whenever needed) the imaginary potential $-iW$, the Hamiltonian acting on $\psi_{\rm flat}$ is the $2N\times2N$ matrix
\be
H =
\begin{bmatrix}
\mathcal L_{\rm sc}+V-iW & C_{\uparrow\downarrow}\\[2pt]
C_{\downarrow\uparrow} & \mathcal L_{\rm sc}+V-iW
\end{bmatrix}
- i\frac{\kappa}{2h_z}
\begin{bmatrix}
P_{\rm roof} & 0\\[2pt]
0 & P_{\rm roof}
\end{bmatrix},
\ee
where $C_{\downarrow\uparrow},C_{\uparrow\downarrow}\in\mathbb{C}^{N\times N}$ are the sparse roof-plane
cross--spin couplings supported only on the top layer $k=N_z-1$, and $P_{\rm roof}$ is the projection onto that
layer.

\paragraph{Time stepping: Crank--Nicolson + Krylov (GMRES).}
Now that we have spatially discretized the TDSE \eqref{eq:Schr\"odinger} into the form \(i\,\partial_t\psi=H\psi\), we turn to the time discretization. One step \(t^n\to t^{n+1}=t^n+\Delta t\) is advanced by the second--order, implicit Crank--Nicolson (CN) discretization
\begin{equation}
\Bigl(I+\tfrac{i\Delta t}{2}\,H\Bigr)\,\psi^{n+1}
\;=\;
\Bigl(I-\tfrac{i\Delta t}{2}\,H\Bigr)\,\psi^{n},
\qquad
\psi^{n}\in\mathbb{C}^{2N} 
\end{equation}
where $ n=0,1,\dots,N_t-1$,  $N_t = \left\lfloor \frac{T}{\Delta t} \right\rfloor, \psi^{\,n} \approx \Psi(\,\cdot\,, t^n) := \psi_{\rm flat}(t^n)$ and $t^n= n\,\Delta t\,$. Define
\begin{equation}
A \;:=\; I+\tfrac{i\Delta t}{2}\,H,
\qquad
B \;:=\; I-\tfrac{i\Delta t}{2}\,H,
\end{equation}
so each step involves solving the linear equation
\begin{equation}
A\,\psi^{n+1} \;=\; B\,\psi^{n}. 
\label{eq:Spars}
\end{equation}
It is convenient to split the discrete Hamiltonian into Hermitian and anti-Hermitian parts
\be
H_{\rm H}:=\tfrac12\!\left(H+H^\dagger\right),\quad
\Gamma:=\tfrac{i}{2}\!\left(H-H^\dagger\right)=\Gamma^\dagger\geq 0,
\quad\text{so that}\quad H=H_{\rm H}-i\,\Gamma.
\ee
When \(\Gamma= 0\) (no absorbers), the CN update
\(U=(I+\tfrac{i\Delta t}{2}H_{\rm H})^{-1}(I-\tfrac{i\Delta t}{2}H_{\rm H})\)
is a Cayley transform and hence unitary (\(U^\dagger U=I\)).
With absorption (\(\Gamma\neq 0\)), CN is contractive,
\(\|\psi^{n+1}\|_2\le \|\psi^{n}\|_2\), and remains second--order accurate in \(\Delta t\).

Both the ABC and the CAP  render the Hamiltonian \(H\) and hence the Crank--Nicolson
matrix \(A\) non-Hermitian, so the Conjugate Gradient method (which requires a Hermitian
positive definite system matrix) cannot be used. We therefore solve \eqref{eq:Spars}
with GMRES, a Krylov subspace method for sparse non-Hermitian systems. GMRES builds the
subspace \(\mathcal{K}_k(A,r_0)=\operatorname{span}\{r_0,\allowbreak Ar_0,\allowbreak \ldots,\allowbreak A^{k-1}r_0\}\), and, at iteration \(k\), computes \(x_k \in x_0+\mathcal{K}_k(A,r_0)\) that minimizes the residual norm
\(\|b-Ax\|_2\) over the affine space \(x_0+\mathcal{K}_k(A,r_0)\), where \(r_0=b-Ax_0\) is the
initial residual. We use a restarted variant with subspace dimension \(M\) and
stop when the relative residual \(\|b-Ax_k\|_2/\|b\|_2\) falls below the prescribed tolerance,
\begin{equation}
\frac{\|r_k\|_2}{\|b\|_2}\le \texttt{tol},
\qquad
r_k=b-Ax_k .
\end{equation}
Each iteration augments the search space with a higher power of \(A\), while the restart
parameter \(M\) limits the size of the stored Krylov basis: after \(M\) inner iterations the
method is restarted with \(x_0\leftarrow x_M\) and a new residual \(r_0\).
For many production runs, each CN step was solved with restarted GMRES using a relative residual tolerance in the range $10^{-7}$ to $10^{-8}$, restart parameter $M=30$, and $\texttt{maxiter}=1000$.

\section{Numerical Bohmian Equations of Motion}
\label{app:Bohm}

We compute Bohmian trajectories from the Pauli current after each TDSE update as described in Appendix~\ref{app:TDSE}. 

\paragraph{Equation of motion and fields.}
With $\Psi=(\psi_\uparrow,\psi_\downarrow)^{\!\top}$,
\be
\rho=\Psi^\dagger\Psi,\quad
\mathbf S=\Psi^\dagger\boldsymbol\sigma\Psi,\quad
\vj^\mathrm{C}=\Im(\Psi^\dagger\nabla\Psi),\quad
\vj^\mathrm{P}=\vj^\mathrm{C}+\tfrac12\,\nabla\times\mathbf S,\quad
\vv=\frac{\vj^\mathrm{P}}{\rho}.
\ee
Each particle position $\vQ(t)=(Q_x(t),Q_y(t),Q_z(t))$ solves $d\vQ/dt=\vv(\vQ(t),t)$
until the first crossing of the counting plane $\Sigma_L=\{z=L\}$.

\paragraph{Discrete derivatives used for $\vj$ and $\nabla\times\mathbf S$.}
Because the interior grid points have enough neighbors on both sides, the $x$- and $y$-derivatives can use a fourth-order-accuracy central stencil in the interior, and standard second-order-accuracy one-sided stencils on the first interior layers next to each boundary. The discretization of the $z$-derivative is kept second-order-accuracy throughout, so that it remains consistent with the second-order-accuracy absorbing roof row and does not require additional ghost layers. That is:
\be
(\partial_x u)_{i,j,k} \approx
\begin{cases}
\dfrac{-u_{i+2,j,k}+8u_{i+1,j,k}-8u_{i-1,j,k}+u_{i-2,j,k}}{12\,h_x},
& 2\le i \le N_x-3, \\[8pt]
\dfrac{-3u_{1,j,k}+4u_{2,j,k}-u_{3,j,k}}{2\,h_x}, & i=1,\\[8pt]
\dfrac{u_{N_x-4,j,k}-4u_{N_x-3,j,k}+3u_{N_x-2,j,k}}{2\,h_x}, & i=N_x-2,\\[8pt]
\dfrac{u_{1,j,k}-u_{0,j,k}}{h_x}, & i=0,\\[8pt]
\dfrac{u_{N_x-1,j,k}-u_{N_x-2,j,k}}{h_x}, & i=N_x-1,
\end{cases}
\ee
with the $y$-derivative obtained by the obvious replacement $i\to j$, $h_x\to h_y$.
In $z$ we use
\be
(\partial_z u)_{i,j,k} \approx
\begin{cases}
\dfrac{u_{i,j,k+1}-u_{i,j,k-1}}{2\,h_z}, & 1\le k \le N_z-2,\\[8pt]
\dfrac{u_{i,j,1}-u_{i,j,0}}{h_z}, & k=0,\\[8pt]
\dfrac{u_{i,j,N_z-1}-u_{i,j,N_z-2}}{h_z}, & k=N_z-1.
\end{cases}
\ee
Dirichlet rows use the enforced boundary values (e.g., $u_0=0$), and the roof
already encodes the spinor absorbing boundary through the TDSE discretization.

\paragraph{Interpolation to particle positions.}
Given grid fields $u_{i,j,k}$, values at particle locations $\vQ$ are obtained by
standard trilinear interpolation within the containing Cartesian cell. The same
interpolant is applied componentwise to $\vj$, to $\rho$, and to
$\nabla\times\mathbf S$. To regularize divisions by very small densities we use a
dynamic floor $ \rho_\varepsilon(\vQ)
=\max\{\rho(\vQ), \,\varepsilon\,\rho_{\max}(t)\}$ with $\varepsilon=10^{-6}$ and $\rho_{\max}(t)=\max_{\Omega}\rho(\cdot,t)$. 

\paragraph{Initial sampling (Born rule).}
Initial particle positions $\vQ(0)$ are drawn independently and $|\Psi(\vr,0)|^2$ distributed:
$Q_x(0),Q_y(0)\sim \mathcal N(L_x/2,\,1/(2\omega))$ (clipped to the box),
$Q_z(0)$ sampled from $2\sin^2(\pi z)$ (taking $d=1$) on $(0,1)$ by 
accept--reject sampler (i.e., propose $u\sim U(0,1)$ and $r\sim U(0,2)$; accept $Q_z(0)=u$ if $r<2\sin^{2}(\pi u)$,
otherwise resample). 
A small offset
$\varepsilon\sim\min(h_x,h_y,h_z)/2$ avoids starting exactly on faces. 

\paragraph{Time integration: RK2 with per--stage CFL capping.} 
 \label{cap}
RK2 is the second-order Runge--Kutta explicit midpoint method (second-order accurate in time).

Because the Bohmian velocity field $\vv(\vr,t)$ is available on a Cartesian grid
(and evaluated by trilinear interpolation), accurate timing of the first arrival on a surface requires that
each update moves a particle by less than about one cell per stage; otherwise a trajectory could
skip the counting plane or sample the field nonlocally. We therefore impose a local displacement
bound (a particle-advection analogue of the Courant--Friedrichs--Lewy condition) and propagate
only particles whose first hit has not yet occurred. Upon the first crossing of the detector
plane $z=L$, we record $\tau$ and remove that trajectory from further updates.

 For each \emph{not--yet--arrived} particle we advance the Bohmian
ODE with explicit midpoint (RK2):
\begin{align}
\vk_1&=\vv(\vQ^n,t^n),
&\vQ^{n+\frac12}=\vQ^n+\frac{\Delta t}{2}\,\widetilde{\vk}_1,\\
\vk_2&=\vv(\vQ^{n+\frac12},t^n+\tfrac{\Delta t}{2}),
&\vQ^{n+1}=\vQ^n+\Delta t\,\widetilde{\vk}_2,
\end{align}
where $\Delta t$ denotes the duration of the current RK2 advection call, \(\vk_1,\vk_2\) are the \emph{stage velocities} sampled from the guidance field, and $\widetilde{\vk}=s\,\vk$ are the capped velocities with
\be
s=\min\!\Bigl\{1,\frac{v_{\max}}{|\vk|}\Bigr\},\qquad
|\vk|=\sqrt{k_x^2+k_y^2+k_z^2},
\ee
this preserves the direction and reduces only the magnitude when needed. Writing $h_{\min}=\min\{h_x,h_y,h_z\}$, we define the (stage--wise) CFL number as the fraction of a cell
that a particle would travel in that stage:
\be
\mathrm{CFL}_{\text{stage}}
=\frac{|\vk|\,\Delta t_{\text{stage}}}{h_{\min}},
\qquad
\Delta t_{\text{stage}}=
\begin{cases}
\Delta t_\mathrm{}/2 & \text{for }\vk_1,\\
\Delta t_\mathrm{}   & \text{for }\vk_2.
\end{cases}
\ee
We enforce $\mathrm{CFL}_{\text{stage}}\le \mathrm{cfl}$ with a user choice $\mathrm{cfl}\in(0,1)$ by taking
\be
v_{\max} = \frac{\mathrm{cfl}\,h_{\min}}{\Delta t_{\text{stage}}}\,.
\ee
With this choice the predictor displacement is bounded by
$\tfrac12\,\mathrm{cfl}\,h_{\min}$ and the corrector by $\mathrm{cfl}\,h_{\min}$, so each RK2 call
moves a particle by at most $1.5\,\mathrm{cfl}\,h_{\min}$ (i.e., $1.2\,h_{\min}$ for $\mathrm{cfl}=0.8$).\footnote{Here, the \emph{predictor} displacement is a first guess of where the particle will go: it is the first RK2 substep
$\Delta\vQ_{\mathrm{pred}} := \vQ^{n+\frac12}-\vQ^n
= \tfrac{\Delta t}{2}\,\widetilde{\vk}_1$.
The \emph{corrector} displacement is a refined version of that guess and is given by the second substep
$\Delta\vQ_{\mathrm{corr}} := \vQ^{n+1}-\vQ^n
= \Delta t\,\widetilde{\vk}_2$.
Because $|\widetilde{\vk}|\le v_{\max}$ by construction, we have
$|\Delta\vQ_{\mathrm{pred}}|\le \tfrac12\,\mathrm{cfl}\,h_{\min}$ and
$|\Delta\vQ_{\mathrm{corr}}|\le \mathrm{cfl}\,h_{\min}$.}

Capping both stages thus ensures sub-cell motion, stable interpolation, and reliable first-hit bracketing at the detector plane. As $\Delta t\to 0$, the cap is rarely active; RK2 retains its $O(\Delta t^2)$ accuracy and the trajectories converge to the continuum Bohmian paths.\footnote{Equivalently, one can view the capping as introducing an \emph{adaptive} effective stage time
$\Delta t_{\mathrm{eff}}=\min\{\Delta t_{\text{stage}},\,\mathrm{cfl}\,h_{\min}/|\vk|\},$
which preserves the guidance direction and only shortens the stage when the proposed displacement would exceed a fraction of a cell. Since $\vv(\cdot,t)$ is frozen over $[t^n,t^{n+1}]$, reducing $\Delta t_{\mathrm{eff}}$ changes only the numerical step length, not the underlying physics.}

\section{Extraction of First Arrival Times}
\label{app:arrival}

We place the counting plane \emph{on grid} at the top layer
\(z_{\mathrm{det}} = z_{N_z-1}\) (the last \(z\) node inside the box).
During each TDSE step \(t^n \to t^{n+1}=t^n+\Delta t_{\text{TDSE}}\),
every particle whose first hit has not yet been recorded is advanced by a single
RK2 advection call of duration \(\Delta t\in\{\Delta t_{\text{TDSE}},\,\tfrac12\Delta t_{\text{TDSE}}\}\)
(full step or substep, as triggered near the roof or at large local Courant number).
Let \(\vQ^{n}\) be the particle position at time \(t^n\) (start of the call),
and let \(\vQ^{n+1}\) be the RK2 corrector at the end of the call.

\paragraph{First-hit test and time stamp.}
With \(L:=z_{\mathrm{det}}\), set \(z_0 = Q^{n}_z\) and \(z_1 = Q^{n+1}_z\).
A first crossing within this RK2 call occurs if and only if
\be
  z_0 < L \quad\text{and}\quad z_1 \ge L.
\ee
In that case we linearly interpolate along the taken segment
\(\vQ^{n} \to \vQ^{n+1}\) to find the hit fraction
\be
  m \;=\; \frac{L - z_0}{\,z_1 - z_0\,} \in (0,1],
\ee
place the particle exactly on the plane, and assign the corresponding timestamp:
\be
  \vQ^{\mathrm{hit}}
  \;=\;
  \vQ^{n} + m\bigl(\vQ^{n+1} - \vQ^{n}\bigr),
  \qquad
  Q^{\mathrm{hit}}_z = L,
  \qquad
  \tau \;=\; t^n + m\,\Delta t .
\ee
Intuitively, the particle hits the plane \(L\) at the fraction \(m\) of the RK2 call; we then read off both the position and the time at that same fraction. After recording \(\tau\), the particle is removed from further propagation (kept pinned on the plane), so only \emph{first} arrivals contribute. If no crossing occurs, we accept the corrector and use \(\vQ^{n+1}\) as the starting position for the next RK2 call.

\paragraph{Ensemble statistics (post-processing).}
Let
\be
\mathcal A=\{\,p:\text{particle $p$ has a recorded first hit}\,\}
\ee
be the set of arrived particles and $M_{\mathrm{arr}}=|\mathcal A|$.
The sample mean and variance of the arrival times are
\be
\bar{\tau}=\frac{1}{M_{\mathrm{arr}}}\sum_{p\in\mathcal A}\tau_p,
\qquad
s_\tau^2=\frac{1}{M_{\mathrm{arr}}-1}\sum_{p\in\mathcal A}(\tau_p-\bar{\tau})^2.
\ee
The arrival time distribution used in the figures is obtained by binning $\{\tau_p\}_{p\in\mathcal A}$.
(All quantities are computed from the stored arrays of first--hit times and arrival flags; no extra simulations are required.)

\paragraph{Remark on numerical convergence.}
We performed numerical convergence tests with respect to the spatial grid spacing $h$, the TDSE time step $\Delta t$, and the RK2 step size employed in the Bohmian trajectory integrator. For each parameter set, we compared the resulting arrival-time histograms, the restricted mean arrival time $\mu^*(\tau)$, and related summary observables under moderate refinement of the discretization. The numerical parameters used in the simulations reported here were chosen from a regime in which further refinement produced no systematic variation beyond the statistical uncertainty associated with the finite Bohmian ensemble, and no visually significant change in the corresponding distributions.

\paragraph{Representative convergence tests.}
As a representative numerical convergence study, we fixed the transverse confinement to $\omega=300$, which is sufficiently large to probe a demanding regime, and monitored the restricted mean arrival time $\mu^*$ under refinement of the temporal and spatial discretization parameters. The relative change is defined as
\be
\varepsilon_{\mathrm{rel}}
=
\frac{\left|\mu^*_{\mathrm{coarse}}-\mu^*_{\mathrm{fine}}\right|}{\left|\mu^*_{\mathrm{fine}}\right|}\times 100\%.
\ee

\begin{table}[H]
\caption{Representative convergence checks for the restricted mean arrival time $\mu^*$ at $\omega=300$, $L_x=L_y=0.6$, $L_z=L=10$.}
\label{tab:convergence_mu}
\centering
\begin{tabular}{lllllll}
\Xhline{1.0pt}
Test & $N_x=N_y$ & $N_z$ & $\Delta t$ & $t_\mathrm{cutoff}$ & $\mu^*$ & $\varepsilon_{\mathrm{rel}}$ \\
\hline
Time step & 100 & 1000 & $2\times10^{-4}$ & 20 & 14.167 & \\
Time step & 100 & 1000 & $5\times10^{-5}$ & 20 & 14.164 & $0.021\%$ \\
\hline
$N_x,N_y$ & 100 & 1000 & $1\times10^{-4}$ & 10 & 8.338 & \\
$N_x,N_y$ & 120 & 1000 & $1\times10^{-4}$ & 10 & 8.339 & $0.012\%$ \\
\hline
$N_z$ & 100 & 1000 & $1\times10^{-4}$ & 20 & 14.165 & \\
$N_z$ & 100 & 2000 & $1\times10^{-4}$ & 20 & 14.162 & $0.021\%$ \\
\Xhline{1.0pt}
\end{tabular}
\end{table}

\begin{table}[H]
\centering
\caption{Representative convergence checks for the restricted mean arrival time $\mu^*$ at $L_z=100$, $N_x=N_y=100$, $t_\mathrm{cutoff}=80$.}
\label{tab:convergence_mu2}
\begin{tabular}{lllllll}
\Xhline{1.0pt}
Test & $\omega$ & $N_z$ & $L_x=L_y$ & $\Delta t$ & $\mu^*$ & $\varepsilon_{\mathrm{rel}}$ \\
\hline
Time Step & 500 & 2000 & 0.45 & $1\times10^{-4}$ & 42.237 & \\
Time Step & 500 & 2000 & 0.45 & $2\times10^{-4}$ & 42.302 & $0.15\%$ \\
\hline
$N_z$ & 1 & 2000 & 10 & $1\times10^{-3}$ & 42.215 & \\
$N_z$ & 1 & 2500 & 10 & $1\times10^{-3}$ & 42.167 & $0.11\%$ \\
\Xhline{1.0pt}
\end{tabular}
\end{table}

Although only $\mu^*$ for representative demanding regime such as $\omega=300$ is reported in Table~\ref{tab:convergence_mu}, we found the same qualitative stability for the corresponding arrival-time histograms and related summary observables under the same refinements.

\paragraph{Code availability.}
All time-dependent simulations were performed with custom Python scripts using CuPy for GPU-accelerated linear algebra, \texttt{cupyx.scipy.sparse} for sparse operators and GMRES, and Matplotlib for visualization. The scripts used for the simulations, together with representative driver files and postprocessing notebooks used to generate the arrival-time histograms, probability-loss curves, figures, and animations of the probability-density evolution, are available at the GitHub repository \cite{JT26_code}.

\section{Derivation of Eq.s \eqref{PTRj} and \eqref{PTRjP}}
\label{app:transform}

Recall that for us $\vn=\hat{\boldsymbol{z}}$. In the spinless case, at any boundary point
\be
\vn\cdot\vj = j_z 
= \tfrac{\hbar}{m} \Im(\Psi^* \partial_z \Psi) 
\stackrel{\eqref{nospinABC}}{=} \tfrac{\hbar}{m} \Im(\Psi^* i\kappa \Psi) = \tfrac{\hbar\kappa}{m} |\Psi|^2\,,
\ee
which proves the equivalence of \eqref{PTRnospin} with \eqref{PTRj}, given \eqref{nospinABC}. In the spin-coupled case, using \eqref{spinABC} multiplied on both sides by $\sigma_z$,
\begin{align}
    \vn\cdot\vj^\mathrm{P} &= j_z^\mathrm{P} \stackrel{\eqref{Paulij}}{=} \tfrac{\hbar}{m} \Im(\Psi^* \partial_z \Psi) + \tfrac{\hbar}{2m} \partial_x(\Psi^\dagger \sigma_y \Psi) -  \tfrac{\hbar}{2m} \partial_y(\Psi^\dagger \sigma_x \Psi)\\
    &\stackrel{\eqref{spinABC}}{=} \tfrac{\hbar}{m} \Im\Bigl(\Psi^\dagger i \kappa \Psi - \Psi^\dagger \underbrace{\sigma_z \sigma_x}_{i\sigma_y} \partial_x \Psi - \Psi^\dagger \underbrace{\sigma_z \sigma_y}_{-i\sigma_x} \partial_y \Psi \Bigr) \nonumber\\
    &\qquad + \tfrac{\hbar}{m}\Re(\Psi^\dagger \sigma_y \partial_x \Psi) - \tfrac{\hbar}{m}\Re(\Psi^\dagger \sigma_x \partial_y \Psi)\\
    &= \tfrac{\hbar\kappa}{m} |\Psi|^2\,,    
\end{align}
which proves the equivalence of \eqref{PTRnospin} with \eqref{PTRjP}, given \eqref{spinABC}.

\section{Relations Between Absorption Rate and Distribution of Absorption Time and Place}
\label{app:rate}

Since generally, to say that a random time $T$ has (possibly time-dependent) rate $h(t)$ means that
\be
\PPP(t\leq T \leq t+dt|T\geq t) = h(t) \, dt\,,
\ee
and since, by the definition of conditional probabilities,
\be
\PPP(t\leq T \leq t+dt|T \geq t) = \frac{\PPP(t\leq T \leq t+dt)}{\PPP(T \geq t)} = \frac{-S'(t)}{S(t)} \, dt
\ee
with $S(t)$ the ``survival probability''
\be\label{Stdef}
S(t) := \PPP(T \geq t)\,,
\ee
one obtains the general formulas
\be\label{Sh}
S(t) = \exp\biggl(-\int_0^t h(t') \, dt' \biggr)
\ee
and
\be\label{rhoS'}
\rho_T(t) = -S'(t) = h(t) \, S(t)
\ee
for the probability density $\rho_T(t)$ of $T$. As a further consequence, the random variable $T$ can be simulated by the following method when the function $h(t)$ (and thus $S(t)$) is known: choose a random, uniformly distributed value $\Xi \in [0,1]$ and define
\be\label{Xi}
T= \inf\{\,t> 0:S(t)\le \Xi\,\} \,.
\ee
Then $T$ has the desired distribution \eqref{rhoS'} because by \eqref{Sh} and $h(t)\geq 0$, $S(t)$ is a decreasing function with values in $[0,1]$, and $\PPP(T\geq t)= \PPP(S(t)\geq \Xi)= \PPP(\Xi \in [0,S(t)])=S(t)$ as desired.

Applying this to Bohmian particles moving in a complex potential $-iW$ and ending with rate $\tfrac{2}{\hbar}W$ at the random time $T$, we find that conditionally on the specific trajectory that started at $\vQ(t=0)$,
\be\label{Lambda}
\PPP(T\geq t|\vQ(0))= \exp\biggl( -\int_0^t \tfrac{2}{\hbar} W(\vQ(t')) \, dt' \biggr)\,.
\ee
In particular, $T$ can be simulated for this trajectory by choosing a random threshold $\Xi$ (uniformly in $[0,1]$) and waiting until the right-hand side of \eqref{Lambda} falls below $\Xi$.

For the following further considerations, we stop conditionalizing on $\vQ(0)$ and take into account that $\vQ(0)$ has distribution density $|\Psi_0(\vr)|^2$.
The density $\rho(\vr,t)$ of the probability that the Bohmian particle has not been absorbed by $t$ and is located in $d^3\vr$ at $t$ evolves according to the modified continuity equation
\be
\partial_t \rho(\vr,t) + \nabla \cdot (\rho(\vr,t) \, \vv(\vr,t))= -h(\vr,t) \, \rho(\vr,t)\,,
\ee
where $h(\vr,t)$ is the absorption rate. (This can be understood by thinking about how amounts of probability would be transported and/or removed.) For the choice $h(\vr,t)=\tfrac{2}{\hbar}W(\vr,t)$ and $\vv$ given by \eqref{vdef}, this equation coincides with the continuity equation \eqref{continuityW}, therefore $\rho(\cdot,t)$ (the distribution density of $\vQ(t)$) agrees with $|\Psi_t(\cdot)|^2$ (equivariance) at every $t\geq 0$, as we claimed at the end of Section~\ref{sec:Bohmian}.

As a further consequence, we can determine the probability distribution of the time $T$ and place $\vR$ where the trajectory ends: First, from the fact that $\rho(\vr,t)=|\Psi_t(\vr)|^2$ it follows that $\PPP(T\geq t) = \|\Psi_t\|^2$, so $T$ has probability density $-\tfrac{d}{dt}\|\Psi_t\|^2$. Second, given that $T=t$, $\vQ(t)$ has distribution density $|\Psi_t(\vr)|^2/\|\Psi_t\|^2$; since the absorption rate is $\tfrac{2}{\hbar}W(\vr)$, the absorption place $\vR$ has distribution  
\be
\PPP(\vR\in d^3\vr|T=t) =
\frac{\tfrac{2}{\hbar}W(\vr) \, |\Psi_t(\vr)|^2/\|\Psi_t\|^2}
{\int d^3\vr' \,\tfrac{2}{\hbar}W(\vr') \, |\Psi_t(\vr')|^2/\|\Psi_t\|^2} d^3\vr = \frac{\tfrac{2}{\hbar}W(\vr) \, |\Psi_t(\vr)|^2}{-(d/dt)\|\Psi_t\|^2} d^3\vr\,,
\ee
so
\be
\PPP(t\leq T\leq t+dt,\vR\in d^3\vr) = \tfrac{2}{\hbar}W(\vr) \, |\Psi_t(\vr)|^2 \, dt \, d^3\vr\,,
\ee
which is the same distribution as used for the detection time and place in \eqref{PTR}.

\section{Shapes of Bohmian Trajectories}
\label{app:Bohmianresults}

\begin{figure}[H]
  \centering

  \begin{minipage}[t]{0.5\linewidth}
    \centering
    \includegraphics[width=\linewidth]{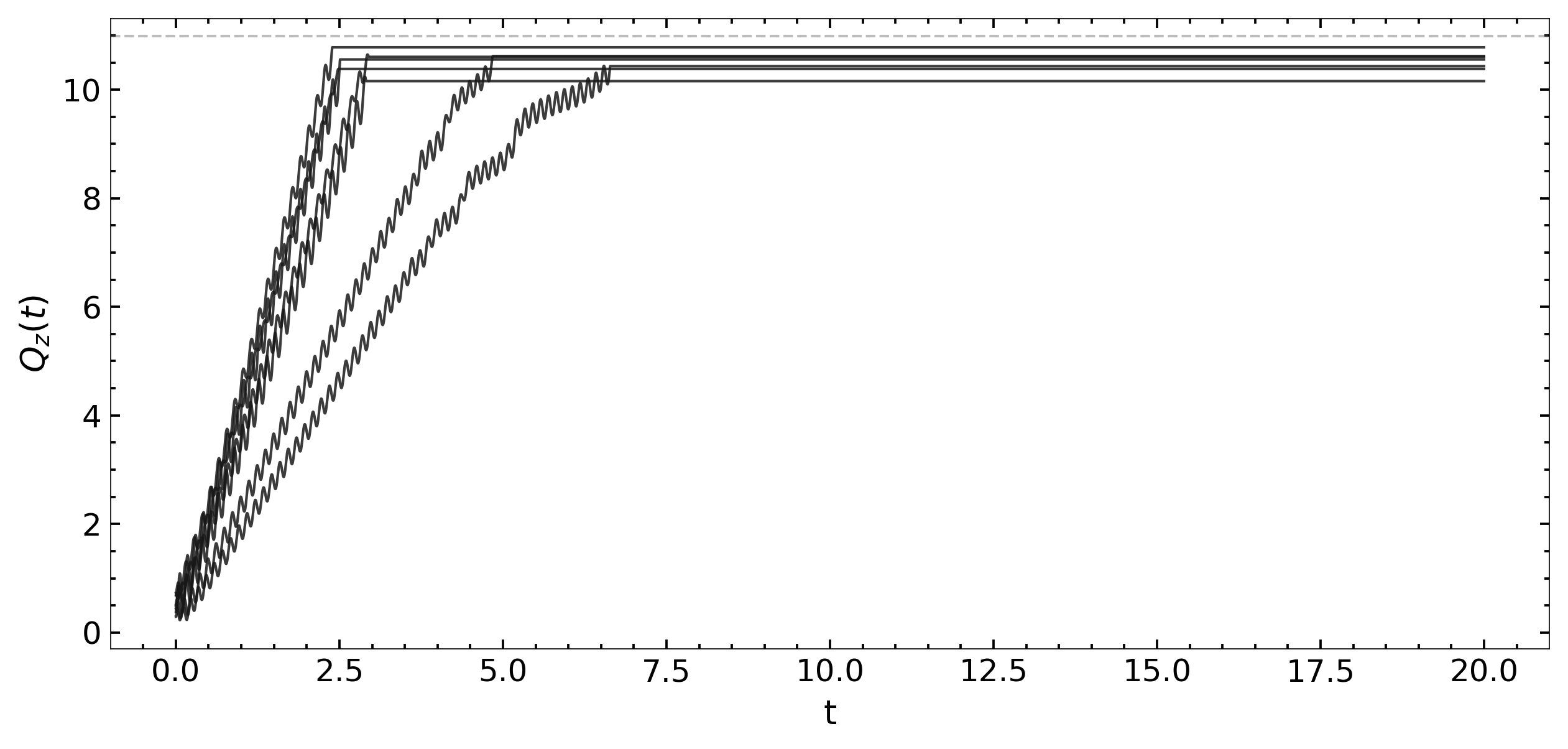}
    \par\vspace{0.3em}{\small (a) Soft-step CAP \eqref{Wcubic}, $w=1$}
  \end{minipage}\hfill
  \begin{minipage}[t]{0.5\linewidth}
    \centering
    \includegraphics[width=\linewidth]{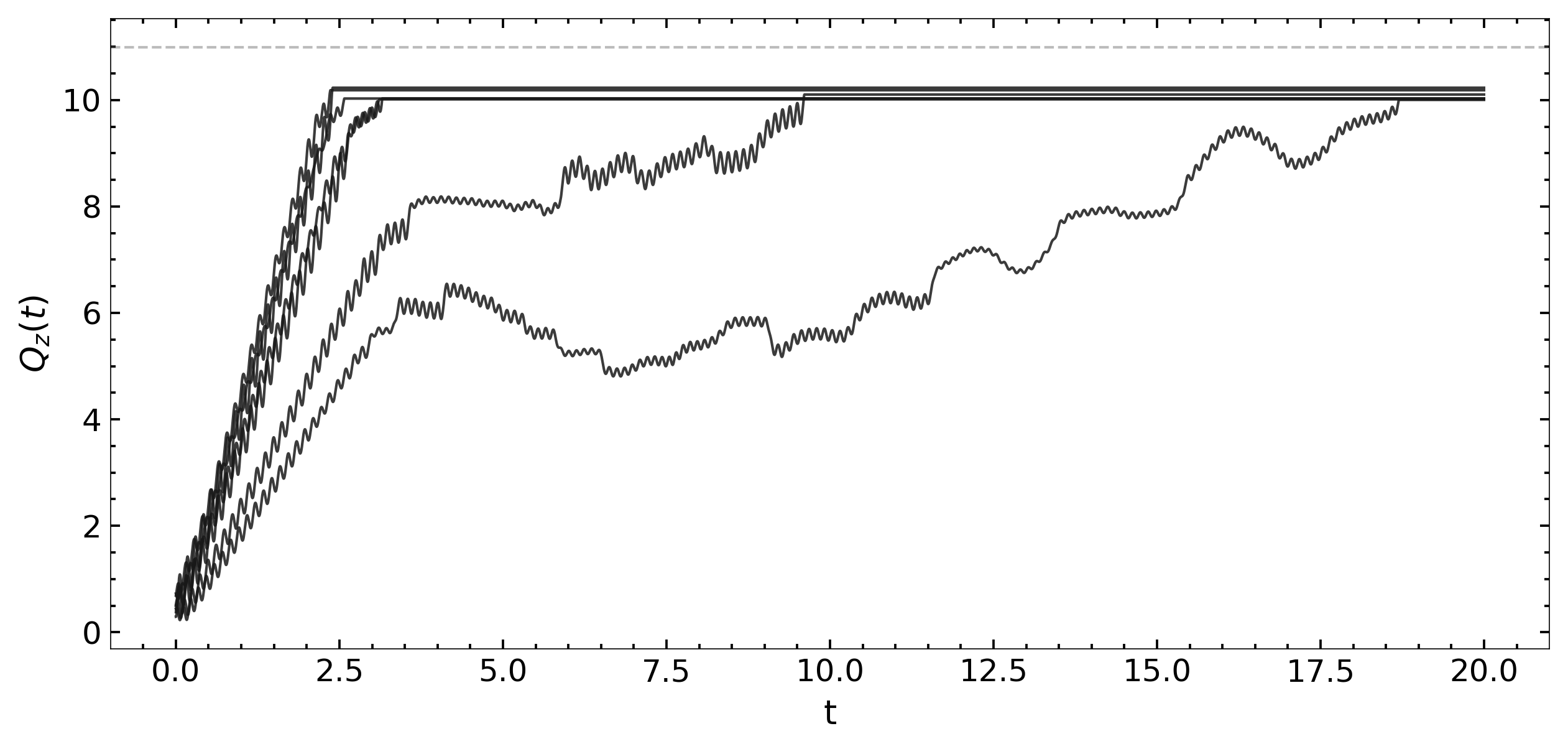}
    \par\vspace{0.3em}{\small (b) Sharp-step CAP \eqref{Wcubic}, $w=0.01$}
  \end{minipage}

  \vspace{0.8em}

  \begin{minipage}[t]{0.48\linewidth}
    \centering
    \includegraphics[width=\linewidth]{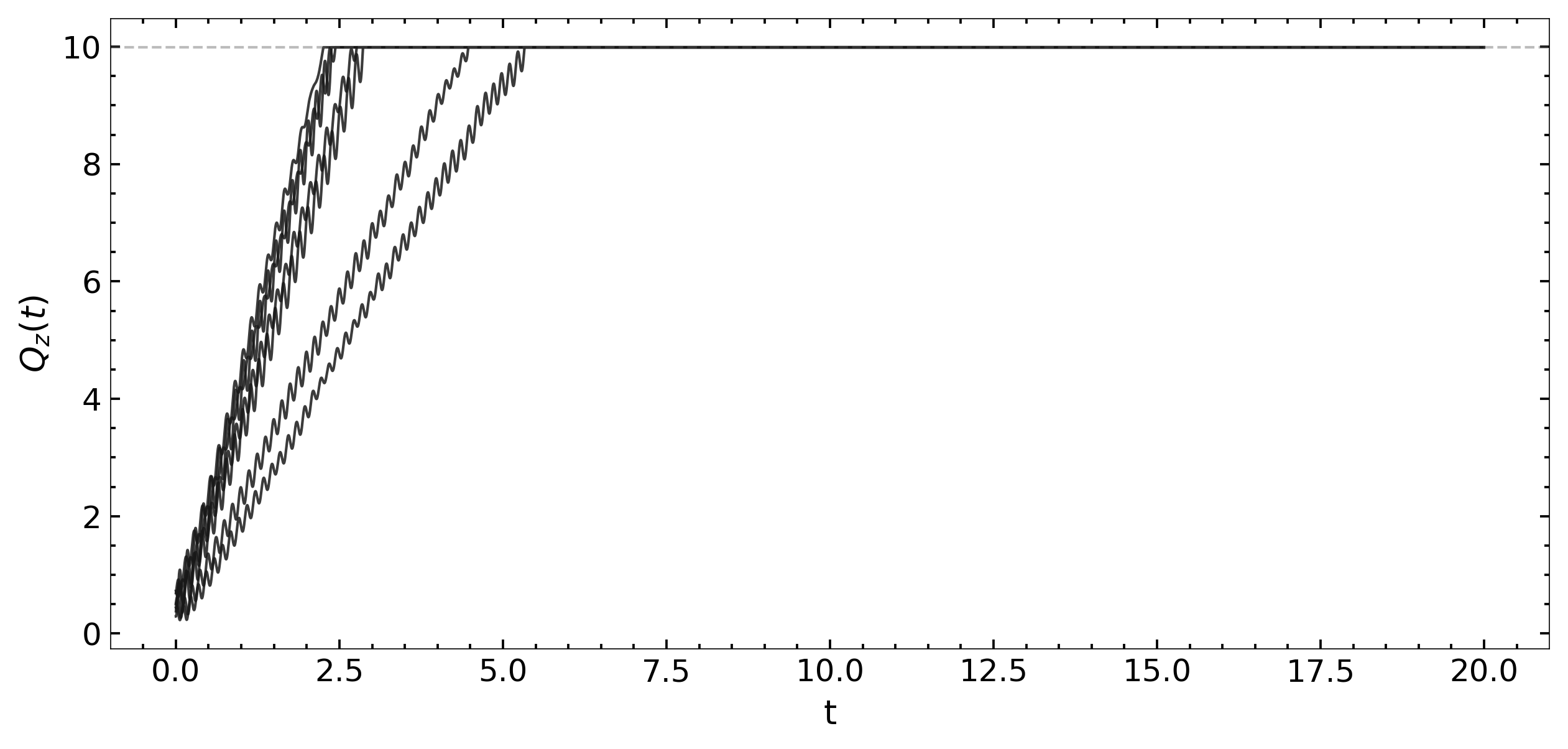}
    \par\vspace{0.3em}{\small (c) Spinless ABC}
  \end{minipage}

    \caption{Random sample of Bohmian trajectories for three setups with $\theta=\pi/3$, $\omega=100$.
    Shown is the $z$ coordinate as a function of $t$, dashed lines show $z=L$. Upon detection, the simulation keeps the position fixed; from that time onward, the trajectory appears in the diagram as horizontal.
    (a) and (b) with complex absorbing potentials in the region $z_0=10\leq z\leq L=11$, roughly similar to Figure~\ref{fig:Arrival_CAP}; (c) with ABC \eqref{nospinABC} at $z=L=10$.}
  \label{fig:Bohmian_CAP}
\end{figure}

The time dependence of the Bohmian motion is shown in Figure~\ref{fig:Bohmian_CAP} for the $z$ coordinate. In (a) and (b), late trajectories are seen to slow down before entering the detector region $[z_0,L]$, presumably due to partial reflection of $\Psi$. One can see in the figure that in different runs of the simulation, the particle gets detected at different times (and, in (a) and (b), different places).

\begin{figure}[H]
  \centering

  \begin{minipage}[c]{0.25\linewidth}
    \centering
    \includegraphics[width=\linewidth]{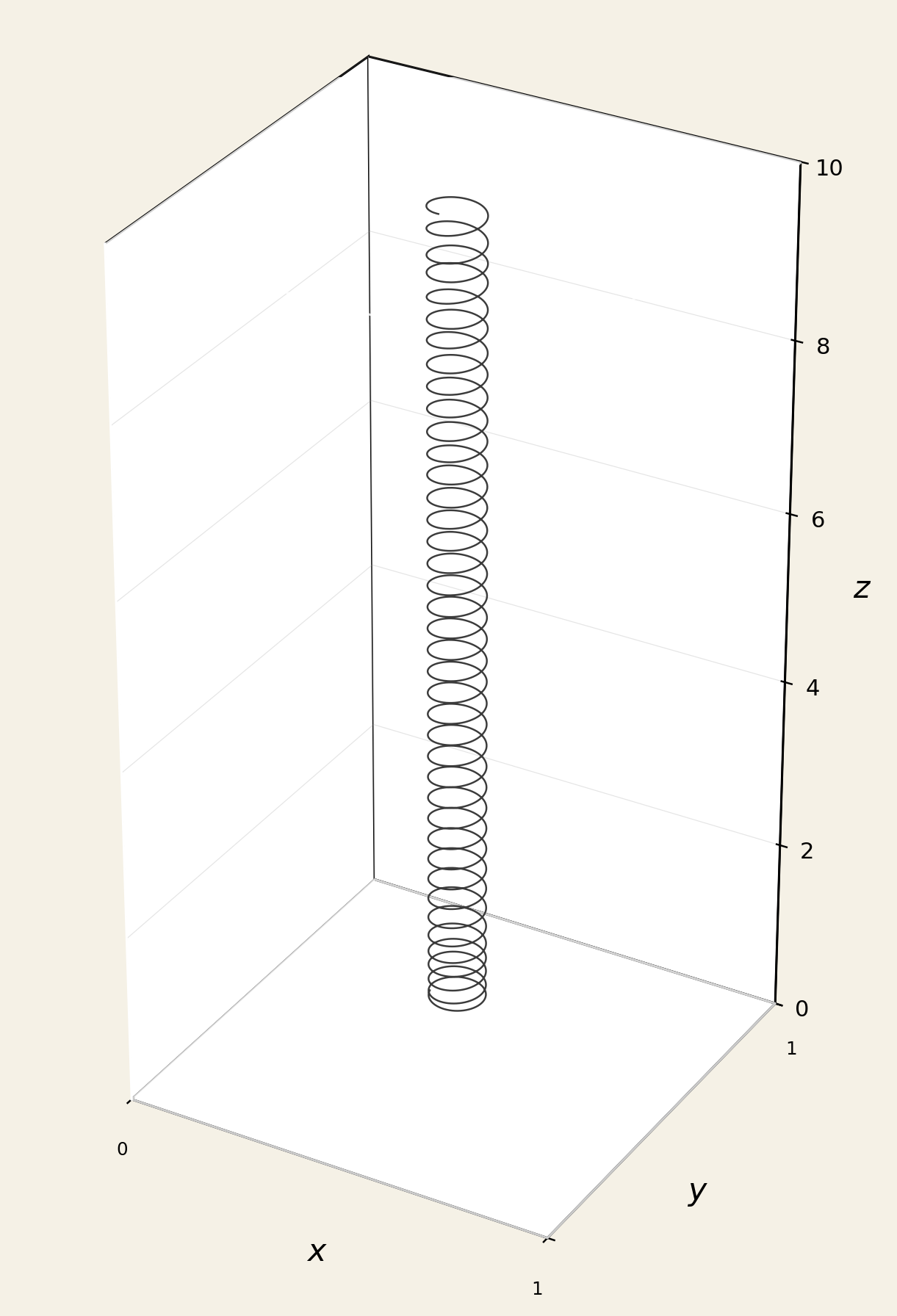}
  \end{minipage}\hfill
  \begin{minipage}[c]{0.65\linewidth}
    \centering
    \includegraphics[width=\linewidth]{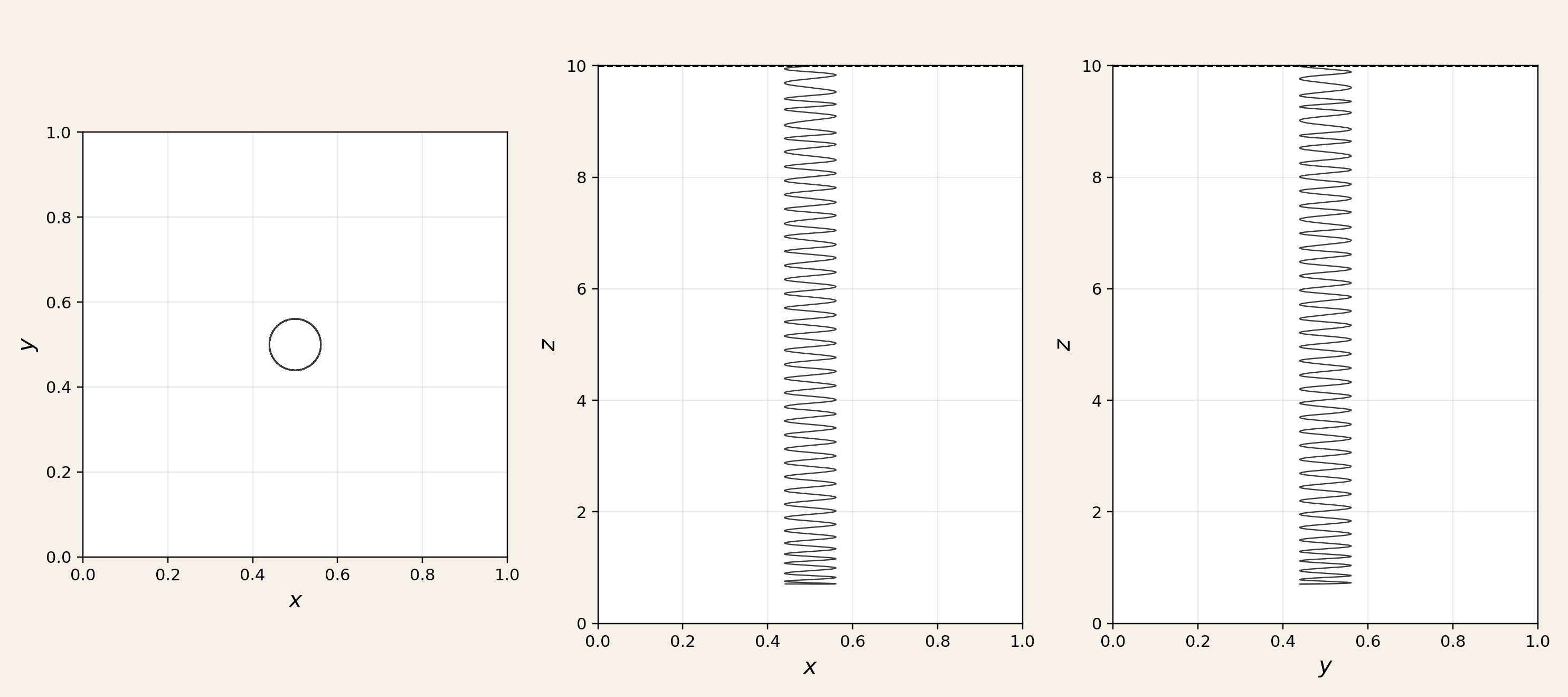}
  \end{minipage}

  \par\vspace{0.3em}
  \begin{minipage}[t]{0.25\linewidth}
    \centering
    {\small (a) $\theta=0$}
  \end{minipage}\hfill
  \begin{minipage}[t]{0.65\linewidth}
    \centering
    {\small (b) $\theta=0$}
  \end{minipage}

  \vspace{1em}

  \begin{minipage}[c]{0.25\linewidth}
    \centering
    \includegraphics[width=\linewidth]{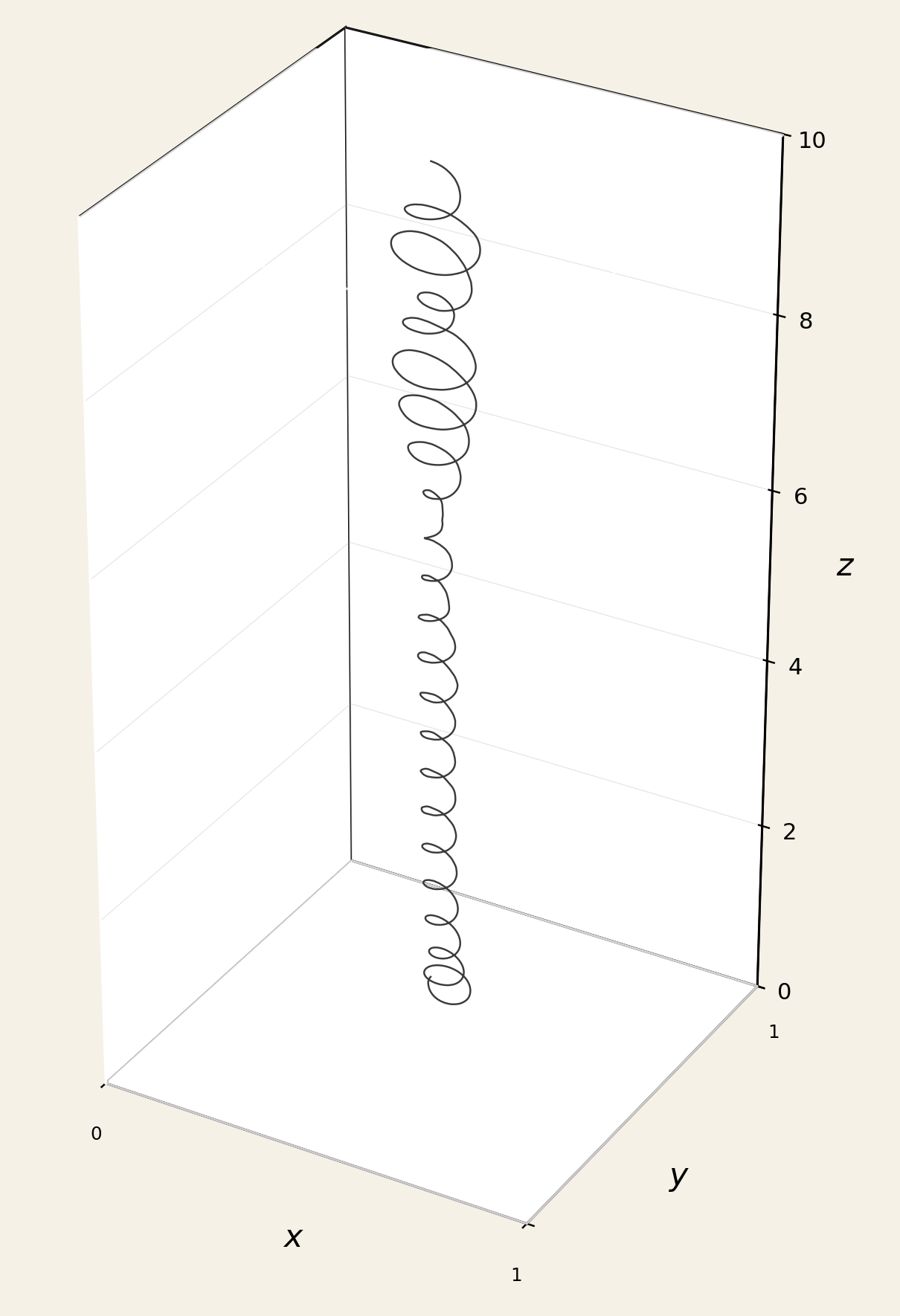}
  \end{minipage}\hfill
  \begin{minipage}[c]{0.65\linewidth}
    \centering
    \includegraphics[width=\linewidth]{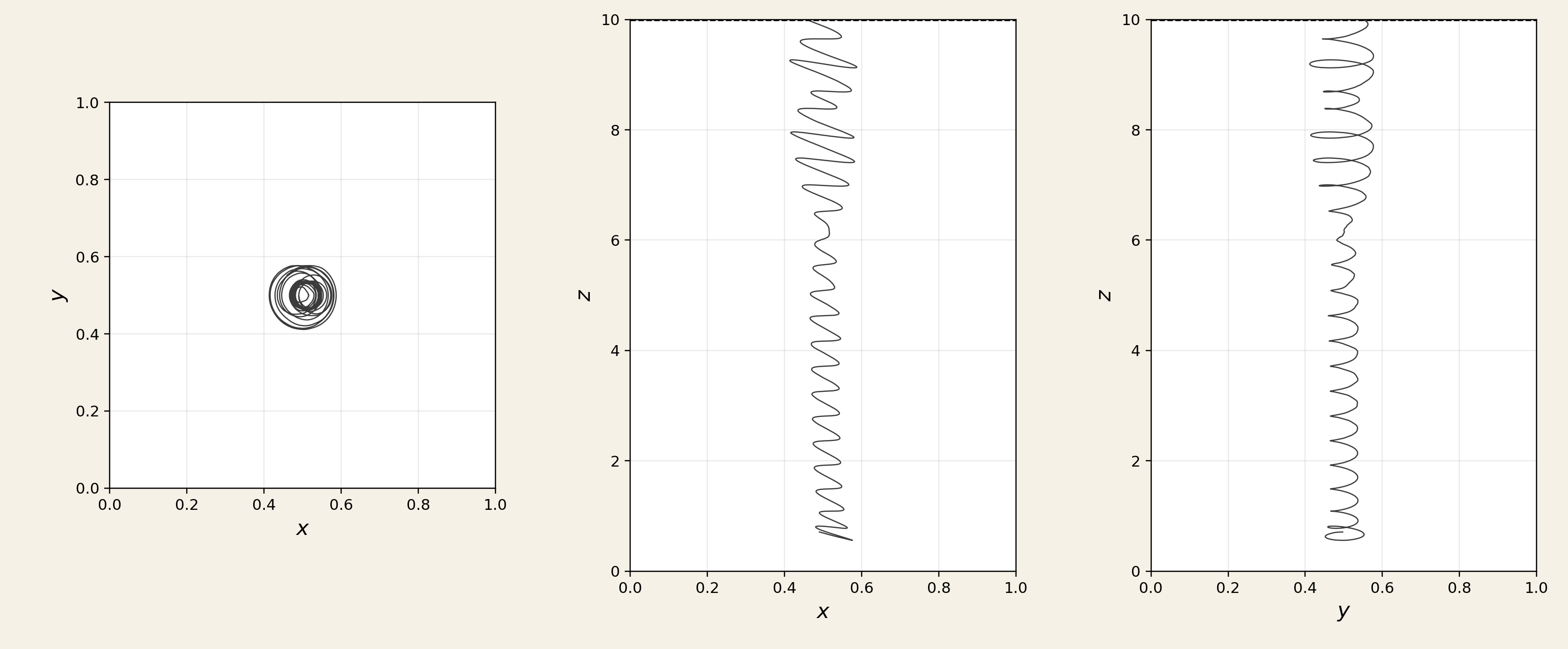}
  \end{minipage}

  \par\vspace{0.3em}
  \begin{minipage}[t]{0.25\linewidth}
    \centering
    {\small (c) $\theta=\pi/3$}
  \end{minipage}\hfill
  \begin{minipage}[t]{0.65\linewidth}
    \centering
    {\small (d) $\theta=\pi/3$}
  \end{minipage}

  \vspace{1em}

  \begin{minipage}[c]{0.25\linewidth}
    \centering
    \includegraphics[width=\linewidth]{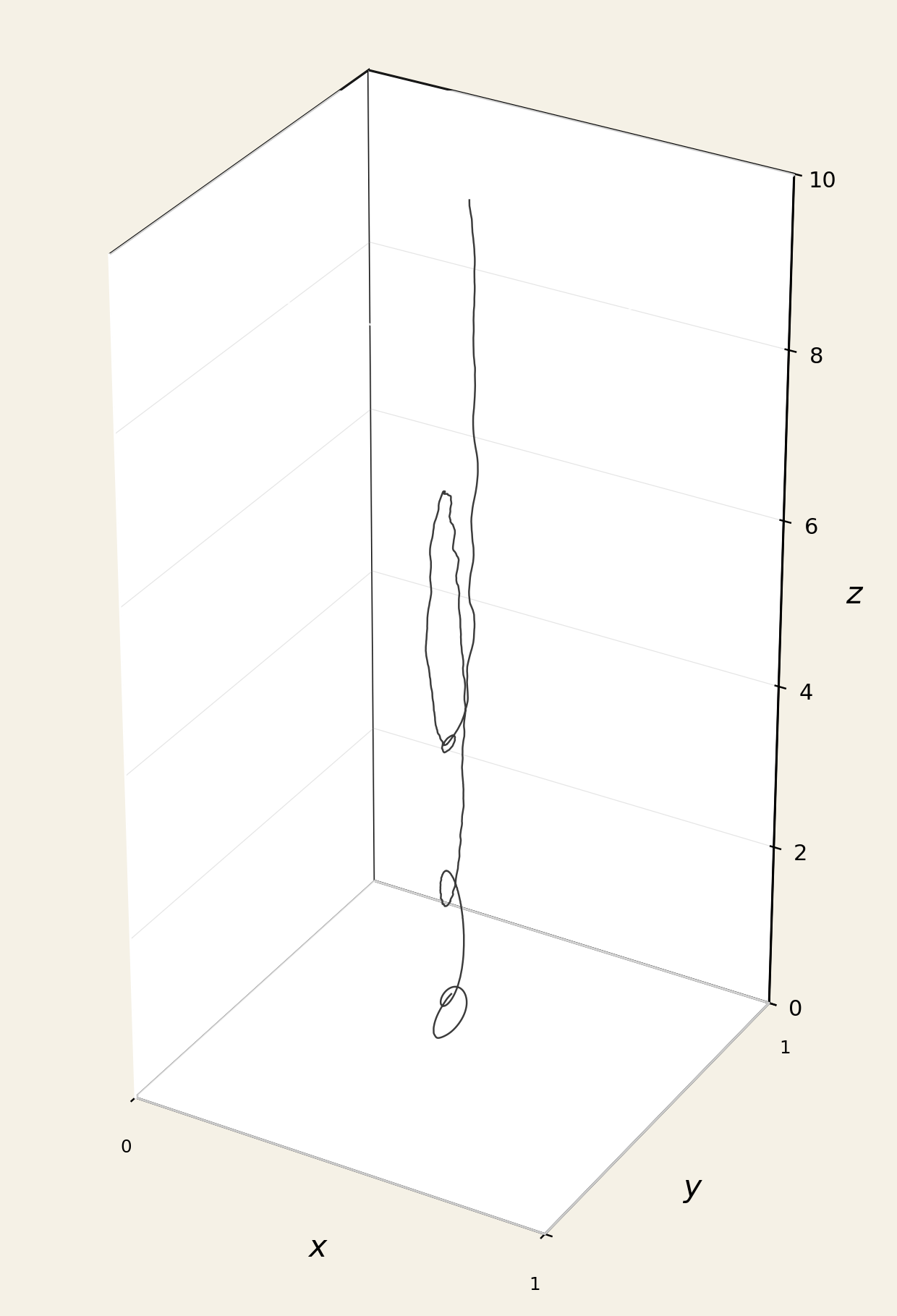}
  \end{minipage}\hfill
  \begin{minipage}[c]{0.65\linewidth}
    \centering
    \includegraphics[width=\linewidth]{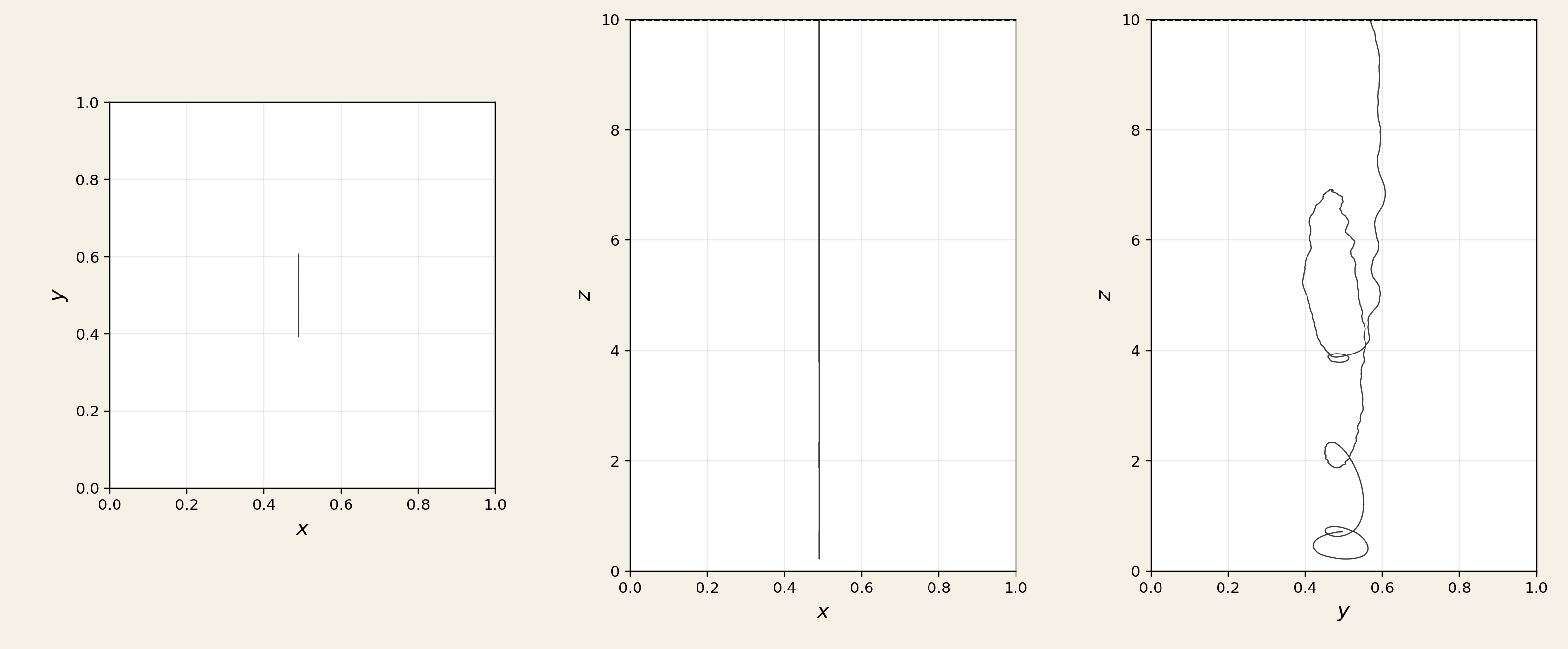}
  \end{minipage}

  \par\vspace{0.3em}
  \begin{minipage}[t]{0.25\linewidth}
    \centering
    {\small (e) $\theta=\pi/2$}
  \end{minipage}\hfill
  \begin{minipage}[t]{0.65\linewidth}
    \centering
    {\small (f) $\theta=\pi/2$}
  \end{minipage}

\caption{Three randomly selected Bohmian trajectories, here for the evolution with spinless ABC, $\kappa=\pi$, $\omega=100$, several choices of $\theta$. Pictures in the same row show the same trajectory, always in a 3d view on the left, and projected to the $xy$, $xz$, and $yz$ planes on the right.}
  \label{fig:traj_omega_theta}
\end{figure}

For a spin-1/2 particle, even when the Hamiltonian does not couple to the spin and the spinor $|\chi\rangle$ is just a constant factor in the wave function, the curl term in \eqref{Paulij} has the consequence that the Bohmian trajectory depends on $|\chi\rangle$. In agreement with \cite{DD19}, we observe that for $\theta=0$, the Bohmian particle orbits the $z$ axis in a circular manner while its $z$ coordinate is slowly increasing; for $\theta=\pi/2$, the particle moves along a loop in a vertical plane through the $z$ axis while the loop is slowly moving in the $z$ direction; and for intermediate $\theta$, the loop is tilted. Behavior of this type is visible in Figure~\ref{fig:traj_omega_theta}, as well as in the small-scale oscillations in Figure~\ref{fig:Bohmian_CAP}. Larger $\omega$ makes this spiraling motion narrower and faster.

\section{Variation of Parameters}
\label{app:variation}

We report here further results on how the distributions of $T$ and $\tau$ vary with the parameters of the setup. 

\begin{figure}[H]
  \centering
  \begin{minipage}[t]{0.48\linewidth}
    \centering
    \includegraphics[width=\linewidth]{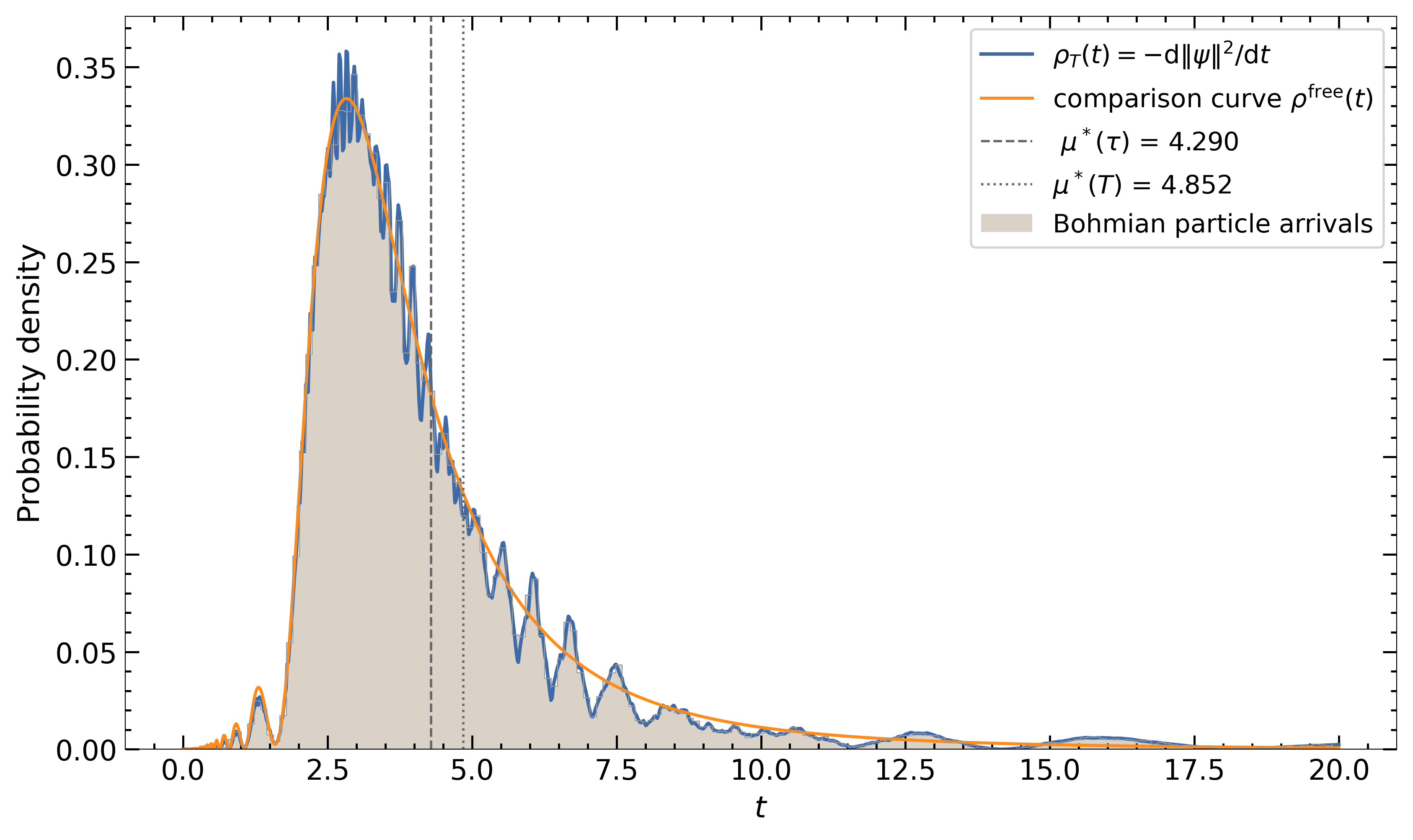}
    \par\vspace{0.3em}
    {\small (a) $\omega=1$. }
  \end{minipage}\hfill
  \begin{minipage}[t]{0.48\linewidth}
    \centering
    \includegraphics[width=\linewidth]{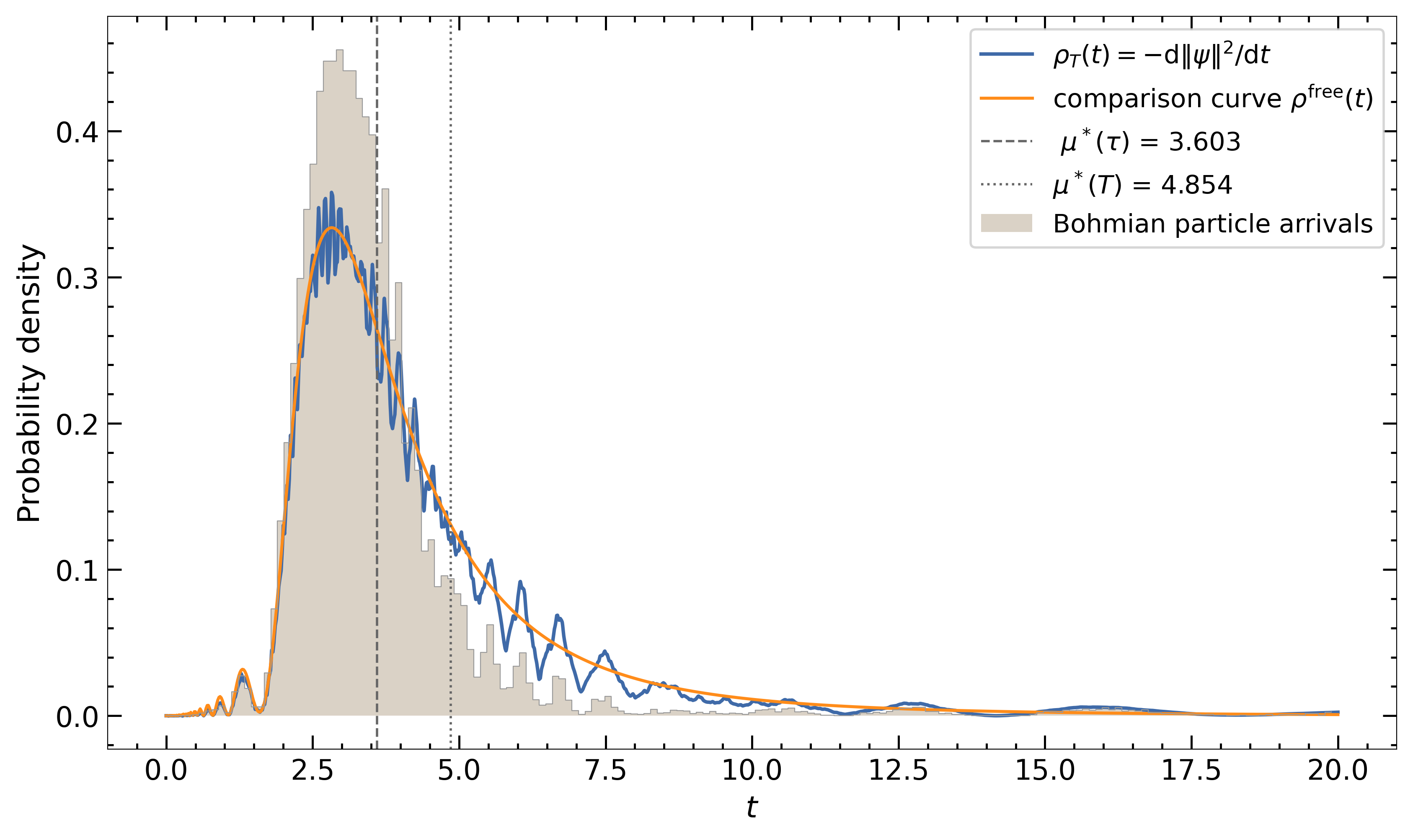}
    \par\vspace{0.3em}
    {\small (b) $\omega=100$}
  \end{minipage} 
    \par\vspace{0.8em}  
  \begin{minipage}[t]{0.48\linewidth}
    \centering
    \includegraphics[width=\linewidth]{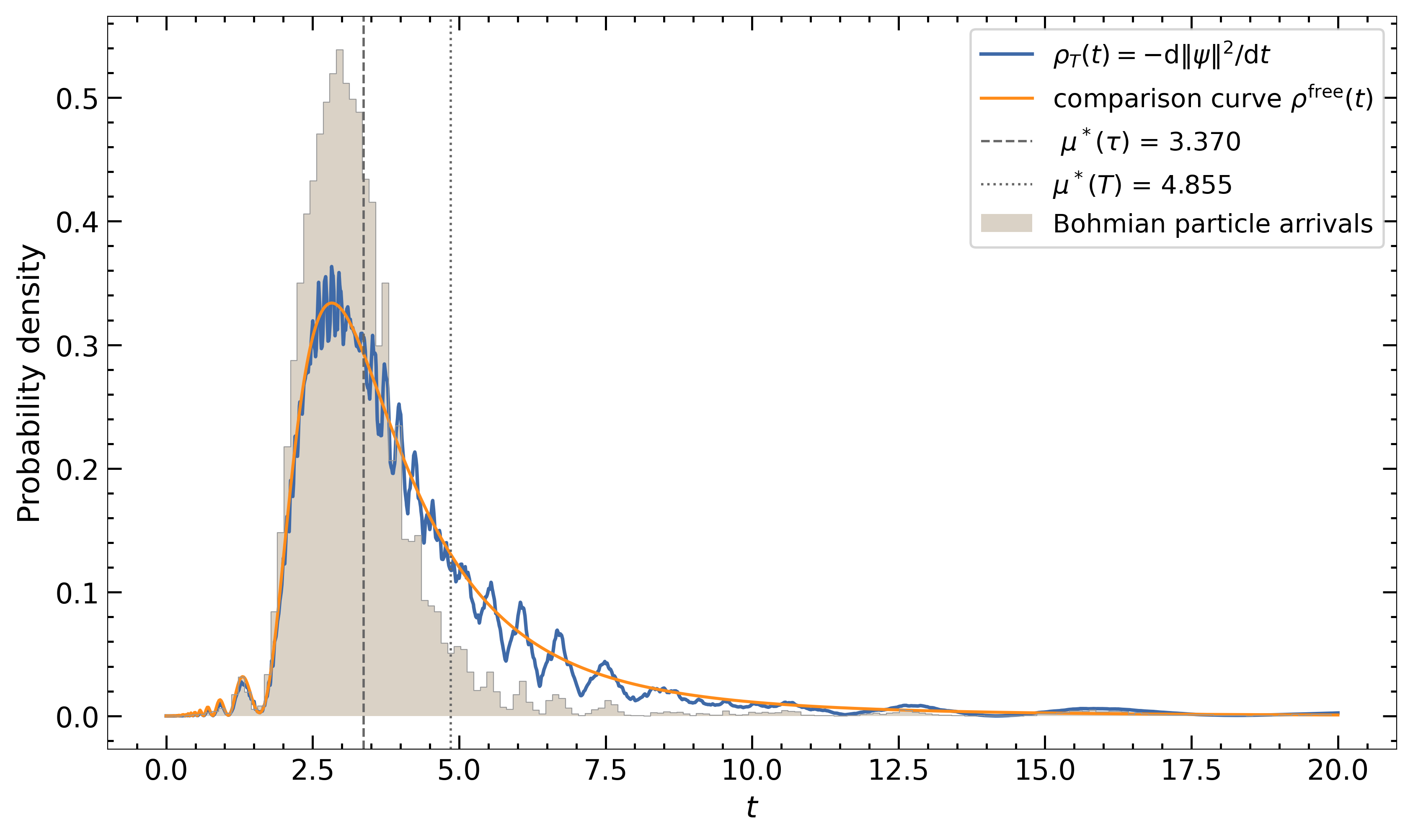}
    \par\vspace{0.3em}
    {\small (c) $\omega=200$ }
  \end{minipage} 
  \begin{minipage}[t]{0.48\linewidth}
    \centering
    \includegraphics[width=\linewidth]{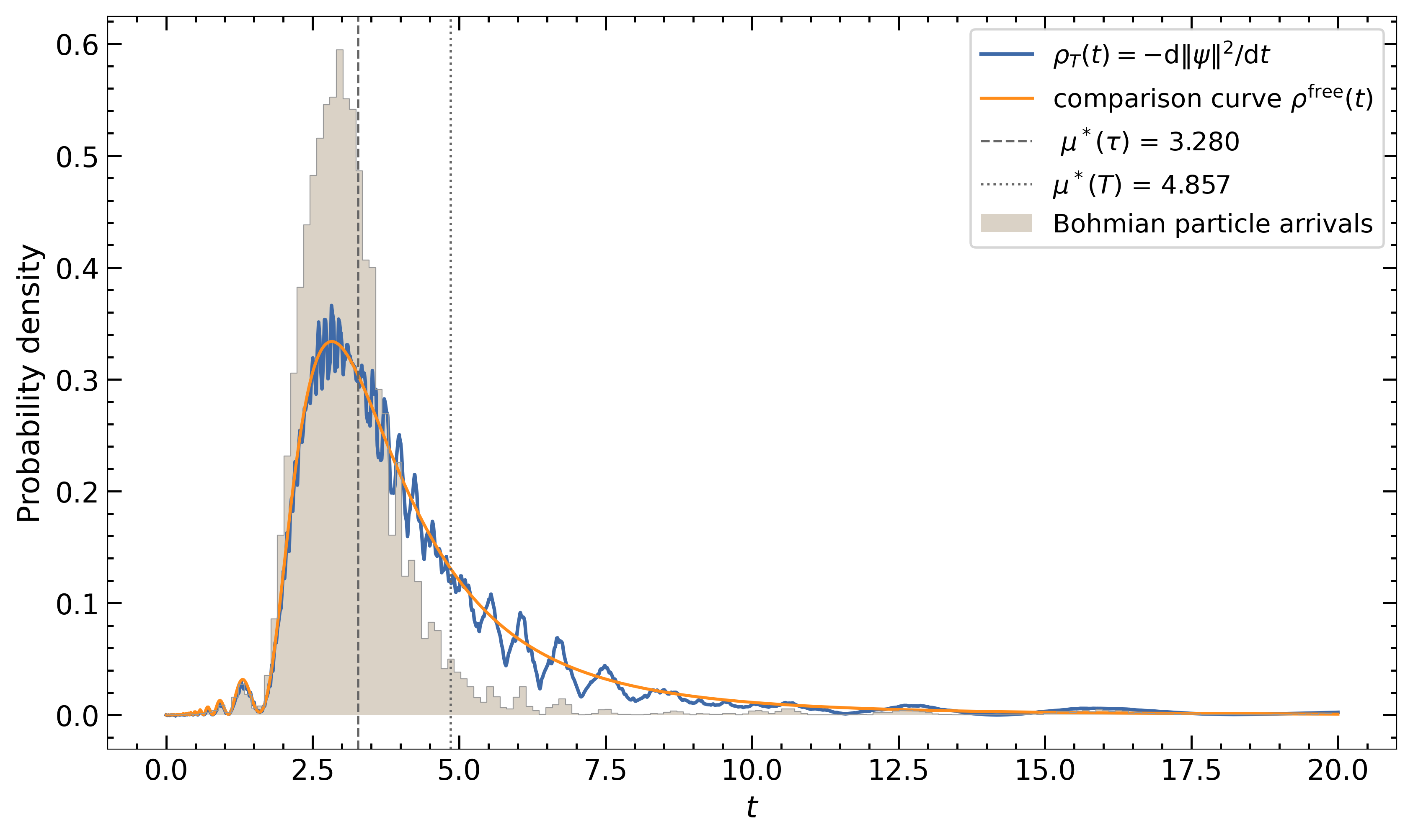}
    \par\vspace{0.3em}
    {\small (c) $\omega=300$ }
  \end{minipage}

  \caption{$\omega$ dependence of distribution (light brown) of arrival time $\tau=T_{WID}$ for $\vj^\mathrm{P}$ as described in Remark~\ref{rem:nospinABCjP} in Section~\ref{subsec:scalarABC}, using a spin-1/2 $\Psi$ subject to the spinless ABC \eqref{nospinABC}. Parameters are $L=10$, $\kappa=\pi$, $\theta=\pi/2$. Dark blue: distribution of $T=T_D$; orange: comparison curve $\rho^\fr(t)$.}
  \label{fig:arrival_hist3}
\end{figure}

\begin{figure}[H]
    \centering
    \includegraphics[width=0.52\linewidth]{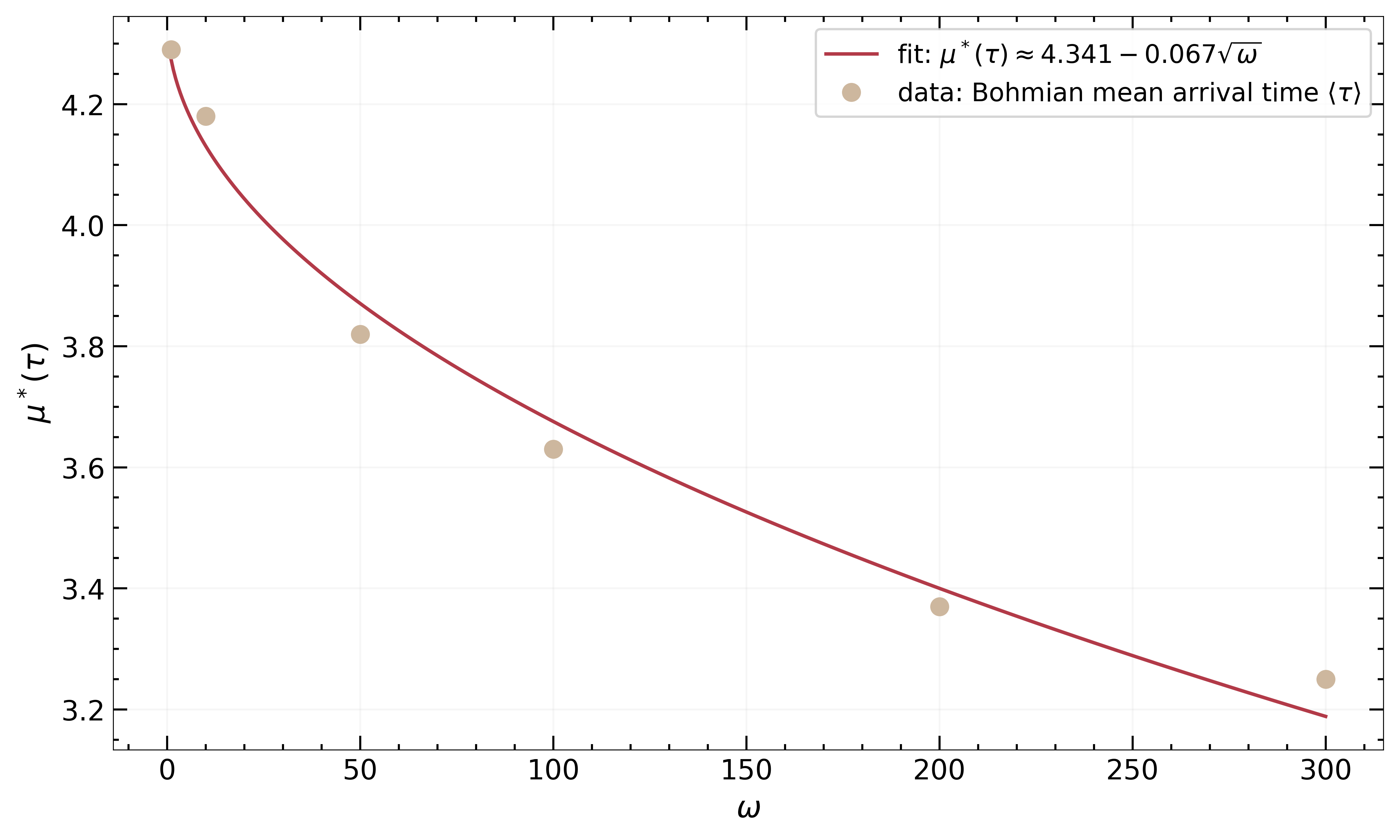}
    \caption{Mean values of $\tau$ (whose distribution is shown as light brown in Figure~\ref{fig:arrival_hist3}, always $\theta=\pi/2$) as a function of $\omega$.}
    \label{fig:Spinless_fit}
\end{figure}

\paragraph{$\omega$ dependence.}Figure~\ref{fig:arrival_hist3} shows, across several plots, the dependence of the distribution of $\tau$ (light brown) on $\omega$, as described in Remark~\ref{rem:nospinABCjP} in Section~\ref{subsec:scalarABC}, while the distribution of $T$ (dark blue) shows no dependence (as already derived theoretically). It is visible that $\tau$ values tend to be smaller than $T$ values, and that this tendency is more pronounced with larger $\omega$ values. Correspondingly, also the mean value of $\tau$ decreases with $\omega$, in fact the numerical data suggest that the dependence is of the form 
\be
\EEE\tau = A_0 -B_0 \sqrt{\omega}
\ee
with positive constants $A_0,B_0$ for not-too-large $\omega$, see Figure~\ref{fig:Spinless_fit} (where we used $\mu^*(\tau)$ as an approximation for $\EEE\tau$). This dependence gets investigated in a separate work \cite{JT26b}.

\begin{figure}[H]
    \centering
     \includegraphics[width=0.6\linewidth]{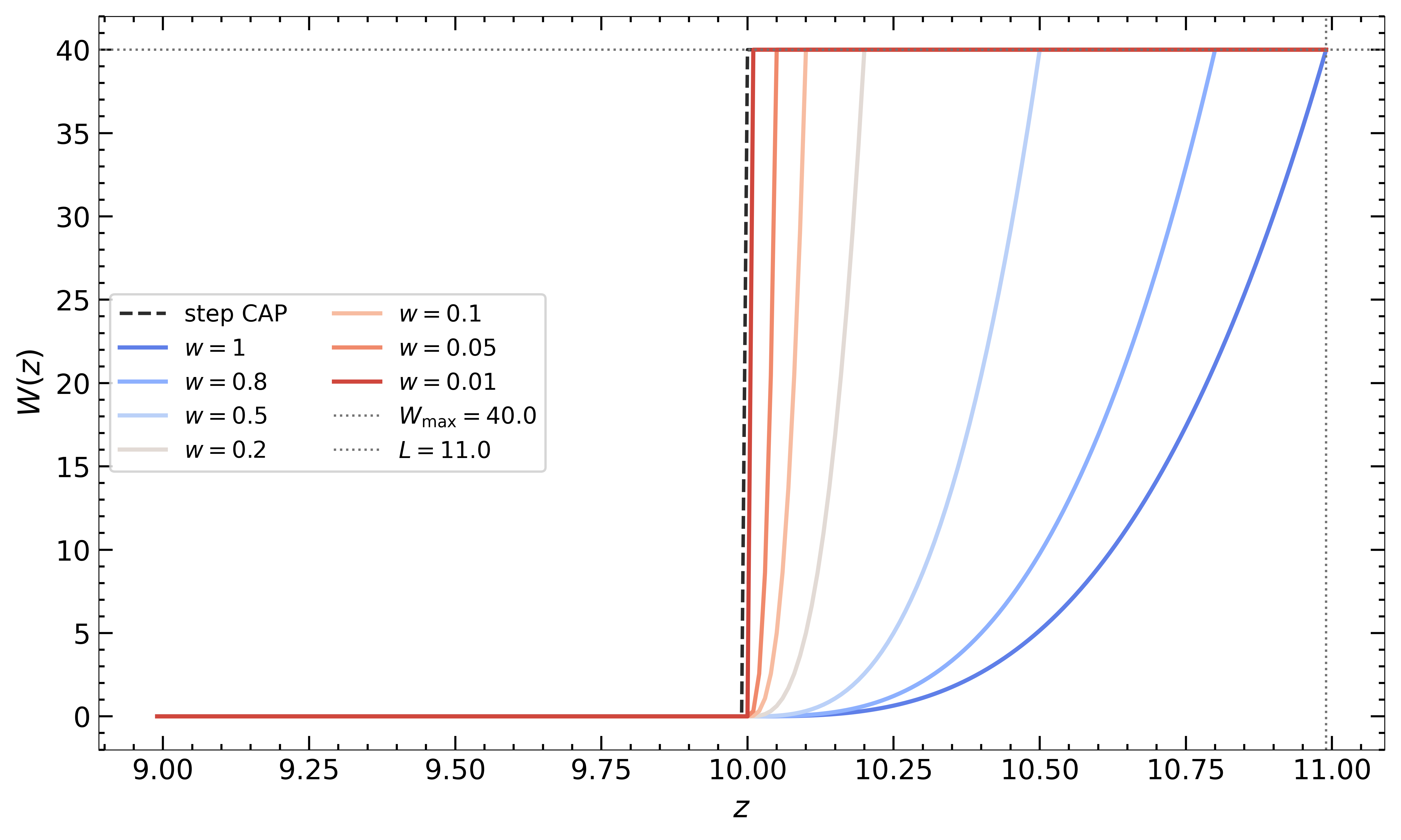}
        
        \caption{Shape of CAP profile $W(z)$ used for studying dependence on the sharpness of the increase: a cubic ramp as given by \eqref{Wcubic} for different step widths $w$, corresponding to different degrees of approximation to the sharp (discontinuous) step \eqref{Wsharp} (black dashed curve).}
        \label{fig:ramp_cap_profile}
\end{figure}

\paragraph{Sharpness of increase in CAP.}
The discontinuous step \eqref{Wsharp} in the profile $W(z)$ of the CAP leads to strong reflection; here we investigate how that varies when the sharpness of the increase in $W$ is changed. Instead of the tanh profile \eqref{Wtanh} used in Figure~\ref{fig:Arrival_CAP}, we consider in this appendix a polynomial ramp CAP localized in the terminal part of the guide,
\be\label{Wcubic}
W(z)=\max\Bigl\{ \Bigl( \frac{z-z_0}{(L-z_0)w} \Bigr)^3 ,1\Bigr\}\, 1_{z>z_0} W_{\max} = \biggl[\operatorname{clip}\!\left(\frac{z-z_0}{(L-z_0)w},\,0,\,1\right) \biggr]^3 W_{\max}
\ee
(with notation clip$(x,a,b):=\max\{a,\min\{x,b\}\}$), which increases continuously from 0 to its maximal strength $W_{\max}$ over an interval of length $w$ (width of the potential step in units of $(L-z_0)$) as shown in Figure~\ref{fig:ramp_cap_profile}.

\begin{figure}[H]
    \centering

    \begin{subfigure}{0.32\textwidth}
        \centering
        \includegraphics[width=\linewidth]{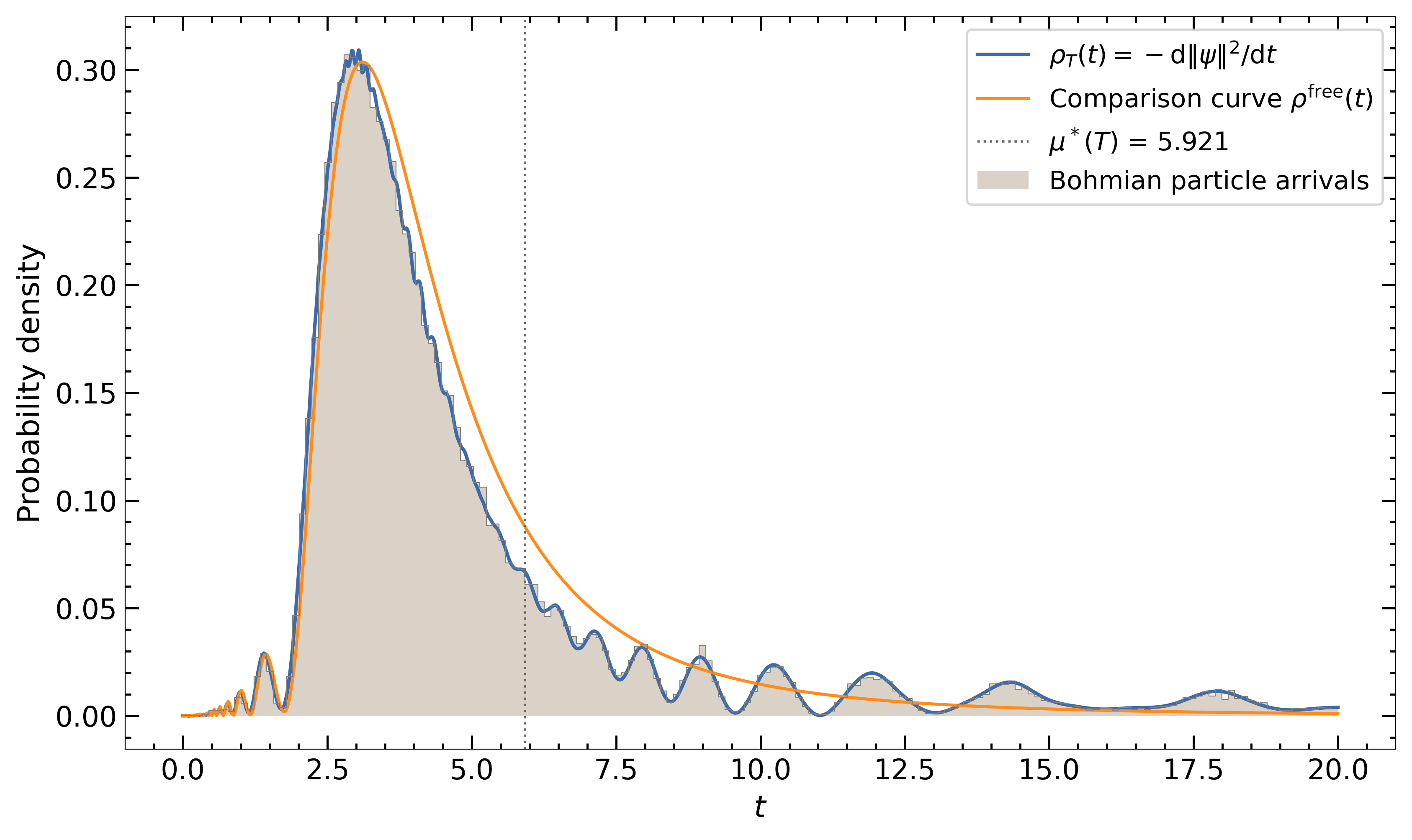}
        \caption{\(w=1\), detection fraction = 93\%}
        \label{fig:ramp_at_w1}
    \end{subfigure}
    \hfill
    \begin{subfigure}{0.32\textwidth}
        \centering
        \includegraphics[width=\linewidth]{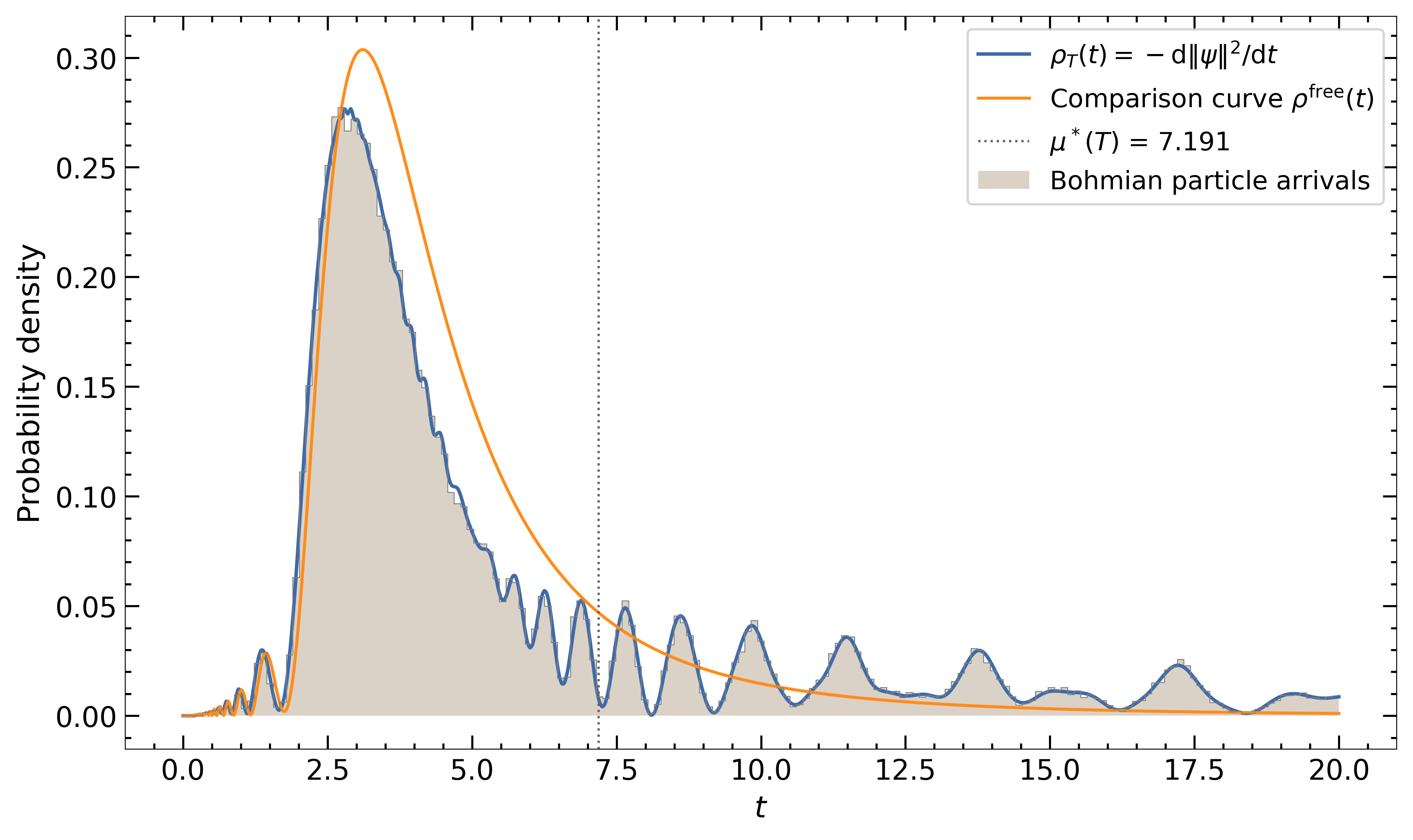}
        \caption{\(w=0.5\), detection fraction = 88\%}
        \label{fig:ramp_at_w05}
    \end{subfigure}
    \hfill
    \begin{subfigure}{0.32\textwidth}
        \centering
        \includegraphics[width=\linewidth]{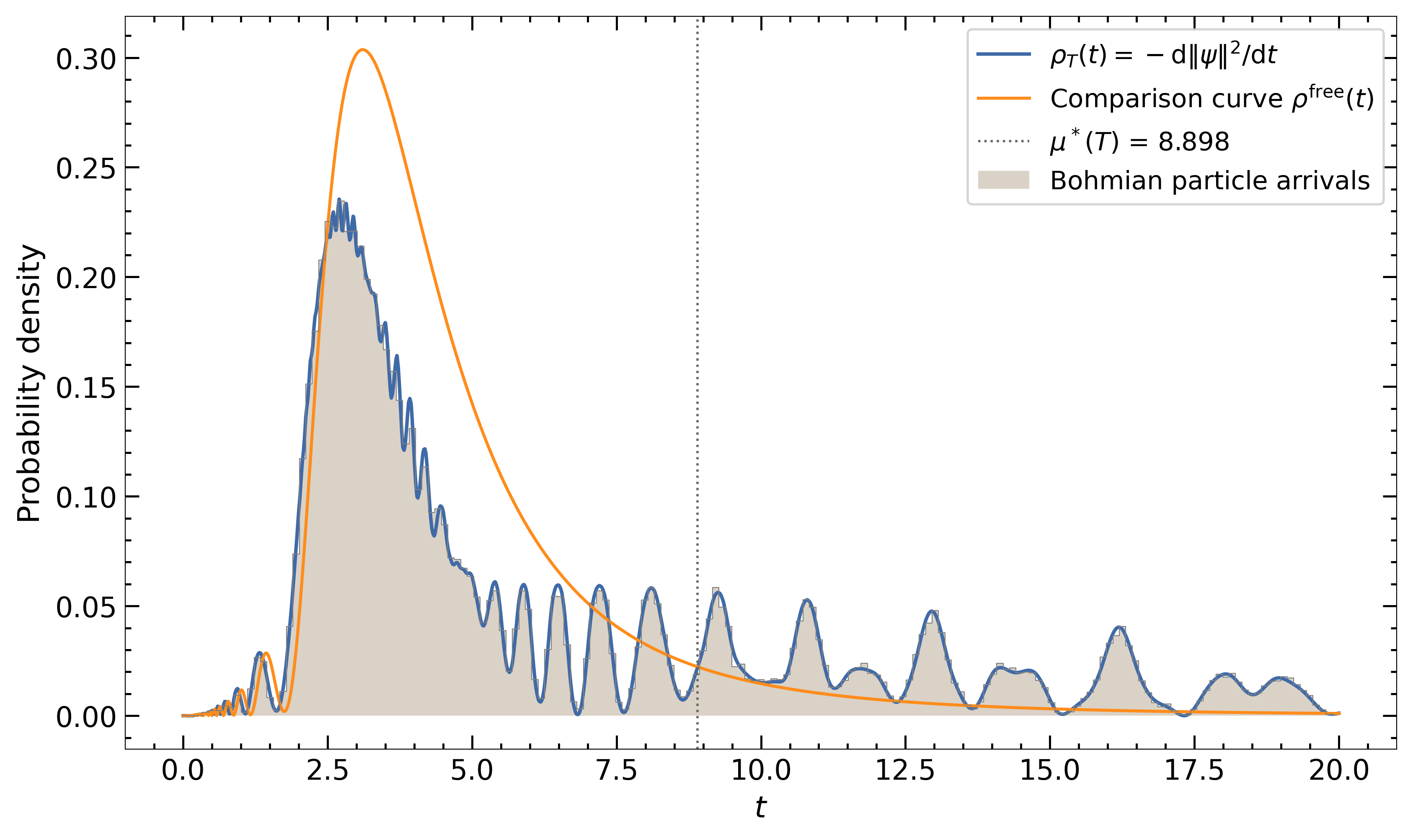}
        \caption{\(w=0.01\), detection fraction = 81\%}
        \label{fig:ramp_at_w001}
    \end{subfigure}

    \caption{Dependence of $\rho_T(t)$ (dark blue curve) on the sharpness of increase in the CAP controlled by the parameter $w$ in the cubic ramp CAP \eqref{Wcubic}.}
    \label{fig:ramp_cap_appendix1}
\end{figure}

The parameter $w$ serves for controlling the sharpness of increase. Distributions of $T$ for several values of $w$ are shown in Figure~\ref{fig:ramp_cap_appendix1}.

\begin{figure}[H]
    \centering
    \includegraphics[width=0.52\linewidth]{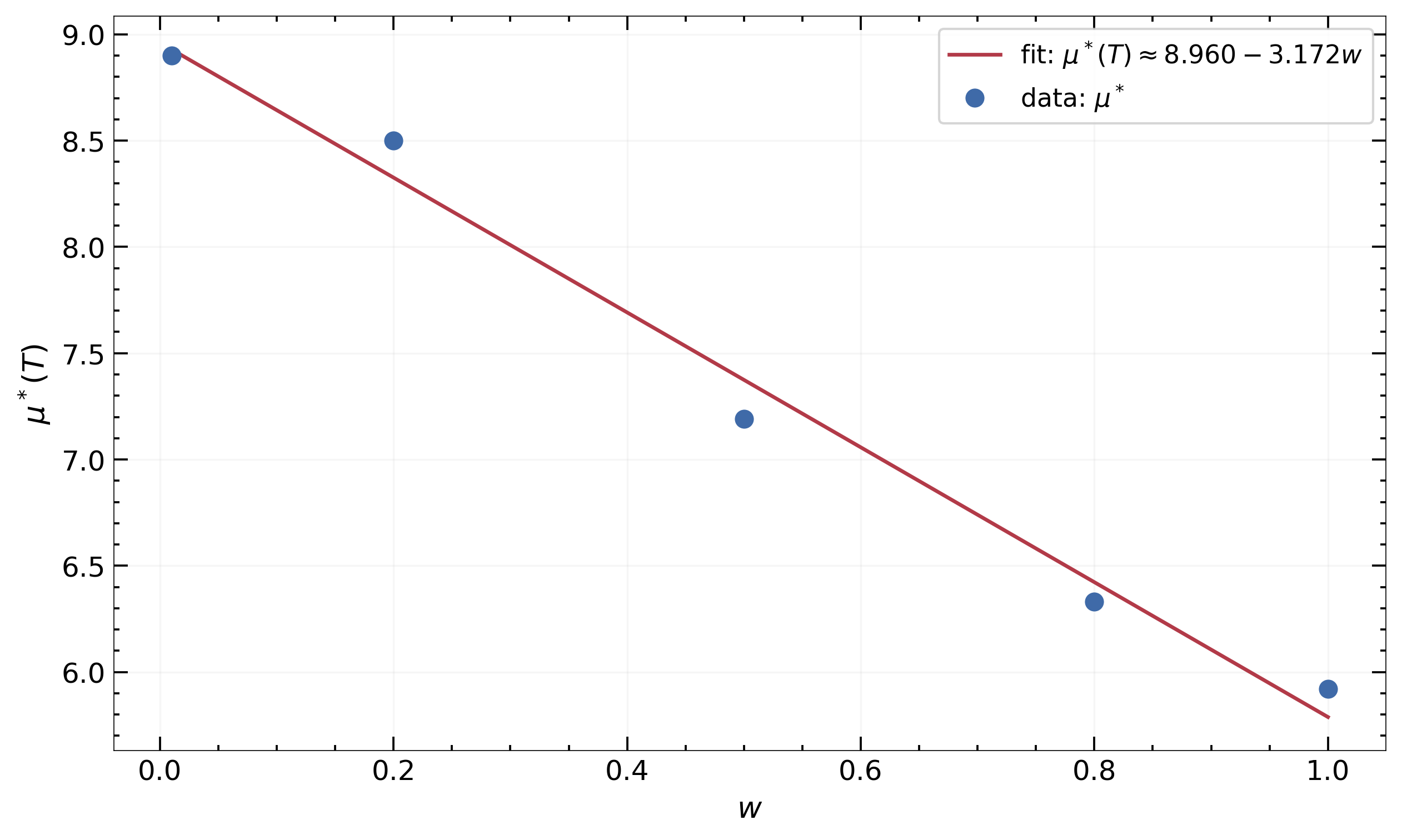}
    \caption{Restricted mean arrival time $\mu^*(T)$ \eqref{mudef} as a function of the width parameter $w$ over the numerically explored range from very smooth to very sharp ramps. The points show the numerical data, and the line is a linear fit.}
    \label{fig:cap_fit}
\end{figure}

As the ramp is sharpened, the late-time oscillatory tail becomes more pronounced, the detected fraction within the fixed observation window decreases, and the restricted mean arrival time $\mu^*$ \eqref{mudef} shifts to later times, see Figure~\ref{fig:cap_fit}.

\begin{figure}[H]
    \centering
    \begin{subfigure}{0.48\textwidth}
        \centering
        \includegraphics[width=\linewidth]{Fig19a.png}
        \caption{$z_0 = 10 \le z \le L = 11$}
        \label{fig:ramp_at_W1}
    \end{subfigure}
    \begin{subfigure}{0.48\textwidth}
        \centering
        \includegraphics[width=\linewidth]{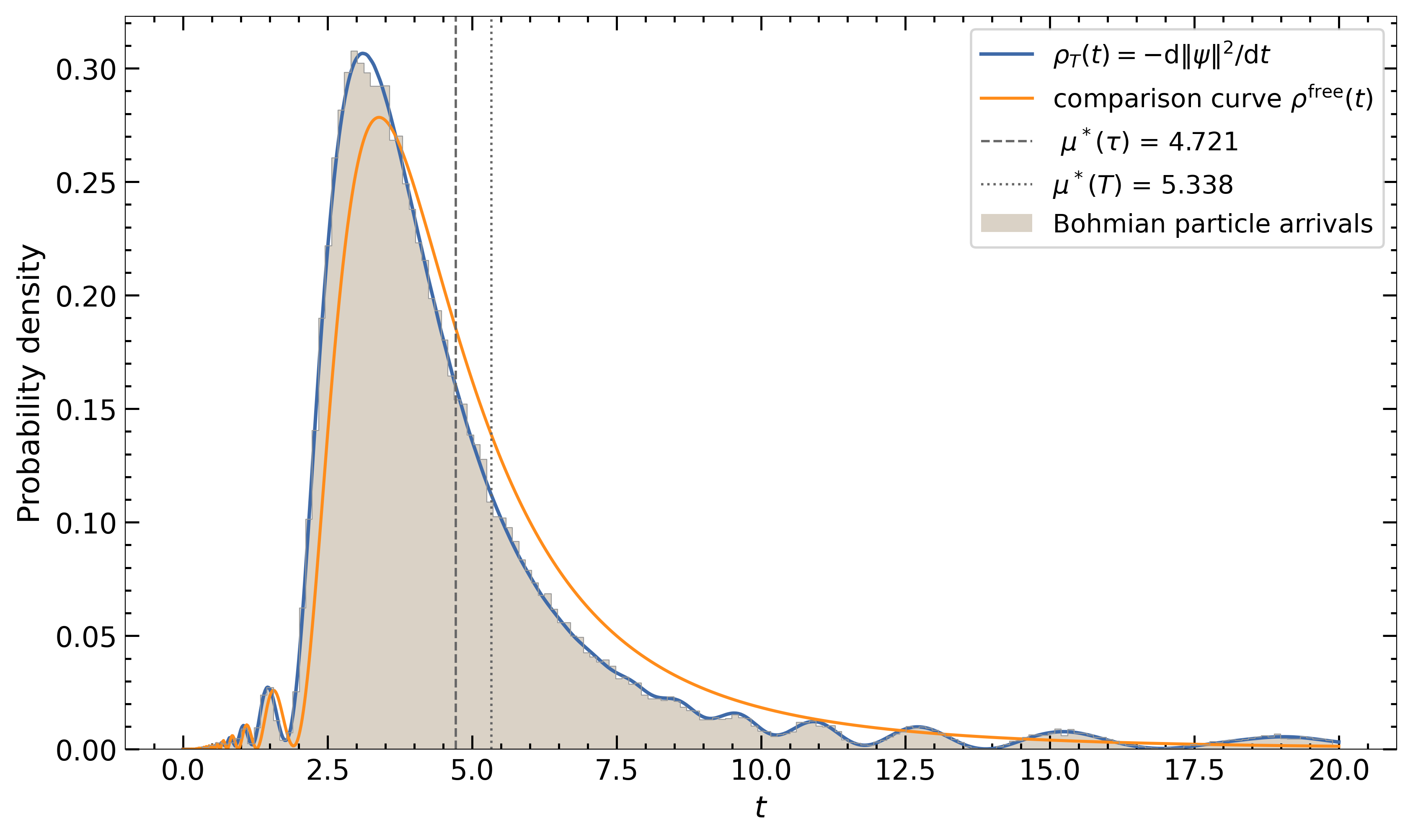}
        \caption{$z_0 = 10 \le z \le L = 12$}
        \label{fig:ramp_at_W2}
    \end{subfigure}
    \caption{Dependence of $\rho_T(t)$ on the size of the region with CAP; both plots use \eqref{Wcubic} with $w=1$.}
    \label{fig:ramp_cap_L}
\end{figure}

Furthermore, we observed in our numerical experiments that, as the CAP region is enlarged---for example from $z_0 = 10 \le z \le L_z = 11$ to $z_0 = 10 \le z \le L_z = 12$, and so on---the late-time reflected tail is progressively reduced, as shown in Figure~\ref{fig:ramp_cap_L}.

\begin{figure}[H]
  \centering

  \begin{minipage}[t]{0.48\linewidth}
    \centering
    \includegraphics[width=\linewidth]{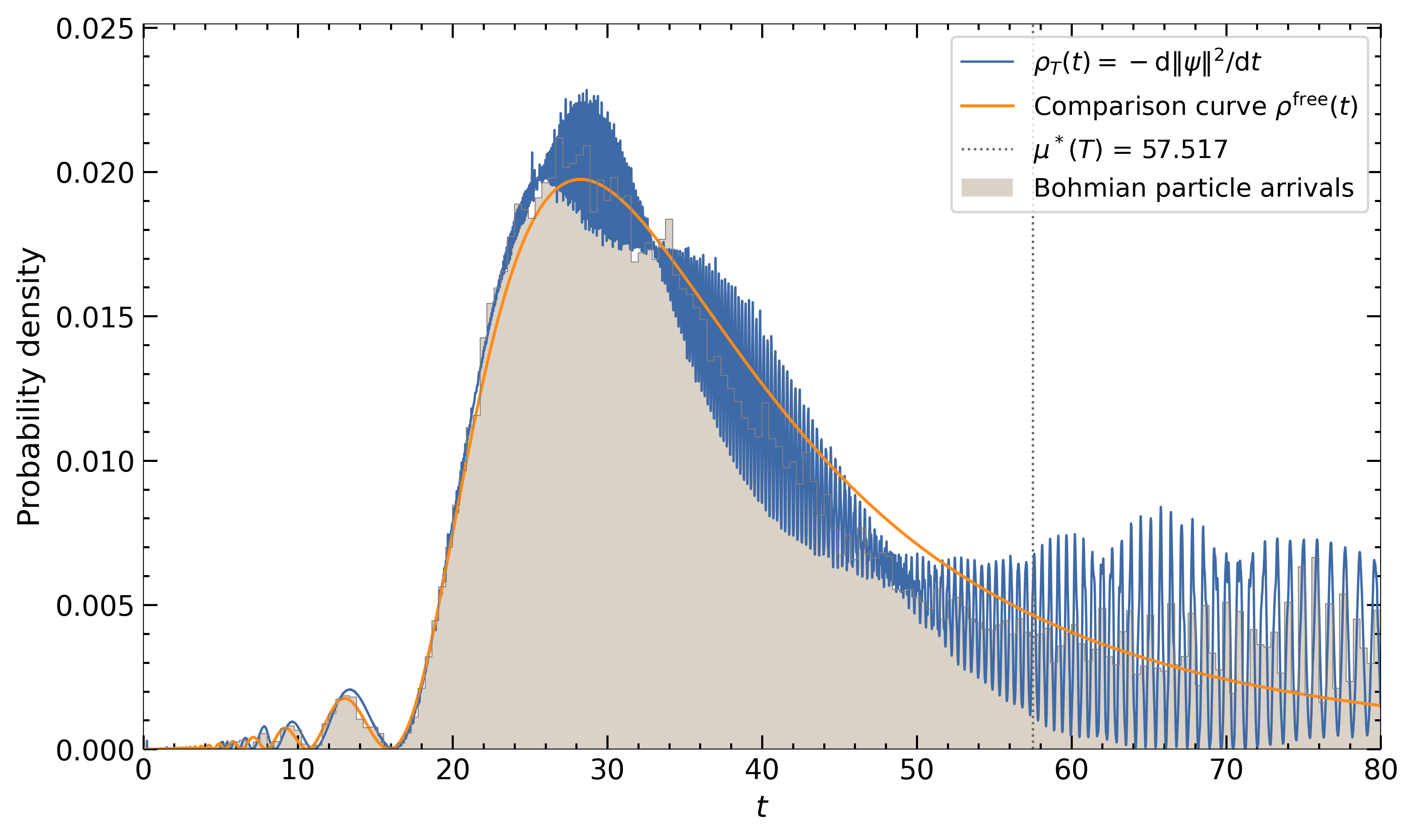}
    \par\vspace{0.3em}{\small (a) $\omega = 100$. detection fraction = 54\%}
  \end{minipage}\hfill
  \begin{minipage}[t]{0.48\linewidth}
    \centering
    \includegraphics[width=\linewidth]{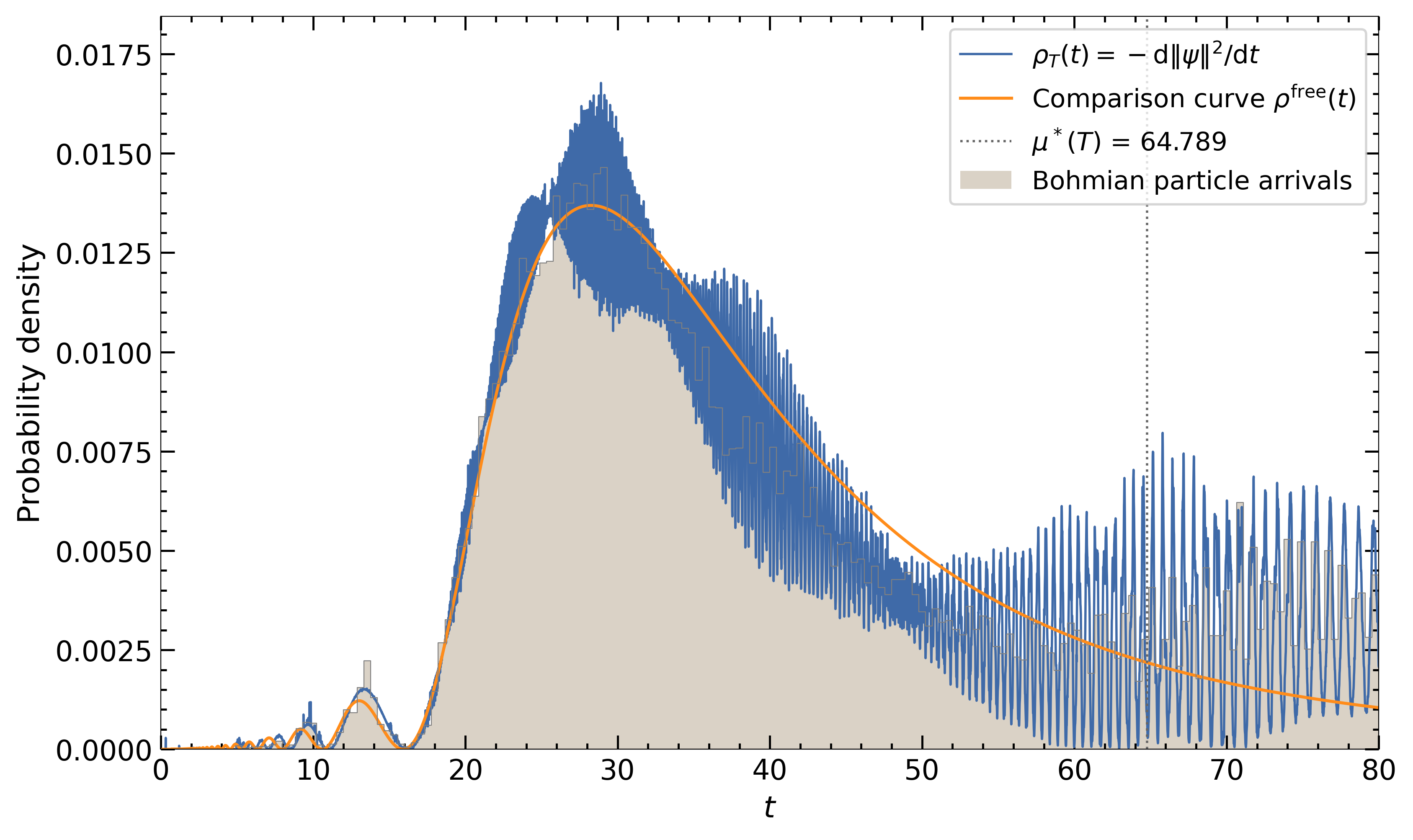}
    \par\vspace{0.3em}
    {\small (b) $\omega=500$. detection fraction = 38\%}
  \end{minipage}
  \caption{Distributions for spinor ABC with $L=100$. For both plots, parameters are $\theta=\pi/2$, $N_z=2000$.}
  \label{fig:LargeL}
\end{figure}

\paragraph{Large $L$.} We report further plots for a larger value $L=100$ and large values of $\omega$ that show the emergence of further patterns: variations of the strength of oscillations in $\rho_T(t)$, as well as secondary maxima or inflection points in the curve at the center of the oscillations, see Figure~\ref{fig:LargeL}.

\bigskip

\paragraph{Declarations.} The authors declare no conflicts of interest.  Artificial intelligence (Open\-AI) has been used for help with creating and improving the computer code for the simulations but not otherwise. 

\paragraph{Acknowledgments.}  The authors acknowledge support by the state of Baden-W\"urttem\-berg through bwHPC and the German Research Foundation (DFG) through grant INST 35/1597-1 FUGG. 
Moreover, we thank Yoann Le H\'enaff, Navid Javid, and Christian Gro\ss\ for helpful discussions.

\end{document}